\documentclass{article}
\usepackage[utf8]{inputenc}
\usepackage{graphicx}
\usepackage{float}
\usepackage[usenames]{color}

\title{Lepton flavor violation in the Littlest Higgs Model with T parity realizing an inverse seesaw}
\author{Iván Pacheco and Pablo Roig\\
Departamento de Física, Centro de Investigación y de Estudios Avanzados\\ del Instituto Politécnico Nacional\\
Apartado Postal 14-740, 07000 Ciudad de México,  México}
\date{}

\begin{document}

\maketitle

\begin{abstract}
We study lepton flavor violation (LFV) within the Littlest Higgs Model with T parity (LHT) realizing an inverse seesaw (ISS) mechanism of type I. With respect to the traditional LHT, there appear new $\mathcal{O}$(\textcolor{black}{10} TeV) Majorana neutrinos,  driving LFV. For $\tau\to\ell\ell'\bar{\ell}''$ (including wrong-sign, $\ell=e, \mu$) decays and $\mu\to e$ conversion in Ti, we get typical rates only one order of magnitude below present bounds ($\ell\to\ell'\gamma$ can reach the current upper limit) and for $Z\to\bar{\tau}\ell$, $\mu\to e e \bar{e}$ and conversion in Au, results are within two orders of magnitude from present limits. Correlations among modes are drastically different to the traditional LHT and other models, which would ease the confrontation of this scenario to eventual measurements of LFV processes involving charged leptons.
\end{abstract}

\section{Introduction}\label{sec:Intro}
\hspace{0.35cm} The long-awaited discovery of the Higgs boson \cite{ATLAS:2012yve, CMS:2012qbp} was the final milestone confirming \cite{ATLAS:2016neq, CMS:2018uag,ATLAS:2019nkf} the Standard Model (SM) of the electroweak interactions \cite{Glashow:1961tr, Weinberg:1967tq, Salam:1968rm}. Composite Higgs models \cite{Arkani-Hamed:2002ikv,Schmaltz:2005ky,Perelstein:2005ka, Panico:2015jxa} are among the most attractive candidates
to solve the corresponding hierarchy problem, associated to the Higgs mass value and its stability against quantum corrections in presence of heavy new physics coupling to the Higgs proportionally to their masses. In this
set of models, the Higgs boson is a pseudo-Nambu-Goldstone boson (pNGB)
of a spontaneously broken global symmetry. Specifically, the Littlest Higgs model with T parity (LHT) \cite{Arkani-Hamed:2002ikv,Arkani-Hamed:2001kyx, Arkani-Hamed:2001nha,  Cheng:2003ju, Cheng:2004yc, Low:2004xc, Cheng:2005as} is one of the most attractive such frameworks. LHT is based upon the spontaneous collective  breaking of a global symmetry group $SU(5)$ down to $SO(5)$, by a vacuum expectation value at a scale of few TeV. The discrete T parity symmetry is possible since the coset space $SU(5)/SO(5)$ is invariant under it. This forbids singly-produced heavy particles (odd under T) and tree level corrections to observables with only SM particles. As a result, direct and indirect constraints on the LHT are significantly relaxed \cite{Hubisz:2004ft, Hubisz:2005tx}. Thus, the LHT remains phenomenologically viable and well-motivated \cite{Hubisz:2005bd, Chen:2006cs,Blanke:2006sb, Buras:2006wk, Belyaev:2006jh,Blanke:2006eb, Blanke:2007db, Hill:2007zv, Goto:2008fj,delAguila:2008zu,Blanke:2009am,delAguila:2010nv,Goto:2010sn,Han:2013ic,Yang:2013lpa,Yang:2014mba,Blanke:2015wba,Yang:2016hrh, delAguila:2017ugt,Dercks:2018hgz,delAguila:2019htj,DelAguila:2019xec,Illana:2021uwu}.

Understanding the tiny values of the neutrino masses is another puzzle that constitutes a very active area of research, as they are the first manifestation of beyond the SM physics.  Within the LHT, it was shown recently \cite{DelAguila:2019xec} that the inverse see-saw of type I \cite{Mohapatra:1986aw, Mohapatra:1986bd, Bernabeu:1987gr} is able to reproduce current data preserving the T symmetry. It is well known that the heavy Majorana masses thus introduced (in the \textcolor{black}{10} TeV scale) impact lepton flavor violating (LFV) processes \cite{Ilakovac:1994kj, Illana:2000ic, Hernandez-Tome:2019lkb}. Here we consider these effects in the following LFV processes: $\ell\to\ell'\gamma$ decays (which were first addressed  in this context in ref. \cite{DelAguila:2019xec}), $Z\to\ell\bar{\ell'}$ decays, $\tau\to3\ell$ decays, with all possible flavor combinations for the charged leptons ($\ell=e,\mu$) in the final state \footnote{Observation of these processes goes beyond the SM extended with massive light neutrinos \cite{Petcov:1976ff,Hernandez-Tome:2018fbq,Blackstone:2019njl}.}, and $\mu\to e$ conversion in nuclei~\footnote{We do not consider $H\to\ell\bar{\ell'}$ as it does not enter as a relevant building block of the studied processes, and it is necessarily below current and near future sensitivities \cite{Hernandez-Tome:2020lmh}. This is a general feature of LH models \cite{Lami:2016mjf, Yang:2016hrh,delAguila:2017ugt}. There are bright future prospects for $\ell$ to $\tau$ conversion in in nuclei \cite{Husek:2020fru, Cirigliano:2021img}, that we plan to study within the LHT elsewhere.}. These will not only allow the wrong-sign decays ($\tau^-\to\ell^+\ell'^-\ell'^-$) but also permit larger upper limits than those typically  predicted in ref. \cite{delAguila:2019htj} for the other  processes. Noteworthy, correlation among the considered processes will be very different to the traditional LHT scenario (without heavy Majorana neutrinos) \cite{DelAguila:2019xec} and other models, which would ease the validation/falsification of this LHT realization, should LFV in the charged lepton sector be discovered and measured in different processes.

This article is structured as follows. In section \ref{sec:numassesinLHT} we review the generation of ISS neutrino masses within the LHT. All the new contributions that we study  in this work arise from this implementation of neutrino masses in the LHT. Next, in section \ref{sec:LFVprocesses}, these new contributions to the LFV processes that we study are presented. Then, their phenomenology  is discussed in section \ref{sec:Phenomenology}. Finally, we present our conclusions in section \ref{sec:Concl}. All necessary loop functions are given in the appendix.

\section{Inverse seesaw neutrino masses in the LHT model}\label{sec:numassesinLHT}
\hspace*{0.5cm} We will summarize next the main aspects first introduced in ref. \cite{DelAguila:2019xec} concerning the implementation of neutrino masses in the model (recovering the ISS scenario). The interested reader is addressed to this reference for further details and references of the topics discussed in this section. As we will concentrate on the corresponding new  contributions to several LFV processes, we will not review here the bulk of the LHT, which is nicely and extensively explained in the reviews \cite{Schmaltz:2005ky,Perelstein:2005ka,Panico:2015jxa}.

The scalar sector of the LHT is a non-linear $\sigma$ model based on the coset space $SU (5)/SO(5)$, with the $SU(5)$ global
symmetry spontaneously broken by the vacuum expectation value (vev) $f$ (of order TeV,  larger that the vev of the Higgs field, $\upsilon$) giving rise to 14 pseudo-Nambu-Goldstone bosons entering the matrix
\begin{equation}
    \Pi = \left(
    \begin{array}{ccccc}
         -\omega^{0}/2 - \eta/\sqrt{20}&-\omega^{+}/\sqrt{2}&-i\pi^{+}/\sqrt{2}&-i\Phi^{++}&-i\frac{\Phi^{+}}{\sqrt{2}}\\
          -\omega^{-}/2 & \omega^{0}/2 - \eta/\sqrt{20}&\frac{\upsilon + h + i\pi^{0}}{2}&-i\Phi^{+}/\sqrt{2}&\frac{-i\Phi^{0} + \Phi^{P}}{\sqrt{2}}\\
           -i\pi^{-}/\sqrt{2} & \left( \upsilon + h - i\pi^{0}\right)/2&\sqrt{\frac{4}{5}}\eta&-i\pi^{+}/\sqrt{2}&\left( \upsilon+h+i\pi^{0} \right)/2\\
           i\Phi^{--}&i\frac{\Phi^{-}}{\sqrt{2}}&i\pi^{-}/\sqrt{2}&-\omega^{0}/2 - \eta/\sqrt{20}&-\omega^{-}/\sqrt{2}\\
           i\frac{\Phi^{-}}{\sqrt{2}}&\frac{i\Phi^{0} + \Phi^{P}}{\sqrt{2}}&\frac{\upsilon+h-i\pi^{0}}{2}&-\omega^{+}/\sqrt{2}&\omega^{0}/2 - \eta/\sqrt{20}
    \end{array}
    \right).
    \label{eqPi}
\end{equation}
It includes the SM Higgs doublet ($h$ and $\pi$ fields), a complex weak isospin triplet $\Phi$, and the longitudinal modes of the heavy $\mathcal{O}$(TeV) gauge fields $\omega^{\pm,0}$ and $\eta$. Although the fields in $\Pi$ transform non-linearly under the symmetry, $\xi=$e$^{i\Pi/f}$ obeys a linear transformation under $SU(5)$. $T$ parity is defined to make $T$-odd all but the SM Higgs doublet, so that the heavy states can only interact pairwise.

In the lepton sector, each SM doublet $l_L=(\nu_L\;\ell_L)^T$ is mirrored (with $_{1,2}$ subindexes, respectively) by introducing two incomplete quintuplets ($\sigma^2$ is the second Pauli matrix) as follows
\begin{equation}
    \Psi_{1} = \left( \begin{array}{c}
         -i\sigma^{2}l_{1L}\\ 0 \\ 0
    \end{array} \right), \quad \Psi_{2} = \left( \begin{array}{c}
         0\\ 0 \\-i\sigma^{2}l_{2L}
    \end{array} \right),
    \label{eqPsi_1,2}
\end{equation}
with $\Psi_2$ transforming with the fundamental $SU(5)$ representation V and $\Psi_2$ with its complex conjugated.

Then, $T$-parity is defined to act on the left-handed (LH) leptons as
\begin{equation}
    \Psi_{1} \longleftrightarrow \Omega \Sigma_{0}\Psi_{2},
    \label{eqtransfPsi1}
\end{equation}
with
\begin{equation}
    \Omega = \mathrm{diag}(-1,-1,1,-1,-1), \quad \Sigma_{0} = \left(\begin{array}{ccc}
        0 & 0 & \textbf{1}_{2 \times 2} \\0&1&0\\
        \textbf{1}_{2 \times 2}&0&0
    \end{array}\right).
    \label{eqOmega&Sigma0}
\end{equation}

The SM doublet, $l_{L} = (l_{1L}-l_{2L})/\sqrt{2}$, will be $T$-even; while its heavy copy, $l_{HL} = \left( \nu_{HL} \ \ell_{HL} \right)^{T} = \left( l_{1L}+l_{2L} \right)/\sqrt{2}$, will be $T$-odd~\footnote{They correspond to $(\Psi_{1}\pm\Omega \Sigma_{0} \Psi_{2})/\sqrt{2}$, respectively.}. This heavy doublet (one per family) will get its mass combining with a right-handed doublet $l_{HR}$ in an $SO(5)$ multiplet $\Psi_R$,
\begin{equation}
    \Psi_{R} = \left( \begin{array}{c}
        \psi^{\prime}_{R} \\ \chi_{R} \\ -i\sigma^{2}l_{HR}
    \end{array} \right), \quad T:\Psi_{R} \leftrightarrow \Omega \Psi_{R},
    \label{eqPsiR}
\end{equation}
getting its large ($\sim f$) mass from the Yukawa Lagrangian
\begin{equation}
    \mathcal{L}_{Y_{H}} = - \kappa f \left( \overline{\Psi}_{2} \xi + \overline{\Psi}_{1} \Sigma_{0} \xi^{\dagger} \right) \Psi_{R} + h.c.,
    \label{eqYukLag}
\end{equation}
where the first term preserves the global symmetry for $\xi \rightarrow V \xi U^{\dagger}$ and the second one is its $T$-transformed for $U=\Omega$. Eq.~(\ref{eqYukLag}) gives a vector-like mass $\sqrt{2}\kappa f$ to $\nu_H$ ($\kappa$ is not a small parameter, as we mention below).

Symmetry allows a large vector-like mass for the lepton singlets $\chi_{R}$ as well, by combining directly with a LH singlet $\chi_{L}$.  This is
\begin{equation}
    \mathcal{L}_{M} = - M \overline{\chi}_{L}\chi_{R} + h.c.
    \label{eqDiracmass}
\end{equation}
$\chi_{L}$ is  $SU(5)$ singlet, so it is natural to include a small Majorana mass for it. We assume Lepton Number (LN) to be broken only by small Majorana masses $\mu$ in the heavy LH neutral sector. Then,
\begin{equation}
    \mathcal{L}_{\mu} = - \frac{\mu}{2} \overline{\chi^{c}_{L}}\chi_{L} + h.c.,
    \label{eqmuterm}
\end{equation}
and the resulting (T-even) neutrino mass matrix reduces to the inverse see-saw one:
\begin{equation}
    \mathcal{L}_{M}^{\nu} = - \frac{1}{2} \left( \overline{\nu^{c}_{l}} \ \overline{\chi_{R}} \ \overline{\chi^{c}_{L}} \right) \mathcal{M}_{\nu}^{T-even} \left( \begin{array}{c}
         \nu_{L} \\ \chi_{R}^{c} \\ \chi_{L}
    \end{array} \right) + h.c.,
    \label{eqISSMasses}
\end{equation}
where 
\begin{equation}
    \mathcal{M}_{\nu}^{T-even} = \left( \begin{array}{ccc}
         0 & i \kappa^{*}f \sin \left( \frac{\upsilon}{\sqrt{2}f} \right) & 0\\
        i \kappa^{\dagger}f \sin \left( \frac{\upsilon}{\sqrt{2}f} \right) & 0 & M^{\dagger} \\
        0 & M^{*} & \mu
    \end{array} \right),
    \label{eqMnuTeven}
\end{equation}
with each entry standing for a $3 \times 3$ matrix accounting for the $3$ lepton families. The $\kappa$ entries are given by the Yukawa Lagrangian in eq.~(\ref{eqYukLag}), $M$ stands for the  heavy Dirac mass matrix from eq.~(\ref{eqDiracmass}), and $\mu$ is the mass matrix of small Majorana masses in eq.~(\ref{eqmuterm}). In the inverse see-saw, the hierarchy $\mu<<\kappa<<M$ holds, with \textcolor{black}{$M\sim4\pi f\sim10$ TeV} ($\kappa\sim\mathcal{O}(1)$ is assumed and $\mu$ needs to be much smaller than a GeV), according to electroweak  precision data~\cite{DelAguila:2019xec}.

Let $\mathcal{U}$ be a unitary transformation that diagonalizes $\mathcal{M}$ and transforms the states $\left( \nu_{L} \ \Psi_{L} \right)$ in the gauge basis to the mass eigenstates $\left(  \nu_{L}^{l} \ \Psi_{L}^{h} \right)$, light and heavy (quasi-Dirac) neutrinos, $l$ and $h$, respectively
\begin{equation}
    \mathcal{U}^{\dagger} \left( \begin{array}{c}
         \nu_{L} \\ \Psi_{L}
    \end{array} \right) = \left( \begin{array}{c}
         \nu_{L}^{l} \\ \Psi_{L}^{h}
    \end{array} \right), \quad \mathrm{with} \quad \Psi_{L} = \left( \begin{array}{c}
         \chi_{R}^{c} \\
         \chi_{L} 
    \end{array} \right).
    \label{eq188}
\end{equation}

The matrix $\mathcal{U}$ can be written as \cite{Grimus:2000vj}
\begin{equation}
    \mathcal{U} = \left( \begin{array}{cc}
        \sqrt{1 - \mathcal{B}\mathcal{B}^{\dagger}} & \mathcal{B} \\
        - \mathcal{B}^{\dagger} & \sqrt{1 - \mathcal{B}^{\dagger} \mathcal{B}}
    \end{array} \right),
    \label{eq189}
\end{equation}
such that $\mathcal{U}$ satisfies 
\begin{equation}
    \mathcal{U}^{T}\mathcal{M}\mathcal{U} = \left( \begin{array}{cc}
        \mathcal{M}_{\nu}^{l} & 0_{3 \times 6} \\
        0_{6 \times 3} & \mathcal{M}_{\chi}^{h} 
    \end{array} \right),
    \label{eq190}
\end{equation}
decoupling the heavy and light neutrino fields. $\mathcal{B}$ is a complex $3 \times 3$ matrix and  $\sqrt{1 - \mathcal{B}\mathcal{B}^{\dagger}}$ shall be expanded for radicand close to one, keeping only order $\mathcal{B}\mathcal{B}^{\dagger}$ terms.

For diagonalizing $\mathcal{M}$, it is convenient to introduce
\begin{equation}
    M_{D} = \left( \begin{array}{c}
        i \kappa^{\dagger} f sin\left( \frac{\upsilon}{\sqrt{2}f} \right) \\ 0
    \end{array} \right), \quad M_{R} = \left( \begin{array}{cc}
        0 & M^{\dagger} \\
        M^{*} & \mu 
    \end{array} \right),
    \label{eq192}
\end{equation}
hence,
\begin{equation}
    \mathcal{M}_{\nu}^{T-even} = \left( \begin{array}{cc}
        0 & M_{D}^{T} \\
        M_{D} & M_{R}
    \end{array} \right).
    \label{eq193}
\end{equation}
A first approximation to $\mathcal{B}$ \cite{Hettmansperger:2011bt} is
\begin{equation}
    \mathcal{B}^{*} = M_{D}^{T}M_{R}^{-1} \rightarrow \mathcal{B} = M_{D}^{\dagger}\left( M_{R}^{-1} \right)^{*} = \left( if \sin\left( \frac{\upsilon}{\sqrt{2}f}\right) \kappa M^{-1} \mu^{*} (M^{T})^{-1} \quad -if \sin\left( \frac{\upsilon}{\sqrt{2}f}\right) \kappa M^{-1} \right).
    \label{eq194}
\end{equation}
Therefore, 
\begin{equation}
    \sqrt{1 - \mathcal{B}\mathcal{B}^{\dagger}} \approx 1 - \frac{1}{2} \mathcal{B}\mathcal{B}^{\dagger} \approx 1 - \frac{1}{2} \theta \theta^{\dagger},
    \label{eq195}
\end{equation}
where we have omitted terms of the order of $\mu$ because $\mu << \kappa << M$ and we redefined $\mathcal{B} \rightarrow \Theta$
\begin{equation}
%\begin{split}
    \Theta  = \left( -\theta \mu^{*} (M^{T})^{-1} \quad \theta \right), \quad \mathrm{with} \quad
    \theta = - i f \sin \left( \frac{\upsilon}{\sqrt{2}f} \right) \kappa M^{-1},
%\end{split}
    \label{eq196}
\end{equation}
in which $\Theta$ is a $3 \times 6$ and $\theta$  a $3 \times 3$ matrix, as in refs.~\cite{DelAguila:2019xec, Nomura:2018ktz}. In this way, the $\mathcal{U}$ matrix reads 
\begin{equation}
    \mathcal{U} = \left( \begin{array}{cc}
        1 - \frac{1}{2}\Theta \Theta^{\dagger} &  \Theta \\
        -\Theta^{\dagger} & 1 - \frac{1}{2} \Theta^{\dagger} \Theta
    \end{array} \right).
    \label{eq197}
\end{equation}
Then, the $\mathcal{M}_{\nu}^{l}$ and $\mathcal{M}_{\chi}^{h}$ matrices in the eq. (\ref{eq190}) are given by \cite{DelAguila:2019xec, Grimus:2000vj, Hettmansperger:2011bt, Nomura:2018ktz}
\begin{equation}
    (\mathcal{M}_{\nu}^{l})_{ij}= - (M_{D}^{T} M_{R}^{-1}M_{D})_{ij} = \theta^{*}_{ik} \mu_{kl} \theta^{\dagger}_{jl}, \qquad \mathcal{M}_{\chi}^{h} = M_{R},
    \label{eq198}
\end{equation}
where we have assumed, without loss of generality, that the $\chi$ mass matrix, $M$, is diagonal and positive definite. The diagonalized (Majorana) mass terms of eq. (\ref{eqISSMasses}) thus read
\begin{equation}
    \mathcal{L}_{M}^{\nu} = -\frac{1}{2} \left( \sum_{i=1}^{3} (\mathcal{M}_{\nu}^{l})_{i} \overline{\nu_{Li}^{l}} \nu_{Ri}^{l} + \sum_{j=4}^{9} (\mathcal{M}_{\chi}^{h})_{j} \overline{\Psi_{Lj}^{h}} \Psi_{Rj}^{h}   \right).
    \label{eq199}
\end{equation}
%We notice $\mathcal{M}_{\nu}^{l}$ is a $3 \times 3$ matrix and $\mathcal{M}_{\chi}^{h}$ is a $6 \times 6$ matrix. 
We can work in the basis where the charged lepton mass matrix is diagonal
\begin{equation}
    \mathcal{M}_{\nu}^{l} = U_{PMNS}
    ^{*}\mathcal{D}_{\nu}^{l}U^{\dagger}_{PMNS}
    ,
    \label{eq200}
\end{equation}
from eq. (\ref{eq198}), 
\begin{equation}
    \mu = (\theta^{*})^{-1} U_{PMNS}
    ^{*}\mathcal{D}_{\nu}^{l}U^{\dagger}_{PMNS}
    (\theta^{\dagger})^{-1},
    \label{eq201}
\end{equation}
where $U_{PMNS}$ in the Pontecorvo-Maki-Nakagawa-Sakata matrix \cite{Pontecorvo:1957qd, Maki:1962mu} (that we will denote simply $U$ in the following) and $\mathcal{D}_{\nu}^{l}$ the diagonal neutrino mass matrix.\\
Applying explicitly $\mathcal{U}^{\dagger}$ to eq. (\ref{eq188})
\begin{equation}
    \left( \begin{array}{cc}
        1 - \frac{1}{2}\Theta \Theta^{\dagger} &  -\Theta \\
        \Theta^{\dagger} & 1 - \frac{1}{2} \Theta^{\dagger} \Theta
    \end{array} \right) \left( \begin{array}{c}
         \nu_{L} \\ \Psi_{L}
    \end{array} \right) = \left( \begin{array}{c}
         \nu_{L}^{l} \\ \Psi_{L}^{h}
    \end{array} \right),
    \label{eq202}
\end{equation}
due to the eq. (\ref{eq200}).
%\begin{equation}
%    \left( \begin{array}{cc}
%        1 - \frac{1}{2}\Theta \Theta^{\dagger} &  -\Theta \\
 %       \Theta^{\dagger} & 1 - \frac{1}{2} \Theta^{\dagger} \Theta
  %  \end{array} \right) \left( \begin{array}{c}
         %\nu_{L} \\ \Psi_{L}
    %\end{array} \right) = \left( \begin{array}{c}
         %U%_{PMNS}
         %\nu_{L}^{l} \\ \Psi_{L}^{h}
    %\end{array} \right),
    %\label{eq203}
%\end{equation}
%h
Hence, the %light and heavy eigenstates are (
mixing relations between flavor and mass eigenstates %)
are
\begin{eqnarray}
%\begin{split}
   \sum_{j=1}^{3} %(
   U%_{PMNS})
   _{ij}\nu_{Lj}^{l} &=& \sum_{j=1}^{3}[\mathbf{1}_{3 \times 3} - \frac{1}{2}(\Theta \Theta^{\dagger})]_{ij}\nu_{Lj} - \sum_{j=4}^{9}\Theta_{ij}\Psi_{Lj},\nonumber \\
    \Psi_{Li}^{h} &=& \sum_{j=4}^{9} [\mathbf{1}_{6 \times 6} - \frac{1}{2}(\Theta^{\dagger} \Theta)]_{ij} \Psi_{Lj} + \sum_{j=1}^{3 }\Theta^{\dagger}_{ij} \nu_{Lj},
%\end{split}
\label{eq204}
\end{eqnarray}
where $\Theta$ matrix elements give the mixing between light and heavy (quasi-Dirac) neutrinos to leading order.

Let $\Phi$ be a flavor eigenstate composed by 
\begin{equation}
    \Phi_{L} = \mathcal{U} \left( \begin{array}{c}
         \nu_{L}^{l}  \\ \Psi_{L}^{h}
    \end{array} \right),
    \label{eq205}
\end{equation}
thus, in terms of the mass eigenstates from the eq. (\ref{eq204}) the SM charged current is modified as follows
\begin{eqnarray}
    %\begin{split}
        \mathcal{L}_{W}  &=& \frac{g}{\sqrt{2}}W_{\mu}^{+} \sum_{i=1}^{9} \sum_{j=1}^{3} \overline{\Phi_{Li}}\gamma^{\mu}\ell_{Lj} \label{eq206} \\
         &=& \frac{g}{\sqrt{2}} W_{\mu}^{+} \sum_{j=1}^{3} \left( \sum_{i=1}^{3} \{ U%_{PMNS}
        ^{\dagger} [\mathbf{1}_{3 \times 3} - \frac{1}{2}(\Theta \Theta^{\dagger})] \}_{ij} \overline{\nu_{Li}^{l}} + \sum_{i=4}^{9} \Theta^{\dagger}_{ij}\overline{\Psi_{Li}^{h}} \right) \gamma^{\mu} \ell_{Lj}. \nonumber
    %\end{split}
\end{eqnarray}
We can split the Lagrangian above in two parts, each one fixing the coupling between the SM leptons and the light and heavy quasi-Dirac neutrinos, respectively,
\begin{eqnarray}
%\begin{split}
   \mathcal{L}_{W}^{l} & = & \frac{g}{\sqrt{2}} W_{\mu}^{+} \sum_{j=1}^{3} \sum_{i=1}^{3} \overline{\nu_{i}^{l}} W_{ij}\gamma^{\mu}P_{L} \ell_{j} + h.c., \quad \mathrm{with} \quad W_{ij} = \{ U%_{PMNS}
   ^{\dagger} [\mathbf{1}_{3 \times 3} - \frac{1}{2}(\Theta \Theta^{\dagger})] \}_{ij},\nonumber\\
   \mathcal{L}_{W}^{lh} & = & \frac{g}{\sqrt{2}} W_{\mu}^{+} \sum_{j=1}^{3} \sum_{i=4}^{9}  \overline{\Psi_{i}^{h}} \Theta^{\dagger}_{ij} \gamma^{\mu}P_{L} \ell_{j} + h.c.
%\end{split}
\label{eq207}
\end{eqnarray}
Due to the presence of $\Theta$, $\mathcal{L}^l_W$ ($\mathcal{L}^{lh}_W$) includes LFV  transitions involving light (heavy) neutrinos.
 
The neutral current coupling to the $Z^0$ gauge boson is written as
\begin{equation}
    %\begin{split}
        \mathcal{L}_{Z} = \frac{g}{2 \cos\theta_{W}} Z_{\mu} \sum_{j=1}^{9} \sum_{i=1}^{9} \overline{\nu_{Li}} \gamma^{\mu} \nu_{Lj}.
    %\end{split}
    \label{eq208}
\end{equation}
We consider  $\Theta\Theta^{\dagger}$ effects to leading order and write down the light and heavy neutral currents as
\begin{eqnarray}
   % \begin{split}
        \mathcal{L}_{Z}^{l} & = & \frac{g}{2 \cos\theta_{W}} Z_{\mu} \sum_{j=1}^{3} \sum_{i=1}^{3} \overline{\nu_{i}^{l}} X_{ij} \gamma^{\mu}P_{L} \nu_{j}^{l}, \quad \mathrm{with} \quad X_{ij} = \{ U%_{PMNS}
        ^{\dagger} [\mathbf{1}_{3 \times 3} - (\Theta \Theta^{\dagger})] U%_{PMNS}
 \}_{ij},\nonumber\\ 
    \mathcal{L}_{Z}^{lh} & = & \frac{g}{2\cos\theta_{W}} Z_{\mu} \sum_{j=1}^{3} \sum_{i=4}^{9} \overline{\Psi_{i}^{h}}(\Theta^{\dagger}U%_{PMNS}
    )_{ij}\gamma^{\mu}P_{L}\nu_{j}^{l} + h.c.,\nonumber\\ \mathcal{L}_{Z}^{h} & = & \frac{g}{2\cos\theta_{W}} Z_{\mu} \sum_{j=4}^{9} \sum_{i=4}^{9} \overline{\Psi_{i}^{h}}(\Theta^{\dagger}\Theta)_{ij}\gamma^{\mu}P_{L}\Psi_{j}^{h}.
  %  \end{split}
    \label{eq209}
\end{eqnarray}
Noteworthy, $\mathcal{L}_Z^l$ includes LFV terms in a purely light neutrino's current. The flavor symmetry is also broken by $\mathcal{L}_Z^{lh}$ and $\mathcal{L}_Z^h$, including heavy neutrinos.

Now, we neglect the $\mu$ term $(\mu \ll \kappa \ll M)$ in the $\Theta$ matrix, $\Theta = \left( 0_{3 \times 3} \quad \theta \right)$. Therefore, the eigenstates in the eq. (\ref{eq204}) transform as \cite{DelAguila:2019xec}
\begin{equation}
%\begin{split}
   \sum_{j=1}^{3} %(
   U
   %_{PMNS})
   _{ij}\nu_{Lj}^{l}  = \sum_{j=1}^{3}[\mathbf{1}_{3 \times 3} - \frac{1}{2}(\theta \theta^{\dagger})]_{ij}\nu_{Lj} - \sum_{j=7}^{9}\theta_{ij}\chi_{Lj}, 
    \chi_{Li}^{h}  = \sum_{j=7}^{9} [\mathbf{1}_{3 \times 3} - \frac{1}{2}(\theta^{\dagger} \theta)]_{ij} \chi_{Lj} + \sum_{j=1}^{3 }\theta^{\dagger}_{ij} \nu_{Lj},
%\end{split}
\label{eq210}
\end{equation}
hence, the Lagrangians from eqs. (\ref{eq207}) and (\ref{eq209}) read
\begin{eqnarray}
%\begin{split}
   \mathcal{L}_{W}^{l} & = & \frac{g}{\sqrt{2}} W_{\mu}^{+} \sum_{i,j=1}^{3} \overline{\nu_{i}^{l}} W_{ij}\gamma^{\mu}P_{L} \ell_{j} + h.c., \quad \mathrm{with} \quad W_{ij} = \sum_{k=1}^{3}  %(
   U%_{PMNS}
   ^{\dagger}%)
   _{ik} [\mathbf{1}_{3 \times 3} - \frac{1}{2}(\theta \theta^{\dagger})]_{kj},\nonumber\\
   \mathcal{L}_{W}^{lh} & = & \frac{g}{\sqrt{2}} W_{\mu}^{+} \sum_{i=7}^{9} \sum_{j=1}^{3} \overline{\chi_{i}^{h}} \theta^{\dagger}_{ij} \gamma^{\mu}P_{L} \ell_{j} + h.c.,
%\end{split}
\label{eq211}
\end{eqnarray}
and
\begin{eqnarray}
   % \begin{split}
        \mathcal{L}_{Z}^{l} & = & \frac{g}{2 \cos\theta_{W}} Z_{\mu} \sum_{i,j=1}^{3} \overline{\nu_{i}^{l}} \gamma^{\mu}(X_{ij}P_{L}-X_{ij}^{\dagger}P_{R}) \nu_{j}^{l}, \; \mathrm{with} \, X_{ij} = \sum_{k=1}^{3} \left(U%_{PMNS}
        ^{\dagger} [\mathbf{1}_{3 \times 3} - (\theta \theta^{\dagger})]\right)_{ik} %(
        U
       %_{PMNS})
        _{kj},\nonumber\\ \mathcal{L}_{Z}^{lh}  &=& \frac{g}{2\cos\theta_{W}} Z_{\mu} \sum_{i=7}^{9} \sum_{j=1}^{3} \overline{\chi_{i}^{h}} \gamma^{\mu} (Y_{ij}P_{L}-Y_{ij}^{\dagger}P_{R})\nu_{j}^{l} + h.c.,\quad \mathrm{with} \quad Y_{ij} = \sum_{k=1}^{3} \theta^{\dagger}_{ik} %(U
        %_{PMNS})
        U_{kj},\nonumber\\ 
    \mathcal{L}_{Z}^{h} & = & \frac{g}{2\cos\theta_{W}} Z_{\mu} \sum_{i,j=7}^{9} \overline{\chi_{i}^{h}}\gamma^{\mu}(S_{ij}P_{L}-S_{ij}^{\dagger}P_{R})\chi_{j}^{h},\quad \mathrm{with} \quad S_{ij} = \sum_{k=1}^{3} \theta^{\dagger}_{ik} \theta_{kj},
 %   \end{split}
    \label{eq212}
\end{eqnarray}
where the dimension of the $W$ and $X$  square mixing matrices is $3 \times 3$. Comparing our charged-current and neutral-current interactions from eqs. (\ref{eq211}) and (\ref{eq212}) with the SM ones, we observe that they differ by the presence of the $\theta$ matrix, which is a consequence of introducing Majorana neutrinos, that allows for both neutral and charged LFV  transitions. We will focus on these new contributions in the remainder of this work.\\
We can define the $B_{ij}$ and $C_{ij}$ matrices  according to SM charged and neutral currents (see eqs. (\ref{eq211}) and (\ref{eq212})) \cite{Ilakovac:1994kj, Illana:2000ic}
\begin{equation}
    B_{ij} = \sum_{k = 1}^{3} U_{ik}\mathcal{U}^{\dagger}_{kj} \quad \mathrm{and} \quad C_{ij} = \sum_{k = 1}^{3} \mathcal{U}_{ki}\mathcal{U}_{kj}^{\dagger},
    \label{eq34}
\end{equation}
where $B$ mixing matrix is $3 \times 9$, whereas $C$ is a $9 \times 9$ matrix. We are grouping both parts of light and heavy Majorana neutrinos. We need to recall that $\mathcal{U}_{ij}$ with $i,j=4,5,6$ entries are suppressed by ISS hierarchy. Eqs.~(\ref{eq4.1.2.11.3}) give unitarity relations among these matrices, which are crucial to verify cancellation of ultraviolet divergences in loop diagrams within this setting.

\section{New contributions to LFV processes}\label{sec:LFVprocesses}
The relevant effective LFV  $V\ell\ell'$ vertices ($V=\gamma,Z$), depicted in fig. \ref{fig:ell_to_ellgamma}, can be written in terms of the allowed Lorentz structures accompanied by  their corresponding form factors~\footnote{We omit the pieces proportional to $q^\nu$ and $\gamma_5q^\nu$, scalar and pseudoscalar form factors, as they do not contribute  for real $V$ and are negligible for virtual $V$ \cite{Hollik:1998vz} in the processes under study.}
\begin{equation}
    \Gamma_\mu^V(q^2)=e\left[\gamma_\mu\left(F^V_L(q^2) P_L+F^V_R(q^2) P_R\right)+2\left(iF^V_M(q^2)+F^V_E(q^2)\gamma_5\right)\sigma_{\mu\nu}q^\nu\right],
\end{equation}
with $q$ the $V$ boson momentum. $F_{L,R}^V(q^2)$ are the monopole form factors of given chirality and $F_{M,E}^V(q^2)$ are the magnetic and electric dipole form factors. 

\subsection{$\ell\to\ell'\gamma$ decays}\label{subsec:Ltolgamma}
The $\ell \rightarrow \ell' \gamma$ vertex reduces to a dipole transition for an on-shell photon,
\begin{equation}
    i \Gamma_{\gamma}^{\mu}(p_{\ell}, p_{\ell'}) = ie \left[ i F_{M}^{\gamma}(Q^{2}) + F_{E}^{\gamma}(Q^{2}) \gamma_{5} \right] \sigma^{\mu\nu}Q_{\nu},
    \label{eq239}
\end{equation}
where $Q_{\nu} = (p_{\ell'} - p_{\ell})_{\nu}$. Neglecting $m_{\ell'} \ll m_{\ell}$,
\begin{equation}
    \Gamma (\ell \rightarrow \ell' \gamma) = \frac{\alpha}{2} m_{\ell}^{3} (|F_{M}^{\gamma}|^{2}+|F_{E}^{\gamma}|^{2}).
    \label{eq240}
\end{equation}
Results below will be simplified using $F^\gamma_M=-iF^\gamma_E$, that holds for all contributions. All computations in this work were done in the 't Hooft-Feynman gauge.

\begin{figure}[!ht]
    \centering
    \includegraphics[scale=0.585]{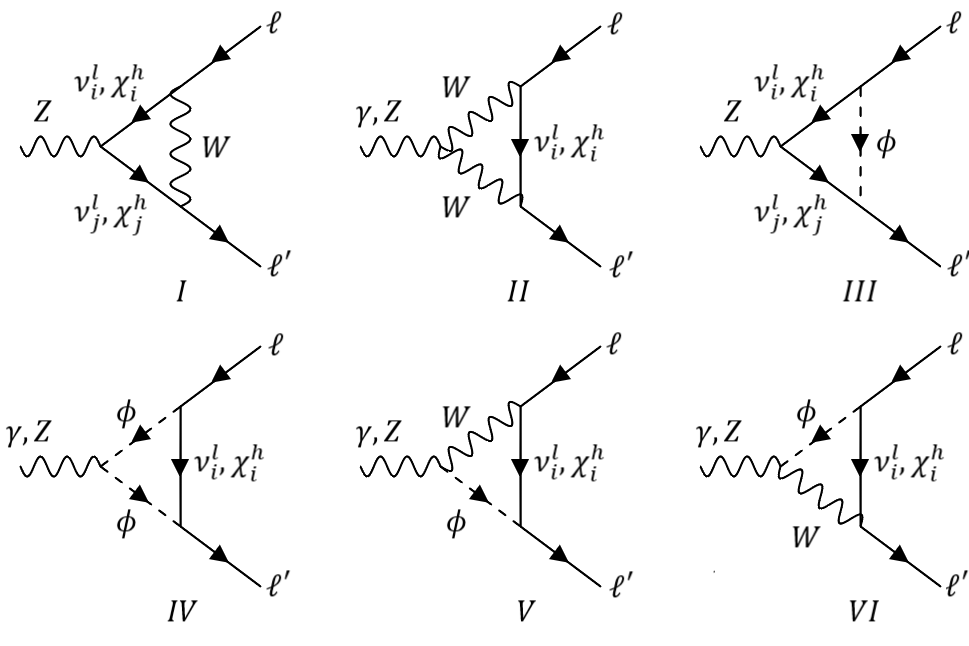}
    \caption{Topologies of the diagrams that contribute to the  $(\gamma/Z)- \bar{\ell}- \ell^{\prime}$ effective vertex.}
    \label{fig:ell_to_ellgamma}
\end{figure}

The active (light) neutrino contribution is analogous to the SM one, just replacing $U$ by $W$ due to eq. (\ref{eq211})~\footnote{Otherwise this branching ratio is unmeasurably small \cite{Petcov:1976ff, Bilenky:1977du, Cheng:1977nv}.}. We note that the $W$ matrix includes the SM contribution, given by $U$,  and the new heavy neutrinos part given by the $\theta \theta^{\dagger}$ term. The corresponding  Feynman diagrams are given by topologies II, IV, V and VI in  figure \ref{fig:ell_to_ellgamma}.
%The same topologies contribute in the heavy neutrino case.

The corresponding result is
\begin{equation}
F_M^\gamma(y_i)\,=\,\frac{\alpha_W}{16\pi}\frac{m_\ell}{M_W^2}\sum_i W^{i\ell'*}W^{i\ell}F_W\left(y_i\right)\,,   
\end{equation}
with $y_i=\frac{m_{\nu_i}^2}{M_W^2}\sim0$, $\alpha_W=\alpha/$sin$^2\theta_W$  and~\footnote{All form factors reported in this subsection can be read from the quoted references. Therefore, we are not giving their expressions in terms of  Passarino-Veltman functions, which can be consulted therein. This information will be provided in the next subsections, which are new results, to our knowledge.}
\begin{equation}
%\begin{split}
    F_{W}(x) = %M_{1}^{2}\left[2 \overline{C}_{1} - 3\overline{C}_{11} -x \left( \overline{C}_{0} + 3 \overline{C}_{1} + \frac{3}{2}\overline{C}_{11} \right) \right] \\
    %& =
    \frac{5}{6} - \frac{3x - 15x^{2}-6x^{3}}{12(1-x)^{3}} + \frac{3x^{3}}{2(1-x)^{4}}\mathrm{ln} \ x.
%\end{split}
\label{eq119}
\end{equation}

Similarly (see eq.~(\ref{eq211})), the $F_{M}^\gamma|_{\chi}(x)$ function turns out to be \cite{Hernandez-Tome:2019lkb}
\begin{equation}
    %\begin{split}
        F_{M}^\gamma|_{\chi}(x) %& = % \frac{-10x^{3} + 33x^{2}-45x+4}{12(1-x)^{3}} + \frac{3x}{2(1-x)^{4}} \mathrm{ln}x &
        %& 
        = \frac{1}{3} - \frac{2x^{3}-7x^{2}+11x}{4(1-x)^{3}} + \frac{3x}{2(1-x)^{4}} \mathrm{ln}x,
        \label{eq247}
    %\end{split}
\end{equation}
with $x_j = \frac{M_{W}^{2}}{M_{j}^{2}}$ and $M_{j}$ the mass of the $j$-th $\chi_{L}^{h}$ state.

Finally, the $\nu_H$ states also contribute through the same topologies than active neutrinos, results are analogous by using  $\lbrace\nu_{H}, W_{H}, \omega\rbrace$ instead of
$\lbrace\nu_{l}, W, \phi\rbrace$:
\begin{equation}
    F_{M}^{\gamma}|_{W_{H}}(x_i)= \frac{\alpha_{W}}{16\pi}\frac{m_{\mu}}{M_{W}^{2}}\frac{v^2}{4f^2} \sum_{i} V_{H\ell}^{ie*}V_{H\ell}^{i\mu} F_{M}^{\nu_{H}}(x_{i}),
    \label{eq118}
\end{equation}
with $x_{i} = \frac{m_{\nu_{Hi}}^{2}}{M_{W_{H}}^{2}}$.%, where

Like the PMNS matrix, $V_{H \ell}$ are $3 \times 3$ unitary mixing matrices parametrizing the misalignment between the SM LH charged leptons $\ell$ and the heavy mirror ones $\ell_{H}$. The observable rotations are
\begin{equation}
    V_{H \nu} \equiv V_{L}^{H\dagger}V_{L}^{\nu}, \quad V_{H \ell} \equiv V_{L}^{H\dagger}V_{L}^{\ell},
\end{equation}
related by $V^{\dagger}_{H \nu}V_{H \ell} = V^{\dagger}_{\mathrm{PMNS}}$ \cite{delAguila:2008zu}.\\
In this way, the new interactions derived in the previous section yield (we omit the upper-index $\gamma$ of the $F^\gamma_M$ form factors below and indicate explicitly instead the type of neutrino appearing in each contribution) \footnote{A similar expression holds for the $\tau\to\ell\gamma$ decays ($\ell=e,\mu$), only accounting for the hadronic tau decay width.} 
\begin{equation}
    \mathrm{Br}(\mu \rightarrow e \gamma) = \frac{3\alpha}{2\pi} \left |W_{e j}W^{*}_{\mu j}F_{M}^{\nu}(y_j)+U_{e j}U^{*}_{\mu j}F_{M}^{\chi}(x_i)+\frac{v^2}{4f^2}V_{H\ell}^{ej*}V_{H\ell}^{\mu j}F_M^{\nu_H}(x_i)\right|^{2},
    \label{eq243}
\end{equation}
where
$x_i=\frac{M_W^2}{M_{N_j}^2} \ll 1$ and $y_j=\frac{m_{\nu_i}^2}{M_W^2}\sim0$. We note that $F^\chi_M(x)=1/3-(11/4)x+\mathcal{O}(x^2)$ and $F^\nu_M(0)=5/6=F^{\nu_H}_M(0)$. The contribution from the third term is suppressed by $v^2/f^2 \ll 1$, so we will neglect it in the following. For analogous reasons we will disregard in the rest of this work the contributions involving T-odd particles, as all of them are suppressed by $v^2/f^2$ in the form factors (explicit expressions can be checked in refs. \cite{delAguila:2008zu,delAguila:2010nv, delAguila:2019htj})~\footnote{We postpone to future work the complete analysis within the LHT keeping the $\mathcal{O}(v^2/f^2)$ contributions to the form factors. This shall be necessary, since the heavy Majorana neutrinos have to have $\mathcal{O}$(TeV) masses, as the T-odd particles, within a gauge invariant theory \cite{Illana:2021uwu}. We will also need to study first  semileptonic LFV tau decays within the LHT (following Refs. \cite{Arganda:2008jj,Celis:2013xja,Lami:2016vrs}).}.

In this way, we get
\begin{equation}
    \mathrm{Br}(\mu \rightarrow e\gamma) = \frac{3\alpha}{2\pi} \left |\theta_{ej}\theta_{\mu j}^{\dagger}F_{M}^{\chi}(x) + \frac{5}{6} W_{ej}W_{\mu j}^{\dagger} \right |^{2}\approx \frac{3\alpha}{8\pi} \left | \theta_{ej} \theta^{\dagger}_{\mu j} \right|^{2}.
    \label{eq252}
\end{equation}

We note that, even for heavy neutrinos as 'light' as $2$ TeV, the correction induced by the $\mathcal{O}(x)$ term to $F^\chi_M(x)$ in observables is at the level of $1\%$ and can be safely neglected.

Then, the $90\%$ C.L. limits $Br(\mu \rightarrow e\gamma) < 4.2 \times 10^{-13}$, $\mathrm{Br}(\tau\rightarrow e \gamma) < 3.3 \times 10^{-8}$ and $\mathrm{BR}(\tau \rightarrow \mu \gamma) < 4.2 \times 10^{-8}$ \cite{ParticleDataGroup:2020ssz, Belle:2021ysv} bind
\begin{equation}
  |\theta_{e j}\theta^{\dagger}_{\mu j}| < 0.14 \times 10^{-4},\quad |\theta_{e j}\theta^{\dagger}_{\tau j}| < 0.95 \times 10^{-2},\quad
  |\theta_{\mu j}\theta^{\dagger}_{\tau j}| < 0.011.
  \label{eq44}
\end{equation}
As discussed in ref. \cite{DelAguila:2019xec}, electroweak precision data \cite{deBlas:2013gla} constrain $|\theta_{e1}|<0.04$, $|\theta_{\mu2}|<0.03$ and $|\theta_{\tau3}|<0.09$, at $95\%$ C.L., assuming that
each heavy neutrino only mixes with one light neutrino of definite flavor and that only one
mixing is non–vanishing at a time. 

\subsection{$Z\to\overline{\ell}\ell'$  decays}\label{subsec:Ztollbar}
\begin{figure}[!ht]
    \centering
    \includegraphics[scale = 0.53]{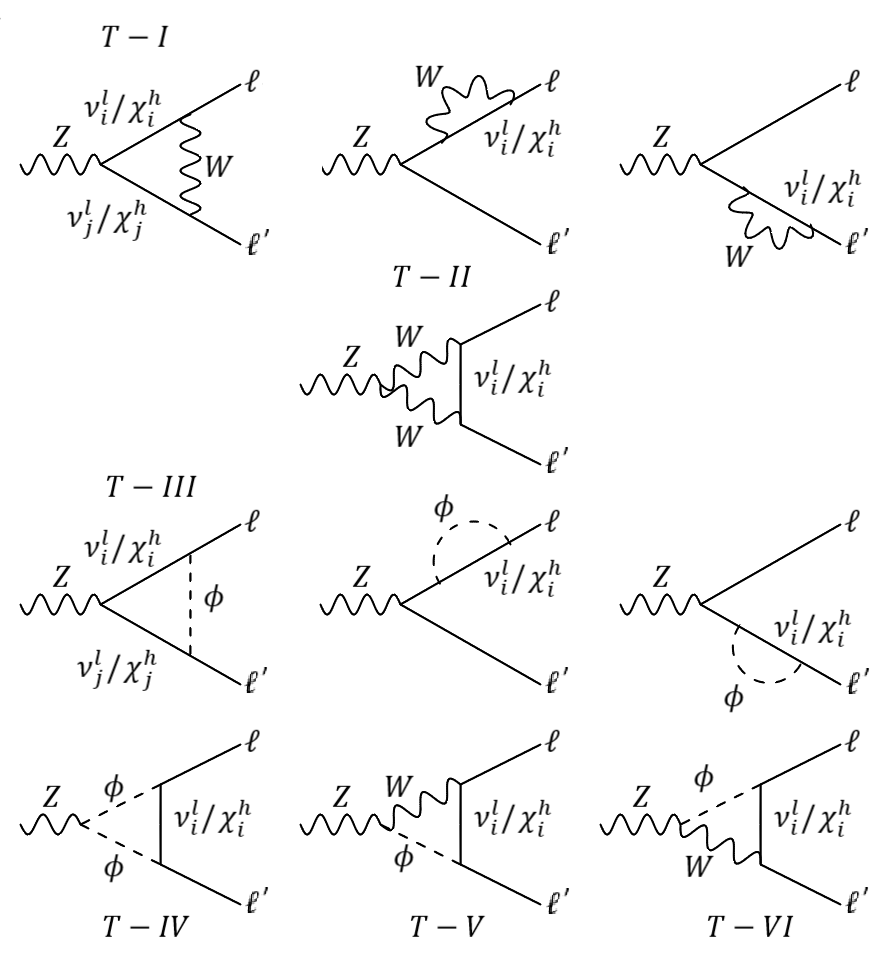}
    \caption{Z penguins diagrams that contribute to the $Z\to\bar{\ell}\ell$ decays. Diagrams corresponding to $T-I$ and $T-III$ allow to mix light and heavy Majorana neutrinos.}
    \label{fig:Z_penguins_Majorana}
\end{figure}

At leading order the $Z \rightarrow \overline{\ell} \ell'$ vertex reduces to
\begin{equation}
    i \Gamma_{Z}^{\mu} (p_{\ell},p_{\ell'}) = ieF_{L}^{Z}(Q^{2})\gamma^{\mu}P_{L}.
    \label{eq4.2.1}
\end{equation}
We work in the approximation of zero light neutrino masses. Therefore, only diagrams with heavy neutrinos contribute to this process. In this type of decay we have that $Q^{2} = M_{Z}^{2}$, so the $Z$ width reads
\begin{equation}
    \Gamma(Z \rightarrow \overline{\ell} \ell') = \frac{\alpha}{3}M_{Z}|F_{L}^{Z}(M_{Z}^{2})|^{2}.
    \label{eq4.2.2}
\end{equation}
There are 10 contributions to $F_L^Z$, which are represented in figure \ref{fig:Z_penguins_Majorana}.
The result can be written \cite{Ilakovac:1994kj,DeRomeri:2016gum,Hernandez-Tome:2019lkb}
\begin{equation}
   %\begin{split}
        F_{L}^{Z}(M_{Z}^{2}) =  \frac{\alpha_{W}}{8\pi c_{W}s_{W}} \sum_{i,j=1}^{3}  \left[ \theta_{\ell' i} \theta^{\dagger}_{\ell i} F^{h}(y_{i};M_{Z}^{2})  + \theta_{\ell' j} S_{ji} \theta^{\dagger}_{\ell i}\left(G^{h}(y_{i},y_{j};M_{Z}^{2}) + \frac{1}{\sqrt{y_{i}y_{j}}} H^{h}(y_{i},y_{j};M_{Z}^{2}) \right) \right],
   %\end{split}
   \label{eq4.2.10}
\end{equation}
where
\begin{eqnarray}
%\begin{split}
    F^{h}(y_{i};M_{Z}^{2}) & = & - 2c_{W}^{2} \left[ M_{Z}^{2} (\overline{C}_{1}+\overline{C}_{2}+\overline{C}_{12}) + 6\overline{C}_{00} - 1 \right] - (1-2s_{W}^{2})\frac{1}{y_{i}}\overline{C}_{00} - 2s_{W}^{2}\frac{1}{y_{i}}M_{W}^{2}\overline{C}_{0}\nonumber\\
    & - & \frac{1}{2}(1-2s_{W}^{2})\left[ \left(2+\frac{1}{y_{i}}\right) \overline{B}_{1} + 1 \right], \nonumber\\
    G^{h} (y_{i},y_{j};M_{Z}^{2}) & = & -M_{Z}^{2}(C_{0}+C_{1}+C_{2}+C_{12})+2C_{00}-1-\frac{1}{2}\frac{1}{y_{i}y_{j}}M_{W}^{2}C_{0} , \nonumber\\  
    H^{h} (y_{i},y_{j};M_{Z}^{2}) & = & M_{W}^{2}C_{0}+\frac{1}{2}M_{Z}^{2}C_{12}-C_{00}+\frac{1}{4},
%\end{split}
\label{eq4.2.11}
\end{eqnarray}
with $y_{i,j} = M_{W}^{2}/M_{i,j}^{2}$ and $M_{i,j}$ heavy neutrino masses~\footnote{Loop functions are given in the appendix.}. Analytic expressions for the functions $F^{h}, G^{h}$, and $H^{h}$ at order $M_{Z}^{2}$ are written
\begin{eqnarray}
    F^{h}(y_{i};M_{Z}^{2}) & = & - \left( \frac{5}{2}-2s_{W}^{2} \right) \Delta_{\epsilon} - \frac{5 \ \mathrm{ln}y_{i}}{2(1-y_{i})^{2}} - \frac{5}{2(1-y_{i})} + \frac{1}{4} \label{eq4.2.12} \\
    & + & \frac{M_{Z}^{2}}{72 M_{W}^{2}} \frac{1}{(1-y_{i})^{4}} ( 6 [24y_{i}^{2}(s_{W}^{2}-1) - 4y_{i}(5s_{W}^{2}-8) - (2s_{W}^{2}-1)]\ \mathrm{ln}y_{i} \nonumber \\
    & - & (1-y_{i})[88y_{i}^{3}(s_{W}^{2}-1) -2y_{i}^{2}(164s_{W}^{2}-171) - y_{i}(297-230s_{W}^{2})-(2s_{W}^{2}+11)] ), \nonumber \\
    G^{h}(y_{i},y_{j};M_{Z}^{2}) & = & \frac{1}{2} \left( \Delta_{\epsilon}-\frac{1}{2} \right) - \frac{1}{2(y_{i}-y_{j})} \left(- \frac{(1-y_{j})\ \mathrm{ln}y_{i}}{(1-y_{i})} + \frac{(1-y_{i}) \ \mathrm{ln}y_{j}}{(1-y_{j})} \right)+ \mathcal{O}\left(\frac{M_{Z}^{2}}{M_{i,j}^{2}}\right), \nonumber \\
    H^{h}(y_{i},y_{j};M_{Z}^{2}) & = & - \frac{1}{4}\left( \Delta_{\epsilon} + \frac{1}{2} \right) - \frac{1}{4(y_{i} - y_{j})} \left( -\frac{(1-4y_{i})y_{j} \ \mathrm{ln}y_{i}}{(1-y_{i})} + \frac{(1 - 4y_{j})y_{i} \ \mathrm{ln}y_{j}}{(1-y_{j})} \right)+\mathcal{O}\left(\frac{M_{Z}^{2}}{M_{i,j}^{2}}\right).
    \nonumber 
\end{eqnarray}
We note that we preferred to write the variable $y_{i,j}$ so that it is small. Then, the neutrino masses are in the denominator, as opposed to the $x_i$ variable employed for light neutrinos (and to previous literature). Taking this into account, our results reproduce those in refs. \cite{Ilakovac:1994kj,DeRomeri:2016gum,Hernandez-Tome:2019lkb}. It is also worth to mention that the $\mathcal{O}(M_Z^2)$ terms are clearly negligible for the $G^h$ and $H^h$ functions. However, the second and third lines of $F^h$ are $\mathcal{O}(M_Z^2/M_W^2)$, which is not small. Still, this correction turns out to be  $\sim26$ times smaller than the (convergent part of the) first line of $F^h$ (and it can be checked that higher order terms in the $M_Z^2$ expansion are further suppressed) ~\footnote{We are not aware this issue was  discussed previously.}. The (generalized) GIM mechanism that applies to the mixing matrices in eq.~(\ref{eq4.2.10})~\cite{Ilakovac:1994kj,Illana:2000ic} cancels all UV divergences encoded in $\Delta_\epsilon=\frac{1}{\epsilon}-\gamma_E+\mathrm{ln}(4\pi)+\mathrm{ln}\left(\frac{\mu^2}{M_W^2}\right)$, regulating them in $4-2\epsilon$ dimensions. Specifically, this happens thanks to the relations \cite{Ilakovac:1994kj, Illana:2000ic}
\begin{eqnarray}
%    \begin{split}
        & & \sum_{k=1}^{9} B_{ik}B_{jk}^{\dagger} = \delta_{ij}, \quad  \sum_{k=1}^{3} B_{ki}^{\dagger}B_{kj} = \sum_{k=1}^{9} C_{ik}C_{jk}^{\dagger} = C_{ij}\,,
        \label{eq4.1.2.11.3} \\
        & & \sum_{k=1}^{9}B_{ik}C_{kj} = B_{ij}, \nonumber\\
        & & \sum_{k=1}^{9}m_{\Phi_{k}}C_{ik}C_{jk} = \sum_{k=1}^{9}m_{\Phi_{k}}B_{ik}C_{kj}^{\dagger} =\sum_{k=1}^{3}m_{\Phi_{k}}B_{ik}B_{jk}^{\dagger} = 0,\nonumber
   % \end{split}
\end{eqnarray}
%where we have split (contrary to common practice) the light ($m_i$) and heavy ($M_j$) neutrino parts.
where $m_{\Phi_{k}} = m_{k}$ with ($k=1,2,3$) and $m_{\Phi_{k}} = M_{k}$ with ($k=7,8,9$) are the light and heavy masses of Majorana neutrinos,  respectively. 

\subsection{$L\to3\ell$ decays}\label{subsec:Lto3ell}
We distinguish three types of three-lepton lepton decays, according to the notation $\ell\to\ell'\ell''\ell'''$:
\begin{enumerate}
    \item $\ell\neq\ell'=\ell''=\ell'''$ (which contains the processes $\mu\to e e\bar{e}$, $\tau\to e e \bar{e}$ and $\tau\to\mu\mu\bar{\mu}$).
    \item $\ell\neq\ell'\neq\ell''=\ell'''$ (including the $\tau\to e \mu\bar{\mu}$  and $\tau\to\mu e\bar{e}$ decays).
    \item $\ell\neq\ell'=\ell''\neq\ell'''$ (constituted by the 'wrong-sign' processes: $\tau\to e e \bar{\mu}$ and $\tau\to \mu\mu\bar{e}$).
\end{enumerate}
We will treat them in turn.
\subsubsection{Type I: $\ell \rightarrow \ell' \ell'' \bar{\ell}'''$ with $\ell \neq \ell' = \ell'' = \ell'''$}
The amplitude for Type I decays gets contributions from $\gamma$ and $Z$ penguin diagrams, as well as from boxes:
\begin{equation}
    \mathcal{M}^{\mathrm{Type\,I}} = \mathcal{M}^{\mathrm{Type\,I}}_{\gamma} + \mathcal{M}^{\mathrm{Type\,I}}_{Z} + \mathcal{M}^{\mathrm{Type\,I}}_{box},
    \label{eq4.1.2.1}
\end{equation}
where each amplitude is defined as follows \cite{delAguila:2019htj}
\begin{eqnarray}
   % \begin{split}
        \mathcal{M}^{{\mathrm{Type\,I}}}_{\gamma} & = & \overline{u}(p_{1})e\left[ iF_{M}^{\gamma}(0)2P_{R}\sigma^{\mu\nu}(p_{1}-p_{\ell})_{\nu}+F_{L}^{\gamma}((p_{1}-p_{\ell})^{2})\gamma^{\mu}P_{L} \right]u(p_{\ell}) \nonumber\\
        & \times & \frac{1}{(p_{1}-p_{\ell})^{2}} \overline{u}(p_{3})\gamma_{\mu}ev(p_{2}) - (p_{1} \leftrightarrow p_{3}),\nonumber\\
        \mathcal{M}^{\mathrm{Type\,I}}_{Z} & = & \overline{u}(p_{1}) \left( -eF_{L}^{Z}(0) \right)\gamma^{\mu}P_{L}u(p_{\ell})\frac{1}{M_{Z}^{2}}\overline{u}(p_{3})\gamma_{\mu}\left( g_{L}^{Z}P_{L}+g_{R}^{Z}P_{R} \right)v(p_{2})\nonumber\\ 
        & -&(p_{1} \leftrightarrow p_{3}),\nonumber\\
        \mathcal{M}^{\mathrm{Type\,I}}_{box} & = & e^{2}B_{L}(0) \overline{u}(p_{1}) \gamma^{\mu}P_{L}u(p_{\ell})\overline{u}(p_{3}) \gamma_{\mu}P_{L}v(p_{2}),
%    \end{split}
    \label{eq4.1.2.2}
\end{eqnarray}
where again  $F_{E}^{\gamma} = iF_{M}^{\gamma}$. The photon magnetic and Z left-handed vector form factors, $F_{M}^{\gamma}(0)$ and
$F_{L}^{Z}(0)$ respectively, are evaluated at $Q^{2} = (p_{1} - p_{\ell})^{2}= 0$ because their leading terms are
momentum independent for small momentum transfer $Q^{2} \sim m^{2}_{\ell}$ whereas the photon left-handed vector form factor, $F_{L}^{\gamma}\left( (p_{1}-p_{\ell})^{2} \right)$, is linear in $Q^{2}$.\\
%The form factors $F_{M}^{\gamma}$ and $F_{E}^{\gamma}$ have the the same expressions than the eqs.(\ref{eq245}) and (\ref{eq248}), where we have supposed $m_{\nu_{i}^{l}} \ll M_{W}$, and $M_{M} \ll M_{j}$ with $m_{\nu_{i}^{l}}$ and $M_{j}$ the mass of light and heavy Majorana neutrinos, respectively. So, t
The complete $F_{M}^{\gamma}$ is given by
\begin{equation}
    F_{M}^{\gamma} = F_{M}^{\nu^{l}} + F_{M}^{\chi^{h}} = \frac{\alpha_{W}}{16\pi} \frac{m_{\ell}}{M_{W}^{2}} \sum_{j=1}^{3} \left( W_{\ell' j}W^{\dagger}_{\ell j}F_{M}^{\nu^{l}}(x_j) + \theta_{\ell' j}\theta^{\dagger}_{\ell j}F_{M}^{\chi^{h}}(y_j)\right).
    \label{eq4.1.2.3}
\end{equation}
%with $x=\frac{m_{\nu}^{2}}{M_{W}^{2}}$ and $y=\frac{M_{W}^{2}}{M_{j}^{2}}$.\\ 

The $F_{L}^{\gamma}$ form factor is obtained from topologies II, IV, V and VI in fig. \ref{fig:ell_to_ellgamma}, 
and it is given by
\begin{equation}
    F_{L}^{\gamma} = F_{L}^{\nu^{l}} + F_{L}^{\chi^{h}} = \frac{\alpha_{W}}{8\pi M_{W}^{2}} \sum_{j=1}^{3} \left( W_{\ell' j}W^{\dagger}_{\ell j}F_{L}^{\nu^{l}}(x_j) + \theta_{\ell' j}\theta^{\dagger}_{\ell j}F_{L}^{\chi^{h}}(y_j) \right),
    \label{eq4.1.2.4}
\end{equation}
where
\begin{eqnarray}
  %  \begin{split}
        F_{L}^{\nu^{l}}(x) & = & 2 M_{W}^{2} \Delta_{\epsilon} + Q^{2} \left(\frac{x^{2}(12-10x+x^{2})\mathrm{ln}x}{6(1-x)^{4}} - \frac{7x^{3}-x^{2}-12x}{12(1-x)^{3}} - \frac{5}{9}\right),\;\; \;\;\;\;\\
        F_{L}^{\chi^{h}}(y) & = & 2M_{W}^{2}\Delta_{\epsilon} +Q^{2} \left( - \frac{(12y^{2}-10y+1) \mathrm{ln}y}{6(1-y)^{4}} + \frac{20y^{3}-96y^{2}+57y+1}{36(1-y)^{3}}\right).\nonumber
  %  \end{split}
    \label{eq4.1.2.5}
\end{eqnarray}

We take in account the Z penguin diagrams that are shown in Figure \ref{fig:Z_penguins_Majorana}. These involve either purely light neutrinos, a mixing between light and heavy neutrinos, or diagrams in which only heavy neutrinos appear. The form factor from  $\nu^{l}$-diagrams in Figure \ref{fig:Z_penguins_Majorana} is given by
\begin{eqnarray}
%   \begin{split}
        F_{L}^{Z-\nu^{l}}(Q^{2}) &=&  \frac{\alpha_{W}}{8\pi c_{W}s_{W}}  \sum_{i,j=1}^{3} \Big[ W_{\ell' i} W^{\dagger}_{\ell i} F^{l}(x_{i}; Q^{2}) \nonumber\\ 
          & + &  W_{\ell' j}X_{ji}W^{\dagger}_{\ell i} \left( G^{l}(x_{i},x_{j};Q^{2}) + \sqrt{x_{i}x_{j}} H^{l}(x_{i},x_{j};Q^{2}) \right)\Big],
  % \end{split}
   \label{eq4.1.2.6}
\end{eqnarray}
 where
\begin{eqnarray}
%\begin{split}
    F^{l}(x_{i};Q^{2}) & = & -2c_{W}^{2} \left[ Q^{2} (\overline{C}_{1}+\overline{C}_{2}+\overline{C}_{12}) + 6\overline{C}_{00} - 1 \right] - (1-2s_{W}^{2})x_{i}\overline{C}_{00}  \nonumber\\
    & & - 2s_{W}^{2}x_{i}M_{W}^{2}\overline{C}_{0} - \frac{1}{2}(1-2s_{W}^{2})\left[ (2+x_{i}) \overline{B}_{1} + 1 \right], \nonumber\\
    G^{l} (x_{i},x_{j};Q^{2}) & = & -Q^{2}(C_{0}+C_{1}+C_{2}+C_{12})+2C_{00}-1-\frac{1}{2}x_{i}x_{j}M_{W}^{2}C_{0} ,\nonumber\\     H^{l} (x_{i},x_{j};Q^{2}) & = & M_{W}^{2}C_{0}+\frac{1}{2}Q^{2}C_{12}-C_{00}+\frac{1}{4}.
%\end{split}
\label{eq4.1.2.7}
\end{eqnarray}

Analytic expressions for the above functions in the low $Q^2$ limit are

\begin{eqnarray}
%    \begin{split}
        F^{l}(x_{i};0) & = & -\left( \frac{5}{2}-2s_{W}^{2} \right) \Delta_{\epsilon} + \frac{5x_{i}^{2} \ \mathrm{ln}x_{i}}{2(x_{i}-1)^{2}} - \frac{5x_{i}}{2(x_{i}-1)} + \frac{1}{4}, \\
    G^{l}(x_{i},x_{j};0) & = & \frac{1}{2} \left( \Delta_{\epsilon}-\frac{1}{2} \right) + \frac{1}{2(x_{i}-x_{j})} \left( \frac{(x_{j}-1)x_{i}^{2} \ \mathrm{ln}x_{i}}{x_{i}-1} - \frac{(x_{i}-1)x_{j}^{2} \ \mathrm{ln}x_{j}}{x_{j}-1} \right),\nonumber\\ 
        H^{l}(x_{i},x_{j};0) & = & - \frac{1}{4}\left( \Delta_{\epsilon} + \frac{1}{2} \right) + \frac{1}{4(x_{i} - x_{j})} \left( \frac{x_{i}(x_{i}-4) \ \mathrm{ln}x_{i}}{x_{i}-1} - \frac{x_{j}(x_{j}-4) \ \mathrm{ln}x_{j}}{x_{j}-1} \right) .\nonumber
 %   \end{split}
    \label{eq4.1.2.7.1}
\end{eqnarray}

The contribution from $\nu^{l}\chi^{h}-$diagrams in Figure \ref{fig:Z_penguins_Majorana} yields
\begin{eqnarray}
%    \begin{split}
        F_{L}^{Z-\nu^{l}\chi^{h}}(Q^{2}) & = & \frac{\alpha_{W}}{8\pi c_{W}s_{W}} \sum_{i,j=1}^{3} %\left
        \Bigg[ \theta_{\ell' j} W_{\ell i}^{\dagger} \left(Y_{ji}G^{lh}_{1}(x_{i},y_{j};Q^{2}) + Y_{ji}^{\dagger}\sqrt{\frac{x_{i}}{y_{j}}}H_{1}^{lh}(x_{i},y_{j};Q^{2}) \right) %\right
        .\nonumber\\  %\left.
        &+& W_{\ell' j} \theta_{\ell i}^{\dagger} \left(Y_{ji}^{\dagger}G^{lh}_{2}(x_{j},y_{i};Q^{2}) + Y_{ji}\sqrt{\frac{x_{j}}{y_{i}}}H_{2}^{lh}(x_{j},y_{i};Q^{2}) \right) %\right
        \Bigg],
 %   \end{split}
    \label{eq4.1.2.8}
\end{eqnarray}
where
\begin{eqnarray}
%    \begin{split}
        G_{1}^{lh} (x_{i},y_{j};Q^{2}) & = & -Q^{2}(C_{0}+C_{1}+C_{2}+C_{12})+2C_{00}-1-\frac{1}{2}\frac{x_{i}}{y_{j}}M_{W}^{2}C_{0} , \nonumber\\
 H_{1}^{lh} (x_{i},y_{j};Q^{2}) & = & M_{W}^{2}C_{0}+\frac{1}{2}Q^{2}C_{12}-C_{00}+\frac{1}{4}, \nonumber\\
        G_{2}^{lh} (x_{j},y_{i};Q^{2}) & = & -Q^{2}(C_{0}+C_{1}+C_{2}+C_{12})+2C_{00}-1-\frac{1}{2}\frac{x_{j}}{y_{i}}M_{W}^{2}C_{0} , \nonumber\\
        H_{2}^{lh}(x_{j},y_{i};Q^{2}) & = & M_{W}^{2}C_{0}+\frac{1}{2}Q^{2}C_{12}-C_{00}+\frac{1}{4}.
   % \end{split}
    \label{eq4.1.2.9}
\end{eqnarray}

These functions can be written in terms of those appearing for the light neutrino case previously:
\begin{equation}
%    \begin{split}
        \left\lbrace \left\lbrace {G,H}\right\rbrace_{1,2}\right\rbrace^{lh}(x_{i},y_{j};0)  = \left\lbrace \left\lbrace {G,H}\right\rbrace_{1,2}\right\rbrace^{l}(x_{i},x_{j};0) \quad \mathrm{with} \quad  \left( x_{j} \rightarrow \frac{1}{y_{j}} \right). \nonumber%\\
%        H^{lh}_{1}(x_{i},y_{j};0) & = H^{l}(x_{i},x_{j};0) \quad \mathrm{with} \quad  \left( x_{j} \rightarrow \frac{1}{y_{j}} \right), &
 %        G^{lh}_{2}(x_{j},y_{i};0) & = G^{l}(x_{i},x_{j};0) \quad \mathrm{with} \quad  \left( x_{i} \rightarrow \frac{1}{y_{i}} \right), &
  %       H^{lh}_{2}(x_{j},y_{i};0) & = H^{l}(x_{i},x_{j};0) \quad \mathrm{with} \quad  \left( x_{i} \rightarrow \frac{1}{y_{i}} \right),
 %   \end{split}
    \label{eq4.1.2.9.2}
\end{equation}

The $F_{L}^{Z-\chi^{h}}$ form factor, which stands for the contribution from $\chi^{h}$-diagrams, yields
\begin{eqnarray}
%   \begin{split}
        F_{L}^{Z-\chi^{h}}(Q^{2})  & = & \frac{\alpha_{W}}{8\pi c_{W}s_{W}} \sum_{i,j=1}^{3}  \Big[ \theta_{\ell' i} \theta^{\dagger}_{\ell i} F^{h}(y_{i};Q^{2}) \label{eq4.1.2.10} %\right. 
        \\ %\left.
        &+& \theta_{\ell' j} S_{ji} \theta^{\dagger}_{\ell i}\left(G^{h}(y_{i},y_{j};Q^{2}) + \frac{1}{\sqrt{y_{i}y_{j}}} H^{h}(y_{i},y_{j};Q^{2}) \right) \Big],%
% %  \end{split}
\nonumber
\end{eqnarray}
including the functions
\begin{eqnarray}
%\begin{split}
    F^{h}(y_{i};Q^{2}) & = & -2c_{W}^{2} \left[ Q^{2} (\overline{C}_{1}+\overline{C}_{2}+\overline{C}_{12}) + 6\overline{C}_{00} - 1 \right] - (1-2s_{W}^{2})\frac{1}{y_{i}}\overline{C}_{00}  \nonumber\\
    & - & 2s_{W}^{2}\frac{1}{y_{i}}M_{W}^{2}\overline{C}_{0} -\frac{1}{2}(1-2s_{W}^{2})\left[ (2+\frac{1}{y_{i}}) \overline{B}_{1} + 1 \right], \nonumber\\
    G^{h} (y_{i},y_{j};Q^{2}) & = & -Q^{2}(C_{0}+C_{1}+C_{2}+C_{12})+2C_{00}-1-\frac{1}{2}\frac{1}{y_{i}y_{j}}M_{W}^{2}C_{0} , \nonumber\\
    H^{h} (y_{i},y_{j};Q^{2}) & = & M_{W}^{2}C_{0}+\frac{1}{2}Q^{2}C_{12}-C_{00}+\frac{1}{4}.
%\end{split}
\label{eq4.1.2.11}
\end{eqnarray}

Their analytic expressions, for low $Q^2$,  are
\begin{eqnarray}
 %   \begin{split}
        F^{h}(y_{i};0) & = & -\left( \frac{5}{2}-2s_{W}^{2} \right) \Delta_{\epsilon} - \frac{5 \ \mathrm{ln}y_{i}}{2(1-y_{i})^{2}} - \frac{5}{2(1-y_{i})} + \frac{1}{4}, \\
        G^{h}(y_{i},y_{j};0) & = & \frac{1}{2} \left( \Delta_{\epsilon}-\frac{1}{2} \right) + \frac{1}{2(y_{j}-y_{i})} \left(- \frac{(1-y_{j})\ \mathrm{ln}y_{i}}{(1-y_{i})} + \frac{(1-y_{i}) \ \mathrm{ln}y_{j}}{(1-y_{j})} \right), \nonumber\\ H^{h}(y_{i},y_{j};0) & = & - \frac{1}{4}\left( \Delta_{\epsilon} + \frac{1}{2} \right) + \frac{1}{4(y_{j} - y_{i})} \left( -\frac{(1-4y_{i})y_{j} \ \mathrm{ln}y_{i}}{(1-y_{i})} + \frac{(1 - 4y_{j})y_{i} \ \mathrm{ln}y_{j}}{(1-y_{j})} \right).\nonumber
   % \end{split}
    \label{eq4.1.2.11.2}
\end{eqnarray}
Ultraviolet divergences cancel, thanks to the relations (\ref{eq4.1.2.11.3}).
\begin{figure}[!ht]
    \centering
    \includegraphics[scale = 0.7]{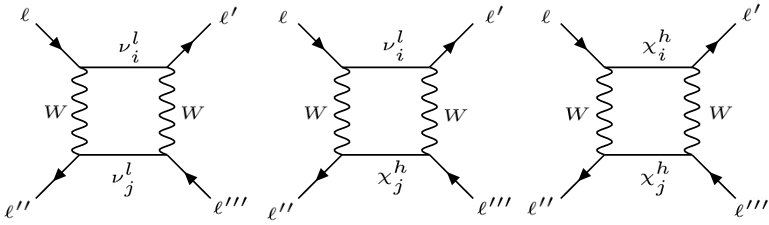}
    \caption{Box diagrams contributing to Type I and II $\ell \rightarrow \ell^{\prime} \ell^{\prime \prime} \bar{\ell}^{\prime \prime \prime}$ decays.}
    \label{fig:box_Majorana}
\end{figure}

The box diagrams, represented in fig. \ref{fig:box_Majorana}, yield
\begin{equation}
    F_{B}^{\nu_{i}^{l}\nu_{j}^{l}} = \frac{\alpha_{W}}{16 \pi M_{W}^{2} s_{W}^{2}} \sum_{i,j=1}^{3}  W_{\ell i}W^{\dagger}_{\ell' i} W^{\dagger}_{\ell' j} W_{\ell'  j} f^{l}_B(y_i,y_j),
    \label{eq214}
\end{equation}
\begin{equation}
    F_{B}^{\nu_{i}^{l}\chi_{j}^{h}} = \frac{\alpha_{W}}{16 \pi M_{W}^{2} s_{W}^{2}} \sum_{i,j=1}^{3}W_{\ell i}W^{\dagger}_{\ell' i} \theta^{\dagger}_{\ell' j} \theta_{\ell'  j} f^{lh}_B(y_i,x_j),
    \label{eq215}
\end{equation}
\begin{equation}
    F_{B}^{\chi_{i}^{h}\chi_{j}^{h}} = \frac{\alpha_{W}}{16 \pi M_{W}^{2} s_{W}^{2}} \sum_{i,j=1}^{3}  \theta^{\dagger}_{\ell i}\theta_{\ell' i} \theta^{\dagger}_{\ell' j} \theta_{\ell'  j}f^{h}_B(x_i,x_j) ,
    \label{eq216}
\end{equation}
with
\begin{eqnarray}
%    \begin{split}
        f_{B}^{l}(y_{i},y_{j}) & = &   \left( 1 + \frac{1}{4}y_{i}y_{j} \right)\bar{d}_{0}(y_{i},y_{j}) - 2y_{i}y_{j}d_{0}(y_{i},y_{j}), \nonumber\\
        f_{B}^{lh}(y_{i},x_{j}) & = &  \left( 1 + \frac{1}{4} \frac{y_{i}}{x_{j}} \right)\bar{d}^{lh}_{0}(y_{i},x_{j}) - 2\frac{y_{i}}{x_{j}}d^{lh}_{0}(y_{i},x_{j}),\nonumber\\
        f_{B}^{h}(x_{i},x_{j}) & = &  \left( 1 + \frac{1}{4} \frac{1}{x_{i}x_{j}} \right)\bar{d}^{h}_{0}(x_{i},x_{j}) - 2\frac{1}{x_{i}x_{j}}d^{h}_{0}(x_{i},x_{j}).
   % \end{split}
    \label{eq217}
\end{eqnarray}

After integrating the three-body phase space the decay width reads \cite{delAguila:2019htj}
\begin{eqnarray}
%    \begin{split}
        \Gamma(\ell \rightarrow \ell' \overline{\ell'} \ell') & = & \frac{\alpha^{2}m_{\ell}^{5}}{96\pi} \Big[ 3 |A_{L}|^{2} +2|A_{R}|^{2} \left( 8 \ \mathrm{ln}\frac{m_{\ell}}{m_{\ell'}} - 13 \right) + 2|F_{LL}|^{2} + |F_{LR}|^{2} + \frac{1}{2}|B_{L}|^{2} \nonumber\\ &-& \left( 6A_{L}A_{R}^{*} - (A_{L}-2A_{R})(2F_{LL}^{*}+F_{LR}^{*}+B_{L}^{*}) - F_{LL}B_{L}^{* } + \mathrm{h.c.}\right) \Big],
 %   \end{split}
    \label{eq4.1.2.12}
\end{eqnarray}
where we have defined
\begin{equation}
    A_{L} = \frac{F_{L}^{\gamma}}{Q^{2}}, \quad A_{R} = \frac{2F_{M}^{\gamma}(0)}{m_{\ell}}, \quad F_{LL} = - \frac{g_{L}F_{L}^{Z}(0)}{eM_{Z}^{2}}, \quad F_{LR} = - \frac{g_{R}F_{L}^{Z}(0)}{eM_{Z}^{2}}, \quad B_{L} = B_{L}(0),
    \label{eq4.1.2.13}
\end{equation}
with $g_{L,R}$ the corresponding Z couplings to the charged lepton $\ell'$.

\subsubsection{Type II: $\ell\to\ell'\ell''\bar{\ell}'''$ with $\ell\neq\ell'\neq\ell''=\ell'''$}
This type of decays can be related to the previous ones, although in this case there are no crossed penguin diagrams contributions. Similarly,  there is no $1/2$ factor in the phase space integration, as all leptons are distinguishable.  Instead, there are additional diagrams for the box contributions at this order, swapping $\ell'$ and $\ell''$. 

With these comments in mind, its decay width reads
\begin{eqnarray}
%    \begin{split}
        \Gamma (\ell \rightarrow \ell' \overline{\ell''} \ell'') & = & \frac{\alpha^{2}m_{\ell}^{5}}{96\pi} \Big[ 2|A_{L}|^{2} + 4|A_{R}|^{2} \left( 4 \ \mathrm{ln}\frac{m_{\ell}}{m_{\ell''}}-7 \right) + |F_{LL}|^{2} + |F_{LR}|^{2} + |B_{L}|^{2} \nonumber\\ &
    -    &  \left( 4A_{L}A_{R}^{*} - (A_{L}-2A_{R})\left( F_{LL}^{*} + F_{LR}^{*} + \frac{B_{L}^{*}}{2} \right) - F_{LL}\frac{B_{L}^{*}}{2} + \mathrm{h.c.} \right) \Big].\;\;\;\;\;\;\;\;
%    \end{split}
    \label{eq4.1.3.1}
\end{eqnarray}

\subsubsection{Type III: $\ell\to\ell'\ell''\bar{\ell}'''$ with $\ell\neq\ell'=\ell''\neq\ell'''$}
These processes only have box contributions. In addition to box diagrams in figure \ref{fig:box_Majorana}, there are contributions coming from box diagrams with LNV vertices shown in figure \ref{fig:LNV}. They are indicated in the next equation and given in turn below:

\begin{figure}[!ht]
    \centering
    \includegraphics[scale = 0.7]{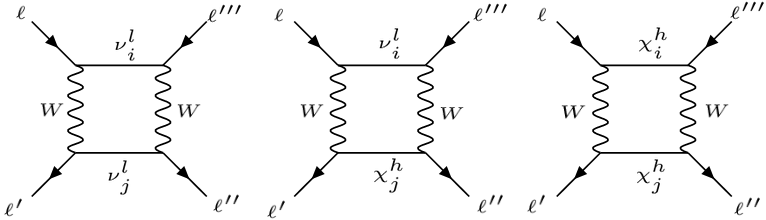}
    \caption{Box contributions to Type III $\ell\to\ell'\ell''\bar{\ell}'''$ decays.}
    \label{fig:LNV}
\end{figure}

\begin{equation}
    F_{B} = F_{B}^{\nu_{i}^{l}\nu_{j}^{l}} + F_{B}^{\nu_{i}^{l}\chi_{j}^{h}} + F_{B}^{\chi_{i}^{h}\chi_{j}^{h}} + F_{B-LNV}^{\nu_{i}^{l}\nu_{j}^{l}} + F_{B-LNV}^{\nu_{i}^{l}\chi_{j}^{h}} + F_{B-LNV}^{\chi_{i}^{h}\chi_{j}^{h}},
    \label{eq262}
\end{equation}
where \cite{Ilakovac:1994kj}
\begin{eqnarray}
%    \begin{split}
        F_{B}^{\nu_{i}^{l}\nu_{j}^{l}} & = & \frac{\alpha_{W}}{16 \pi M_{W}^{2} s_{W}^{2}} \sum_{i,j=1}^{3} \{ W_{\ell i}W^{\dagger}_{\ell' i} W_{\ell''' j} W_{\ell''  j}^{\dagger} + (\ell' \leftrightarrow \ell'') \} f_{B}^{l}(y_{i},y_{j}),\nonumber\\ F_{B}^{\nu_{i}^{l}\chi_{j}^{h}} & = & \frac{\alpha_{W}}{16 \pi M_{W}^{2} s_{W}^{2}} \sum_{i,j=1}^{3} \{ W_{\ell i}W^{\dagger}_{\ell' i} \theta^{\dagger}_{\ell''' j} \theta_{\ell''  j} + (\ell' \leftrightarrow \ell'') \} f_{B}^{lh}(y_{i},x_{j}),\nonumber\\ F_{B}^{\chi_{i}^{h}\chi_{j}^{h}} & = & \frac{\alpha_{W}}{16 \pi M_{W}^{2} s_{W}^{2}} \sum_{i,j=1}^{3} \{ \theta^{\dagger}_{\ell i}\theta_{\ell' i} \theta^{\dagger}_{\ell''' j} \theta_{\ell''  j} + (\ell' \leftrightarrow \ell'') \} f^{h}_{B}(x_{i},x_{j}),\nonumber\\ F_{B-LNV}^{\nu_{i}^{l}\nu_{j}^{l}} & = & \frac{\alpha_{W}}{16 \pi M_{W}^{2}s_{W}^{2}} \sum_{i,j}^{3} W_{\ell i}W_{\ell' j}^{\dagger}W_{\ell''' i}W_{\ell'' j}^{\dagger} f_{B}^{l-LNV}(y_{i},y_{j}),\nonumber\\ 
 F_{B-LNV}^{\nu_{i}^{l}\chi_{j}^{h}} & = & \frac{\alpha_{W}}{16 \pi M_{W}^{2} s_{W}^{2}} \sum_{i.j=1}^{3} W_{\ell i}\theta_{\ell' j} W_{\ell''' i} \theta_{\ell''  j}  f_{B}^{lh-LNV}(y_{i},x_{j}), \nonumber\\
   F_{B-LNV}^{\chi_{i}^{h}\chi_{j}^{h}} & = & \frac{\alpha_{W}}{16 \pi M_{W}^{2} s_{W}^{2}} \sum_{i,j=1}^{3}  \theta^{\dagger}_{\ell i}\theta_{\ell' j} \theta^{\dagger}_{\ell''' i} \theta_{\ell''  j} f^{h-LNV}_{B}(x_{i},x_{j}).
%    \end{split}
    \label{eq263}
\end{eqnarray}
For small (large) light (heavy) neutrino masses, i. e. $x_i,y_j\to0$, the previous expressions simplify to
\begin{equation}
    F_{B}^{\nu_{i}^{l}\nu_{j}^{l}} \approx - \frac{\alpha_{W}}{16 \pi M_{W}^{2} s_{W}^{2}} \sum_{i,j=1}^{3} \{ W_{\ell i}W^{\dagger}_{\ell' i} W_{\ell''' j} W_{\ell''  j}^{\dagger} + (\ell' \leftrightarrow \ell'') \} \left[1 + (y_{j} + y_{i}) \left( 1 + \mathrm{ln}y_{j} \right) \right],
    \label{eq265}
\end{equation}
\begin{equation}
  F_{B}^{\nu_{i}^{l}\chi_{j}^{h}} \approx \frac{\alpha_{W}}{16 \pi M_{W}^{2} s_{W}^{2}} \sum_{i,j=1}^{3} \{ W_{\ell i}W^{*}_{\ell' i} \theta^{\dagger}_{\ell''' j} \theta_{\ell''  j} + (\ell' \leftrightarrow \ell'') \} \left[ x_{j} (1 + \mathrm{ln}x_{j}) + \frac{1}{4} y_{i} (\mathrm{ln}x_{j} - 7) \right],
  \label{eq267}
\end{equation}
\begin{equation}
    F_{B-LNV}^{\nu_{i}^{l}\nu_{j}^{l}} = \frac{\alpha_{W}}{16 \pi M_{W}^{2}s_{W}^{2}} \sum_{i,j=1}^{3} \{ W_{\ell i}W_{\ell' j}^{\dagger}W_{\ell''' i}W_{\ell'' j}^{\dagger} \} \left[ 2 \sqrt{y_{i}y_{j}} (1 + 2 \ \mathrm{ln}y_{j}) \right],
    \label{eq270}
\end{equation}
\begin{equation}
    F_{B-LNV}^{\nu_{i}^{l}\chi_{j}^{h}}  \approx \frac{\alpha_{W}}{32 \pi M_{W}^{2} s_{W}^{2}} \sum_{i,j=1}^{3} \{ W_{\ell i}\theta_{\ell' j} W_{\ell''' i} \theta_{\ell''  j} \} \left[ 2 \sqrt{y_{i}x_{j}} (\mathrm{ln}x_{j} - 1) \right],
    \label{eq272}
\end{equation}
and
\begin{eqnarray}
%    \begin{split}
        \left. f_{B}^{h-LNV}(x_{i},x_{j}) \right|_{x_{i},x_{j} \rightarrow 0} &\approx& \sqrt{\frac{x_{i}}{x_{j}^{3}}}  \mathrm{ln} \left( \frac{x_{j}}{x_{i}} \right) + \sqrt{\frac{x_{i}}{x_{j}}} (2 \ \mathrm{ln}x_{i} + 1) + 3 \sqrt{x_{i} x_{j}} (\mathrm{ln}x_{j} + 1)\nonumber\\ &+& \frac{1}{\sqrt{x_{i}x_{j}}} (\mathrm{ln}x_{j} + 1) + \sqrt{\frac{x_{j}}{x_{i}}} (2 \ \mathrm{ln}x_{j} + 1).
 %   \end{split}
    \label{eq273}
\end{eqnarray}
We recall that the $y_j\to0$ limit is not physical (as perturbative unitarity limits the maximum value of $M_{N_j}$ to some \textcolor{black}{tenths of} TeVs), so that the previous expressions are of course free of infrared singularities.
The corresponding total decay width is just given by
\begin{equation}
    \Gamma (\ell \rightarrow \ell' \ell'' \bar{\ell}''') = \frac{\alpha^{2}m_{\ell}^{5}}{192 \pi} \left | F_{B} \right |^{2}.
    \label{eq274}
\end{equation}

\subsection{$\mu\to e$ conversion in nuclei}\label{subsec:mutoeinnuc}
The $\mu - e$ conversion in nuclei has penguin and box contributions as $\ell \rightarrow \ell' \ell'' \bar{\ell}'''$ decays, replacing the last two leptons by a q = u or d quark. It has no crossed penguin diagrams because the lower fermionic line, where the gauge boson is attached, is now a coherent sum of quarks composing the probed nucleus. There is also no crossed box contribution due to the exchange of leptons.\\

The matrix element is
\begin{equation}
    \mathcal{M}^{\mu q \rightarrow eq} = \mathcal{M}^{\mu q \rightarrow eq}_{\gamma} + \mathcal{M}^{\mu q \rightarrow eq}_{Z} + \mathcal{M}^{\mu q \rightarrow eq}_{box},
    \label{eq4.3.1}
\end{equation}
with the amplitudes defined as
%In these processes there appear a couple of new box diagrams in which $q=u$ and $d$ quarks are involved, as can be seen in fig \ref{fig:mue_conversion}. It has the same $\gamma-$ and $Z-$penguin diagram contributions  computed before.
%\begin{figure}[h!]
%    \centering
%    \includegraphics[scale=0.5]{Figures/mu_e_conversion/mu_e_conversion.png}
%    \caption{Box diagrams contributing to $\mu-e$ conversion in nuclei considering light-heavy Majorana neutrinos.}
 %   \label{fig:mue_conversion}
%\end{figure}
%
%In this case the box contributions, with $q=u,d$
%\begin{equation}
%    \mathcal{M}^{\mu q \rightarrow eq} =  \mathcal{M}^{\mu q \rightarrow eq}_{box},
 %   \label{eq4.3.1}
%\end{equation}
%have the following  amplitude 
\cite{delAguila:2019htj}
\begin{eqnarray}
%    \begin{split}
        \mathcal{M}^{\mu q \rightarrow eq}_{\gamma} & = & \overline{u}(p_{1})e\left[ iF_{M}^{\gamma}(0)2P_{R}\sigma^{\mu\nu}(p_{1}-p_{\ell})_{\nu}+F_{L}^{\gamma}((p_{1}-p_{\ell})^{2})\gamma^{\mu}P_{L} \right]u(p_{\ell}) \nonumber\\ \times &&\frac{1}{(p_{1}-p_{\ell})^{2}} \overline{u}(p_{3})\gamma_{\mu}(g_{Lq}^{\gamma}P_{L}+g_{Rq}^{\gamma}P_{R})v(p_{2}),\nonumber\\
        \mathcal{M}^{\mu q \rightarrow eq}_{Z} & = & \overline{u}(p_{1}) \left( -eF_{L}^{Z}(0) \right)\gamma^{\mu}P_{L}u(p_{\ell})\frac{1}{M_{Z}^{2}}\overline{u}(p_{3})\gamma_{\mu}\left(g_{Lq}^{Z}P_{L}+g_{Rq}^{Z}P_{R} \right)v(p_{2}), \nonumber\\
        \mathcal{M}^{\mu q \rightarrow eq}_{box} & = & e^{2}B_{L}^{q}(0) \overline{u}(p_{1}) \gamma^{\mu}P_{L}u(p_{\ell})\overline{u}(p_{3}) \gamma_{\mu}P_{L}v(p_{2}),
%    \end{split}
    \label{eq4.3.2}
\end{eqnarray}
given in terms of the form factors
\begin{eqnarray}
    %\begin{split}
        A_{1L} & = & \frac{F_{L}^{\gamma}(Q^{2})}{Q^{2}} = \frac{\alpha_{W}}{8\pi M_{W}^{2}} \sum_{i = 1}^{3} \theta_{e i}\theta_{\mu i}^{\dagger} \left( - \frac{(12y_{i}^{2} -10y_{i} +1)\ \mathrm{ln}y_{i}}{6(1-y_{i})^{4}} + \frac{20y_{i}^{3} - 96y_{i}^{2} + 57y_{i} +1}{36(1-y_{i})^{3}} \right),\nonumber\\
A_{2R} & = & \frac{2 F_{M}^{\gamma}(0)}{m_{\mu}} = \frac{\alpha_{W}}{8\pi M_{W}^{2}} \sum_{i = 1}^{3} \theta_{e i}\theta_{\mu i}^{\dagger} \left( - \frac{2y_{i}^{3} - 7y_{i}^{2}+11y_{i}}{4(1-y_{i})^{3}} + \frac{3y_{i} \ \mathrm{ln}y_{i}}{2(1-y_{i})^{4}} \right), \nonumber\\        F_{LL}^{u} & = & - \frac{F_{L}^{Z}(0)g_{Lu}^{Z}}{M_{Z}^{2}} = - \frac{\alpha_{W}}{16\pi M_{W}^{2} s_{W}^{2}} \left(1-\frac{4}{3}s_{W}^{2}\right) \sum_{i,j=1}^{3} \Bigg[ \theta_{ei}\theta_{\mu i}^{\dagger} \left( -\frac{5 \ \mathrm{ln}y_{i}}{2(1-y_{i})^{2}} - \frac{5}{2(1-y_{i})} \right)    \nonumber\\  && + \theta_{ej}S_{ji}\theta_{\mu i}^{\dagger} \Bigg( -\frac{1}{2(y_{i}-y_{j})} \Bigg(-\frac{(1-y_{j}) \ \mathrm{ln}y_{i}}{(1-y_{i})} + \frac{(1-y_{i}) \ \mathrm{ln}y_{j}}{(1-y_{j})}  \nonumber\\
 && - \frac{1}{\sqrt{y_{i}y_{j}}}  \frac{1}{4(y_{i}-y_{j})} \left( -\frac{y_{j}(1-4y_{i}) \ \mathrm{ln}y_{i}}{1-y_{i}} + \frac{y_{i}(1-4y_{j}) \ \mathrm{ln}y_{j}}{1-y_{j}}\right) \Bigg)\Bigg) \Bigg],\nonumber\\  F_{LR}^{u} & = & - \frac{F_{L}^{Z}(0)g_{Ru}^{Z}}{M_{Z}^{2}} = \frac{\alpha_{W}}{12 \pi M_{W}^{2}} \sum_{i,j=1}^{3} \Bigg[ \theta_{ei}\theta_{\mu i}^{\dagger} \left( -\frac{5 \ \mathrm{ln}y_{i}}{2(1-y_{i})^{2}} - \frac{5}{2(1-y_{i})} \right) \nonumber\\  &&+ \theta_{ej}S_{ji}\theta_{\mu i}^{\dagger} \Bigg( -\frac{1}{2(y_{i}-y_{j})} \left(-\frac{(1-y_{j}) \ \mathrm{ln}y_{i}}{(1-y_{i})} + \frac{(1-y_{i}) \ \mathrm{ln}y_{j}}{(1-y_{j})} \right)  \nonumber\\ && - \frac{1}{\sqrt{y_{i}y_{j}}}  \frac{1}{4(y_{i}-y_{j})} \left( -\frac{y_{j}(1-4y_{i}) \ \mathrm{ln}y_{i}}{1-y_{i}} + \frac{y_{i}(1-4y_{j}) \ \mathrm{ln}y_{j}}{1-y_{j}}\right) \Bigg) \Bigg], \nonumber\\
F_{LL}^{d} & = & - \frac{F_{L}^{Z}(0)g_{Lu}^{Z}}{M_{Z}^{2}} = - \frac{\alpha_{W}}{16\pi M_{W}^{2} s_{W}^{2}} \left(-1+\frac{2}{3}s_{W}^{2}\right) \sum_{i,j=1}^{3} \Bigg[ \theta_{ei}\theta_{\mu i}^{\dagger} \left( -\frac{5 \ \mathrm{ln}y_{i}}{2(1-y_{i})^{2}} - \frac{5}{2(1-y_{i})} \right) \nonumber\\
  && + \theta_{ej}S_{ji}\theta_{\mu i}^{\dagger} \Bigg( -\frac{1}{2(y_{i}-y_{j})} \left(-\frac{(1-y_{j}) \ \mathrm{ln}y_{i}}{(1-y_{i})} + \frac{(1-y_{i}) \ \mathrm{ln}y_{j}}{(1-y_{j})} \right)   \nonumber\\   &&- \frac{1}{\sqrt{y_{i}y_{j}}}  \frac{1}{4(y_{i}-y_{j})} \left( -\frac{y_{j}(1-4y_{i}) \ \mathrm{ln}y_{i}}{1-y_{i}} + \frac{y_{i}(1-4y_{j}) \ \mathrm{ln}y_{j}}{1-y_{j}}\right) \Bigg) \Bigg], \nonumber\\ F_{LR}^{d} & = & - \frac{F_{L}^{Z}(0)g_{Ru}^{Z}}{M_{Z}^{2}} = - \frac{\alpha_{W}}{24 \pi M_{W}^{2}} \sum_{i,j=1}^{3} \Bigg[ \theta_{ei}\theta_{\mu i}^{\dagger} \left( -\frac{5 \ \mathrm{ln}y_{i}}{2(1-y_{i})^{2}} - \frac{5}{2(1-y_{i})} \right) \nonumber\\  &&+ \theta_{ej}S_{ji}\theta_{\mu i}^{\dagger} \Bigg( -\frac{1}{2(y_{i}-y_{j})} \left(-\frac{(1-y_{j}) \ \mathrm{ln}y_{i}}{(1-y_{i})} + \frac{(1-y_{i}) \ \mathrm{ln}y_{j}}{(1-y_{j})} \right)   \nonumber\\   &&- \frac{1}{\sqrt{y_{i}y_{j}}}  \frac{1}{4(y_{i}-y_{j})} \left( -\frac{y_{j}(1-4y_{i}) \ \mathrm{ln}y_{i}}{1-y_{i}} + \frac{y_{i}(1-4y_{j}) \ \mathrm{ln}y_{j}}{1-y_{j}}\right) \Bigg) \Bigg], \nonumber\\
B_{L}^{d} & = & \frac{\alpha_{W}}{16 \pi M_{W}^{2} s_{W}^{2}} \sum_{i=1}^{3} \theta^{\dagger}_{\mu i}\theta_{e i} \left( |V_{td}|^{2} \left[ f_{B_{d}}(y_{i},x_{t}) - f_{B_{d}}(y_{i},0)  \right] - f_{B_{d}}(y_{i},0) \right), \nonumber\\   B_{L}^{u} & = & \frac{\alpha_{W}}{16 \pi M_{W}^{2} s_{W}^{2}} \sum_{i=1}^{3} \theta^{\dagger}_{\mu i}\theta_{e i} f_{B_{u}}(y_{i},0),
  %  \end{split}
    \label{5.7.1}
\end{eqnarray}
where
\begin{eqnarray}
        %\begin{split}
            f_{B_{d}}(y_{i},x_{t}) & = & \left( 1 + \frac{1}{4} \frac{x_{t}}{y_{i}} \right) \left[ \frac{y_{i} \ \mathrm{ln}y_{i}}{(1-y_{i})^{2}(1-y_{i}x_{t})} + \frac{y_{i}x_{t}^{2} \ \mathrm{ln}x_{t}}{(1-x_{t})^{2}(1-y_{i}x_{t})} + \frac{y_{i}}{(1-y_{i})(1-x_{t})} \right] \nonumber\\
            && - 2\frac{x_{t}}{y_{i}}\left[ \frac{y_{i}^{2} \ \mathrm{ln}y_{i}}{(1-y_{i})^{2}(1-y_{i}x_{t})} + \frac{x_{t}y_{i} \  \mathrm{ln}x_{t}}{(1-x_{t})^{2}(1-y_{i}x_{t})} + \frac{y_{i}}{(1-y_{i})(1-x_{t})} \right], \nonumber\\    f_{B_{d}}(y_{i},0) & = & \bar{d}^{lh}_{0}(y_{i},0) = \frac{y_{i} \ \mathrm{ln}y_{i}}{(1-y_{i})^{2}} + \frac{y_{i}}{(1-y_{i})},\nonumber\\
  f_{B_{u}}(y_{i},0) & = & - 4 \bar{d}^{lh}_{0}(y_{i},0) = -4 \left( \frac{y_{i} \ \mathrm{ln}y_{i}}{(1-y_{i})^{2}} + \frac{y_{i}}{(1-y_{i})} \right),
        %\end{split}
        \label{5.7.2}
\end{eqnarray}
with $x_{t} = m_{t}^{2}/M_{W}^{2}$, being $m_t$ the top quark mass.\\
The form factors $F_{M}^{\gamma}(0)$, $F_{L}^{\gamma}$ and $F_{L}^{Z}(0)$ are given by (\ref{eq4.1.2.3}), (\ref{eq4.1.2.4}), and (\ref{eq4.1.2.10}), respectively while the couplings $g_{L(R)q}^{Z}$  read \cite{Hernandez-Tome:2019lkb}
\begin{equation}
    %\begin{split}
        g_{Lu}^{Z} = \frac{1 - \frac{4}{3}s_{W}^{2}}{2s_{W}c_{W}}, \qquad g_{Ru}^{Z} = - \frac{2s_{W}}{3c_{W}},  \qquad
        g_{Ld}^{Z}  = \frac{-1 + \frac{2}{3}s_{W}^{2}}{2s_{W}c_{W}}, \qquad g_{Rd}^{Z} = \frac{s_{W}}{3c_{W}}.
    %\end{split}
    \label{eq4.3.3}
\end{equation}
%Conventionally, the $\gamma$ and $Z$ penguin form factors are rewritten in terms of the $A_{1L}$, $A_{2R}$, $F_{LL}$ and $F_{LR}$ functions:
%\begin{eqnarray}
 %    F_{L}^{\gamma}(Q^{2}) &\equiv& Q^{2}A_{1L}, \quad F_{M}^{\gamma}(Q^{2}) \approx F_{M}^{\gamma}(0) \equiv \frac{m_{\ell}}{2}A_{2R},\nonumber \\
  %   F_{LL}^{\gamma} &=& - \frac{F_{L}^{Z}(0)g_{Lq}^{Z}}{M_{Z}^{2}}, \quad F_{LR}^{\gamma} = - \frac{F_{L}^{Z}(0)g_{Rq}^{Z}}{M_{Z}^{2}}
%\end{eqnarray}
In terms of the latter, the $\mu - e$ conversion rate in a nucleus with Z protons and $N = A-Z$ neutrons yields \cite{delAguila:2019htj, Hernandez-Tome:2019lkb}
\begin{eqnarray}
    %\begin{split}
        \mathcal{R} & = & \frac{\alpha^{5}Z_{\mathrm{eff}}^{4}}{\Gamma_{\mathrm{Capt}}Z} F_{P}^{2}m_{\mu}^{5} \left | 2Z (A_{1L}+A_{2R}) - (2Z + N)(F_{LL}^{u}+F_{LR}^{u}+B_{L}^{u})\right. \label{eq4.3.12} \nonumber \\ 
        &- & \left. (Z + 2N)(F_{LL}^{d}+F_{LR}^{d}+B_{L}^{d}) \right|^{2}, 
    %\end{split}
\end{eqnarray}
where $Z_{\mathrm{eff}}$ is the nucleus effective charge for the muon and $F_{P}$ the associated form factor. In Table \ref{tab:inputs_mu_e_conversion} we gather the input parameters for Al, Ti and Au \cite{delAguila:2019htj,Hernandez-Tome:2019lkb, R Kitano,T. Suzuki}.  
\begin{table}[!ht]
    \centering
    \begin{tabular}{l||c c c c c}
    \hline
    \hline
    Nucleus & N & Z & $\mathrm{Z}_{\mathrm{eff}}$ & $\mathrm{F}_{\mathrm{P}}$ & $\Gamma_{\mathrm{Capt.}} [\mathrm{GeV}]$ \\
    \hline
    $^{27}_{13}$Al & 14 & 13 & 11.5 & 0.64 & 4.6 $\times 10^{-19}$ \\
    $^{48}_{22}$Ti & 26 & 22 & 17.6 & 0.54 & 1.7 $\times 10^{-18}$ \\
    $^{197}_{79}$Au & 118 & 79 & 33.5 & 0.16 & 8.6 $\times 10^{-18}$ \\
    \hline
    \end{tabular}
    \caption{Input parameters for different nuclei.}
    \label{tab:inputs_mu_e_conversion}
\end{table}

\section{Phenomenology}\label{sec:Phenomenology}
In this section we show the numerical analysis for each LFV processes described previously through Monte Carlo simulations. We will follow the light Majorana neutrinos massless approximation. Therefore, only diagrams involving heavy Majorana neutrinos contribute. We begin the discussion with a joint analysis considering $Z\to\bar{\ell}\ell'$, Type I and II $\ell\to\ell'\ell''\bar{\ell}'''$ decays and $\mu-e$ conversion rate in nuclei, as they share the same free parameters: three heavy neutrinos masses $\{M_{i}\}_{i=1,2,3}$ and neutral couplings given by the $(\theta S \theta^{\dagger})$ entries. Afterwards, we focus in LFV Type III (well known as `wrong sign' processes) that will bind directly the LNV couplings. In the following subsections we are using the limits of $(\theta \theta^{\dagger})_{\ell' \ell}$ previously obtained from the $\ell \rightarrow \ell' \gamma$ decays,  (\ref{eq44}).\\
All processes analyzed have 3 common free parameters which are the heavy neutrino masses $M_{i}$ that will run from %$3
$\textcolor{black}{15}$ to $%5
\textcolor{black}{20}$ TeV. We decided to take this interval based on the experience gained by doing  simplified analysis for each process separately~\footnote{We do not present them here and only quote that the results on the individual processes agree with the joint analysis that we will discuss next.}.  \textcolor{black}{This range of $\mathrm{M_i}$ values corresponds to $f\in[1.2,1.6]$ TeV, which is currently allowed (see e. g. Ref.~\cite{delAguila:2019htj}).}

We mention at this point that the LNV contributions that we are studying within the LHT also induce LNV  semileptonic tau decays (analogously to neutrinoless double beta decays, but also with LFV). Of course their rates are very much suppressed as there is no resonant enhancement of the Majorana neutrino exchanges. Specifically, for typical values of the relevant parameters (that are allowed considering all other processes that we analyze in the remainder of this section) we get
\begin{eqnarray}
  & & BR(\tau^-\to e^+\pi^-\pi^-)\leq%3\times10^{-27}
  \textcolor{black}{1.6\times 10^{-29}}\,,\quad BR(\tau^-\to \mu^+\pi^-\pi^-)\leq%2\times10^{-26}
  \textcolor{black}{8.9\times 10^{-29}}\,, \nonumber\\
  & & BR(\tau^-\to e^+K^-K^-)\leq%5\times10^{-30}
  \textcolor{black}{2.5\times 10^{-32}}\,,\quad BR(\tau^-\to \mu^+K^-K^-)\leq%2\times10^{-29}
  \textcolor{black}{1.4\times 10^{-31}}\,, \nonumber\\
  & & BR(\tau^-\to e^+\pi^-K^-)\leq%1\times10^{-28}
  \textcolor{black}{6.7\times 10^{-31}}\,,\quad BR(\tau^-\to \mu^+\pi^-K^-)\leq%6\times10^{-28}
  \textcolor{black}{3.8\times 10^{-30}}\,, \nonumber\\
\end{eqnarray}
which are \textcolor{black}{more than} twenty orders of magnitude below current limits \cite{ParticleDataGroup:2020ssz}. Much more interesting are the processes presented in the next subsections. For some of them, average values of the branching ratios or conversion rates are within one order of magnitude of current upper limits, as we will see.

\subsection{Joint Analysis for $Z\to\bar{\ell}\ell'$, Type I and II $\ell\to\ell'\ell''\bar{\ell}'''$ decays and $\mu \to e$ conversion in nuclei}\label{sec:Joint}
In this part we do a global analysis of the following 10 processes: LFV Z decays $Z \rightarrow \bar{\mu}e$, $Z \rightarrow \bar{\tau}e$, and $Z \rightarrow \bar{\tau}\mu$; LFV Type I $\mu \rightarrow ee \bar{e}$, $\tau \rightarrow ee\bar{e}$ and $\tau \rightarrow \mu \mu \bar{\mu}$; LFV Type II $\tau \rightarrow e \mu \bar{\mu}$ and $\tau \rightarrow \mu e\bar{e}$; $\mu - e$ conversion in nuclei $^{48}_{22}$Ti and $^{197}_{79}$Au. \\
We do the analysis through a single Monte Carlo simulation where the 10 processes are run simultaneously. The peculiarity of all these LFV processes is that they share the same free parameters: three heavy neutrino masses $\{\mathrm{M_{i}}\}_{i=1,2,3}$  and the neutral couplings given by $(\theta S \theta^{\dagger})$ matrix.\\
Every process has to respect  its own upper limit reported by PDG \cite{ParticleDataGroup:2020ssz} (see also \cite{HFLAV:2019otj}), though the conditions on the heavy neutrinos masses and neutral couplings of heavy Majorana neutrinos are common to all.\\
These LFV processes receive two types of contributions: one is coming from charged couplings $(\theta \theta^{\dagger})$ and the other one from neutral couplings $(\theta S \theta^{\dagger})$. As a result, there is an interference between them. Therefore, we are able to determine the relative sign between the entries of the $(\theta \theta^{\dagger})$ (which were bound in (\ref{eq44})) and $(\theta S \theta^{\dagger})$ matrices, which turns out to be negative.\\
The Monte Carlo simulation finds combinations of the free parameters values that return allowed results for each branching ratio and conversion rate %that is less than the experimentally measured upper limit
\cite{ParticleDataGroup:2020ssz}. In the following Table we show the mean values of our simulations that respect all experimental bounds. According to the current upper limits and our mean values, the $\tau\to e e \bar{e}$, $\tau\to \mu \mu \bar{\mu}$, $\tau\to e \mu \bar{\mu}$, $\tau\to \mu e \bar{e}$ and $\mu\to e$ conversion in Ti seem to be more promising in the near future than the LFV $Z$ decays, the $\mu\to e e \bar{e}$ decays and $\mu\to e$ conversion in Au. However, prospects for future sensitivities on the latter processes (see e.g. table I in \cite{Hernandez-Tome:2019lkb} and refs. therein and ref. \cite{Calibbi:2021pyh} focusing on $Z\to\tau\ell$) all go below our mean values.\\
%The neutral couplings are written in a matrix representation and they are given at $(\mathrm{C.L.} = 90\%)$.  
\begin{table}[!ht]
\begin{center}
\begin{tabular}{|c||c||c|}
\hline
LFV Z decays & Our mean values %(\mathrm{C.L.} = 95\%)
& Present limits \cite{ParticleDataGroup:2020ssz}\\
\hline
$\mathrm{Br}(Z \rightarrow \bar{\mu}e)$ & $%1.21%04
%\times 10^{-14} 
\textcolor{black}{1.20\times 10^{-14}} %\pm 1.3144\times10^{-16} 
$ & $3.7\times10^{-7}$\\
\hline
$\mathrm{Br}(Z \rightarrow \bar{\tau}e)$ & $%1.66%01 
%\times 10^{-8} 
\textcolor{black}{1.46\times 10^{-8}} %\pm 2.2071 \times 10^{-10}
$ & $4.9\times10^{-6}$\\
\hline
$\mathrm{Br}(Z \rightarrow \bar{\tau}\mu)$ & $%1.13%45 
%\times 10^{-8} 
\textcolor{black}{1.09\times 10^{-8}} %\pm 1.7103 \times 10^{-10}
$ & $0.6\times10^{-5}$\\
\hline
& & \\
\hline
LFV Type I & %(\mathrm{C.L.} = 90\%) 
& \\
\hline
$\mathrm{Br}(\mu \rightarrow ee\bar{e})$ & $%1.30%2968
%\times 10^{-14} 
\textcolor{black}{1.85\times 10^{-14}} %\pm 1.1561 \times 10^{-16}
$ &1.0$\times10^{-12}$\\
\hline
$\mathrm{Br}(\tau \rightarrow ee\bar{e})$ & $%4.08 %796
%\times 10^{-9} 
\textcolor{black}{4.16\times 10^{-9}} %\pm 4.0906 \times 10^{-11}
$ & $2.7\times10^{-8}$\\
\hline
$\mathrm{Br}(\tau \rightarrow \mu \mu \bar{\mu})$ & $%4.15%462
%\times 10^{-9} 
\textcolor{black}{4.24 \times 10^{-9}}%\pm 3.1010 \times 10^{-11}
$ & $2.1\times10^{-8}$\\
\hline
& & \\
\hline
LFV Type II & %(\mathrm{C.L.} = 90\%) 
& \\
\hline
$\mathrm{Br}(\tau \rightarrow e \mu\bar{\mu})$ & $%3.61%30
%\times 10^{-9} 
\textcolor{black}{3.60 \times 10^{-9}}%\pm 3.1365 \times 10^{-11}
$ & $2.7\times10^{-8}$\\
\hline
$\mathrm{Br}(\tau \rightarrow \mu e \bar{e})$ & $%2.21%11
%\times 10^{-9} 
\textcolor{black}{2.48\times 10^{-9}}%\pm 2.3952 \times 10^{-11}
$ & $1.8\times10^{-8}$\\
\hline
& & \\
\hline
$\mu - e$ conversion rate & %(\mathrm{C.L.} = 90\%) 
& \\
\hline
$\mathcal{R}(\mathrm{Ti})$ & $%5.84%28 
%\times 10^{-14} 
\textcolor{black}{6.21\times 10^{-14}} %\pm 6.5696 \times 10^{-16}
$ & $4.3\times10^{-13}$\\
\hline
$\mathcal{R}(\mathrm{Au})$ & $%7.83%38 
%\times 10^{-14} 
\textcolor{black}{7.82\times 10^{-14}}%\pm 8.3552 \times 10^{-16}
$& $7.0\times10^{-12}$\\
\hline
& & \\
\hline
Heavy neutrino masses & %(\mathrm{C.L.} = 90\%)
& \\
\hline
$\mathrm{M}_{1}$ (TeV) & $%4.049 
\textcolor{black}{17.186}%89 %\pm 0.0056
$ & 
\\
\hline
$\mathrm{M}_{2}$ (TeV) & $%4.050 
\textcolor{black}{17.185}%499 %\pm 0.0055
$ &
\\
\hline
$\mathrm{M}_{3}$ (TeV) & $%4.044
\textcolor{black}{17.187}%4 %\pm 0.0056
$ & \\
\hline
\end{tabular}
\caption{Mean values for branching ratios, conversion rates and three heavy neutrino masses compared to current upper limits (at $95\%$ confidence level for the Z decays and at $90\%$ for all other processes). Statistical errors are at the $1\%$ level and order permille for the heavy neutrino masses.}
\end{center}
\end{table}

The modulus of the $(\theta S \theta^{\dagger})_{e \mu}$ elements are all smaller than $%1\times10^{-8} (
\textcolor{black}{7.5\times 10^{-10}}%)
$, while for the other flavor combinations we get $|(\theta S \theta^{\dagger})_{e \tau}|<%8\times10^{-6}(
\textcolor{black}{5.1\times 10^{-7}}%)
$ and $|(\theta S \theta^{\dagger})_{\mu \tau}|<%1\times10^{-6}(
\textcolor{black}{6.2\times 10^{-7}}%)
$.

\begin{figure}[!ht]
    \centering
    \includegraphics[scale = 0.6]{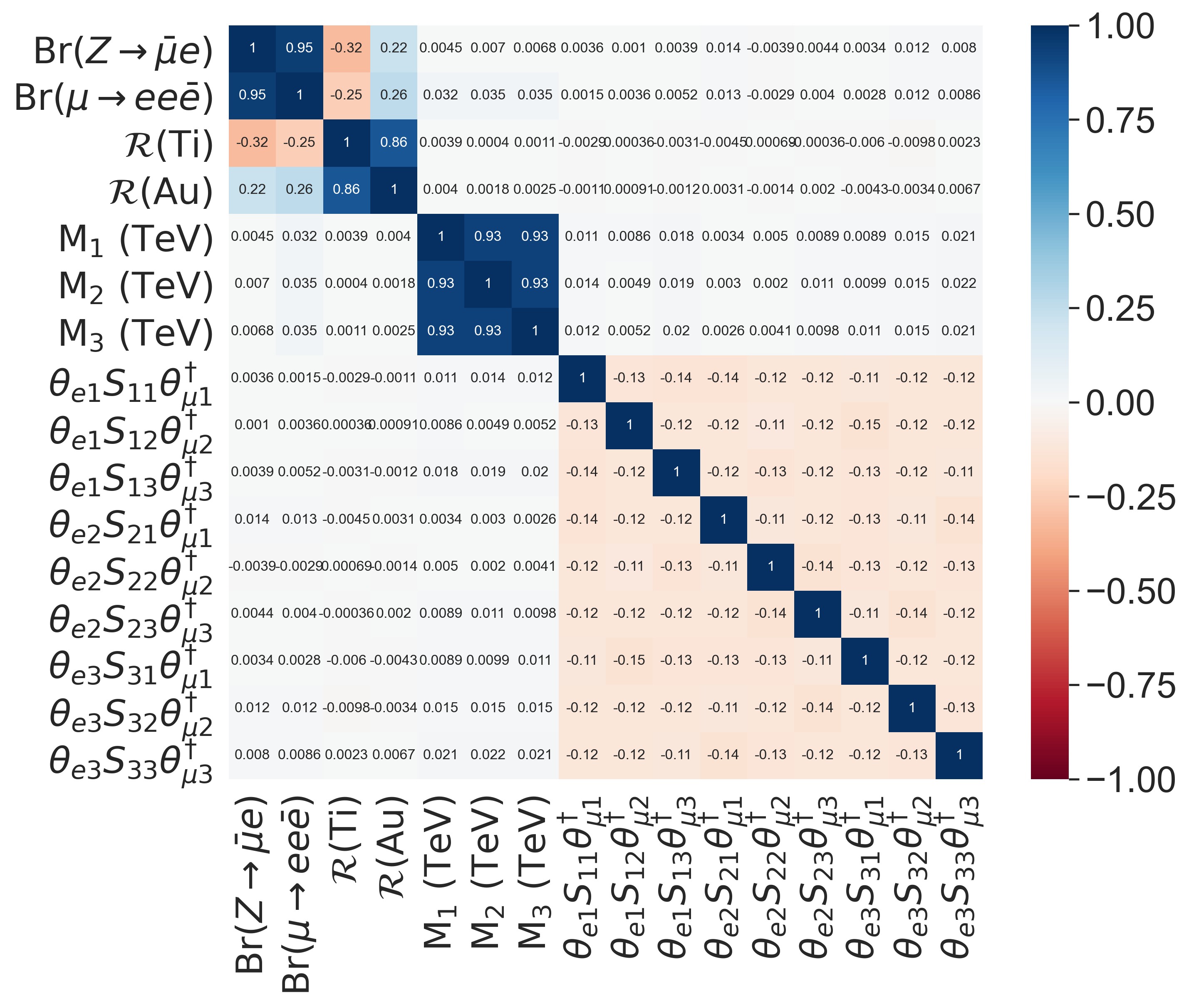}
    \caption{Heat map that stands for the correlation matrix among $(\theta S \theta^{\dagger})_{e \mu}-$processes: $Z \rightarrow \bar{\mu}e$, $\mu \rightarrow ee\bar{e}$, $\mu - e$ conversion in nuclei $^{48}_{22}$Ti and $^{197}_{79}$Au, and their free parameters.}
    \label{fig:emu}
\end{figure}

\begin{figure}[!ht]
\centering
    \begin{minipage}{0.45\linewidth}
    \centering
        \includegraphics[width=\linewidth]{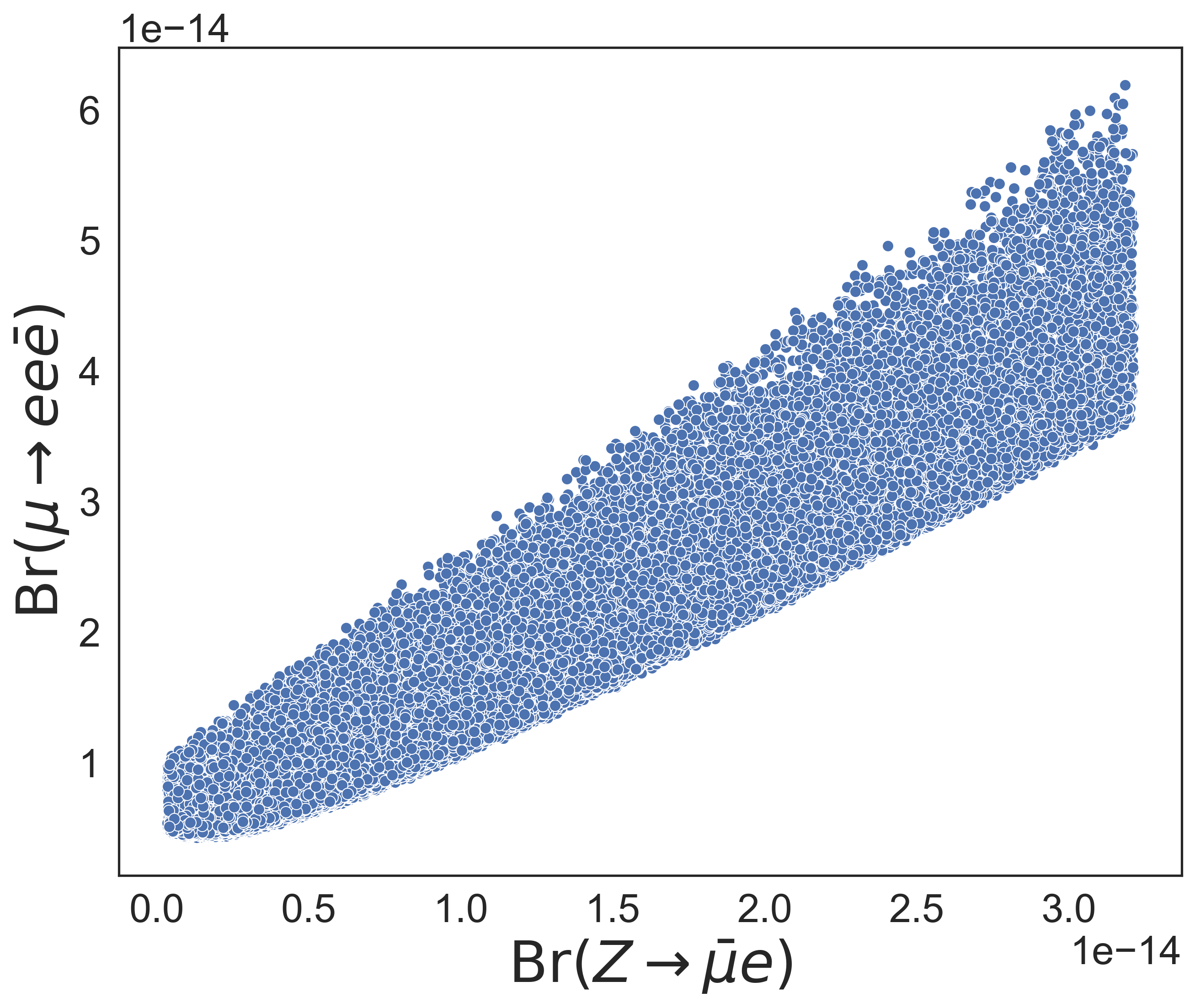}
        \caption{Scatter plot $\mathrm{Br}(Z \rightarrow \bar{\mu}e)$ vs. $\mathrm{Br}(\mu \rightarrow ee \bar{e})$.}
        \label{fig:emu_1}
    \end{minipage}
    \hspace{1cm}
    \begin{minipage}{0.45\linewidth}
        \centering
        \includegraphics[width=\linewidth]{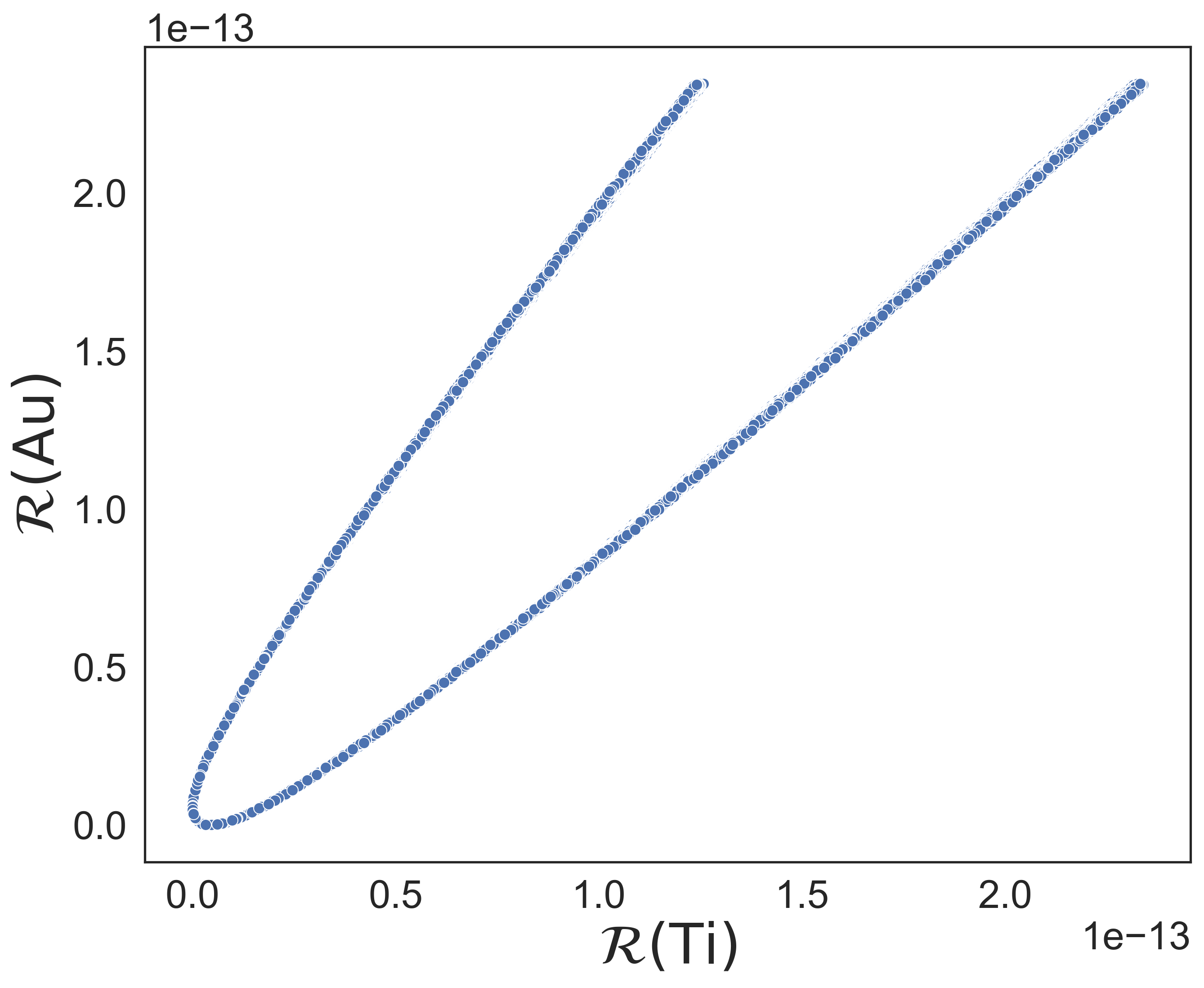}
        \caption{Scatter plot $\mathcal{R}$(Ti) vs. $\mathcal{R}$(Au).}
        \label{fig:emu_2}
    \end{minipage}
\end{figure}

\begin{figure}[!ht]
    \centering
    \includegraphics[scale = 0.6]{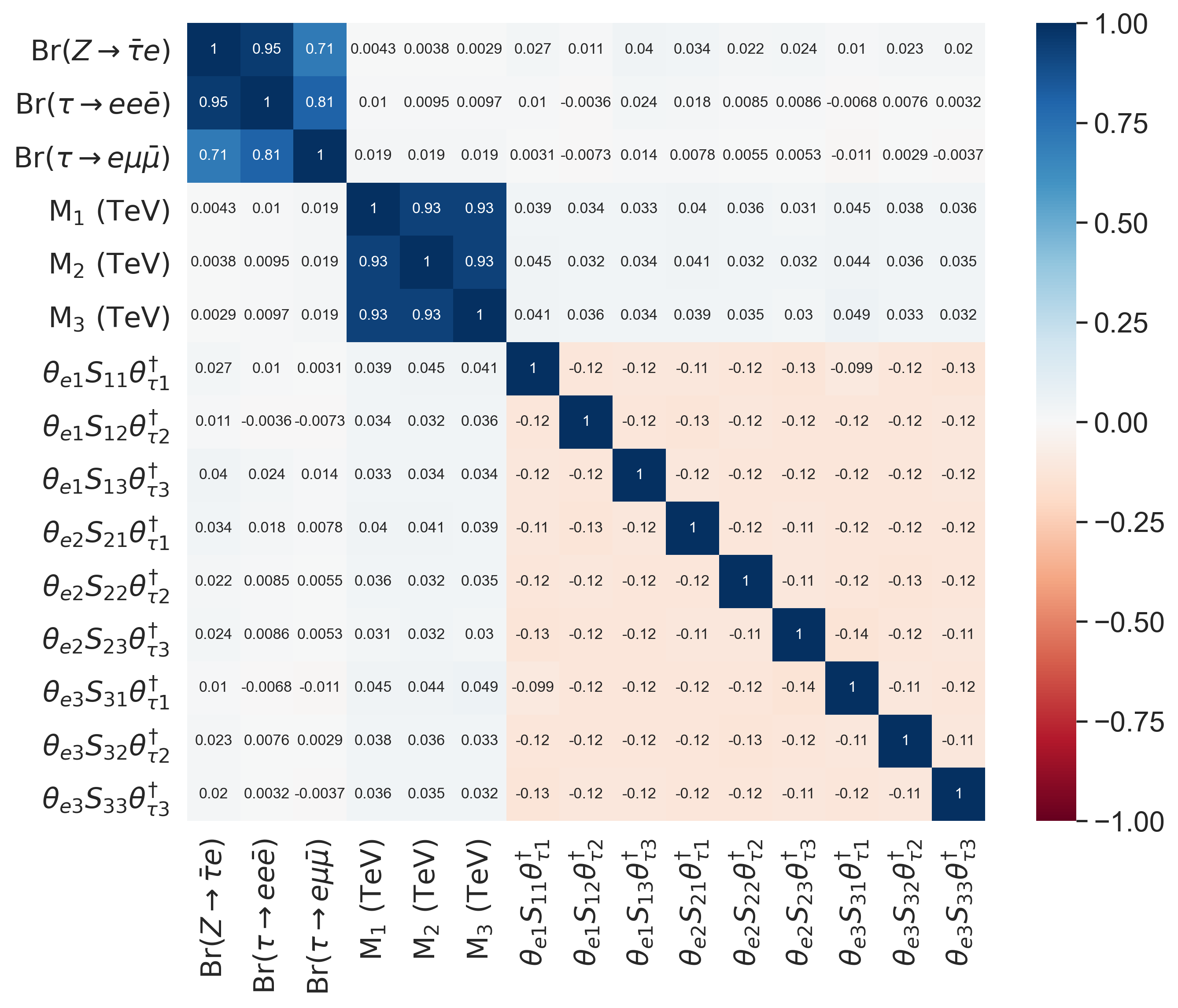}
    \caption{Heat map that stands for the correlation matrix among $(\theta S \theta^{\dagger})_{e \tau}-$processes: $Z \rightarrow \bar{\tau}e$, $\tau \rightarrow ee\bar{e}$, $\tau \rightarrow e \mu \bar{\mu}$, and their free parameters.}
    \label{fig:etau}
\end{figure}

\begin{figure}[!ht]
\centering
    \begin{minipage}{0.45\linewidth}
    \centering
        \includegraphics[width=\linewidth]{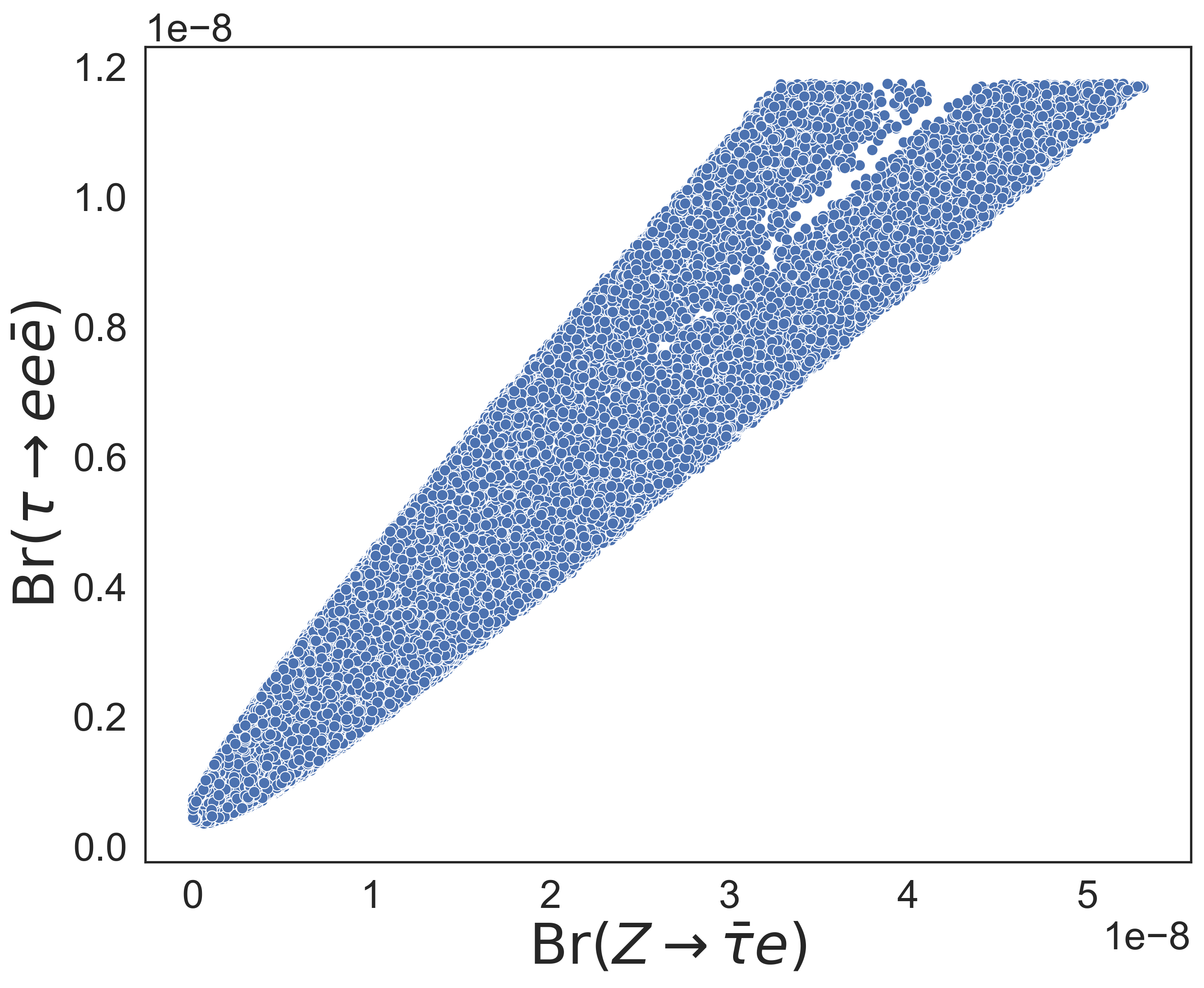}
        \caption{Scatter plot $\mathrm{Br}(Z \rightarrow \bar{\tau}e)$ vs. $\mathrm{Br}(\tau \rightarrow ee \bar{e})$.}
        \label{fig:etau_1}
    \end{minipage}
    \hspace{1cm}
    \begin{minipage}{0.45\linewidth}
        \centering
        \includegraphics[width=\linewidth]{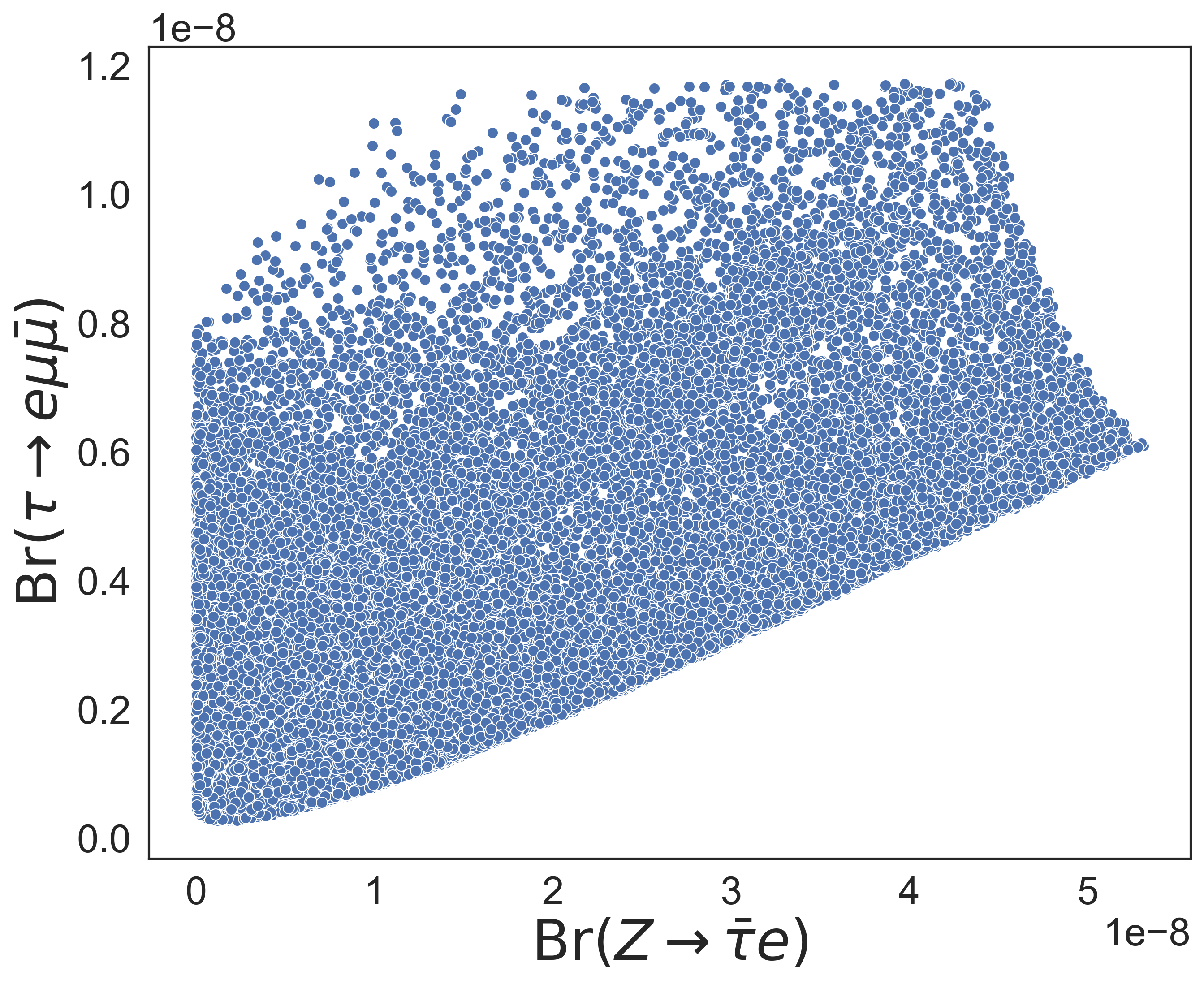}
        \caption{Scatter plot $\mathrm{Br}(Z \rightarrow \bar{\tau}e)$ vs. $\mathrm{Br}(\tau \rightarrow e \mu \bar{\mu})$.}
        \label{fig:etau_2}
    \end{minipage}
\end{figure}

\begin{figure}[!ht]
    \centering
    \includegraphics[scale = 0.25]{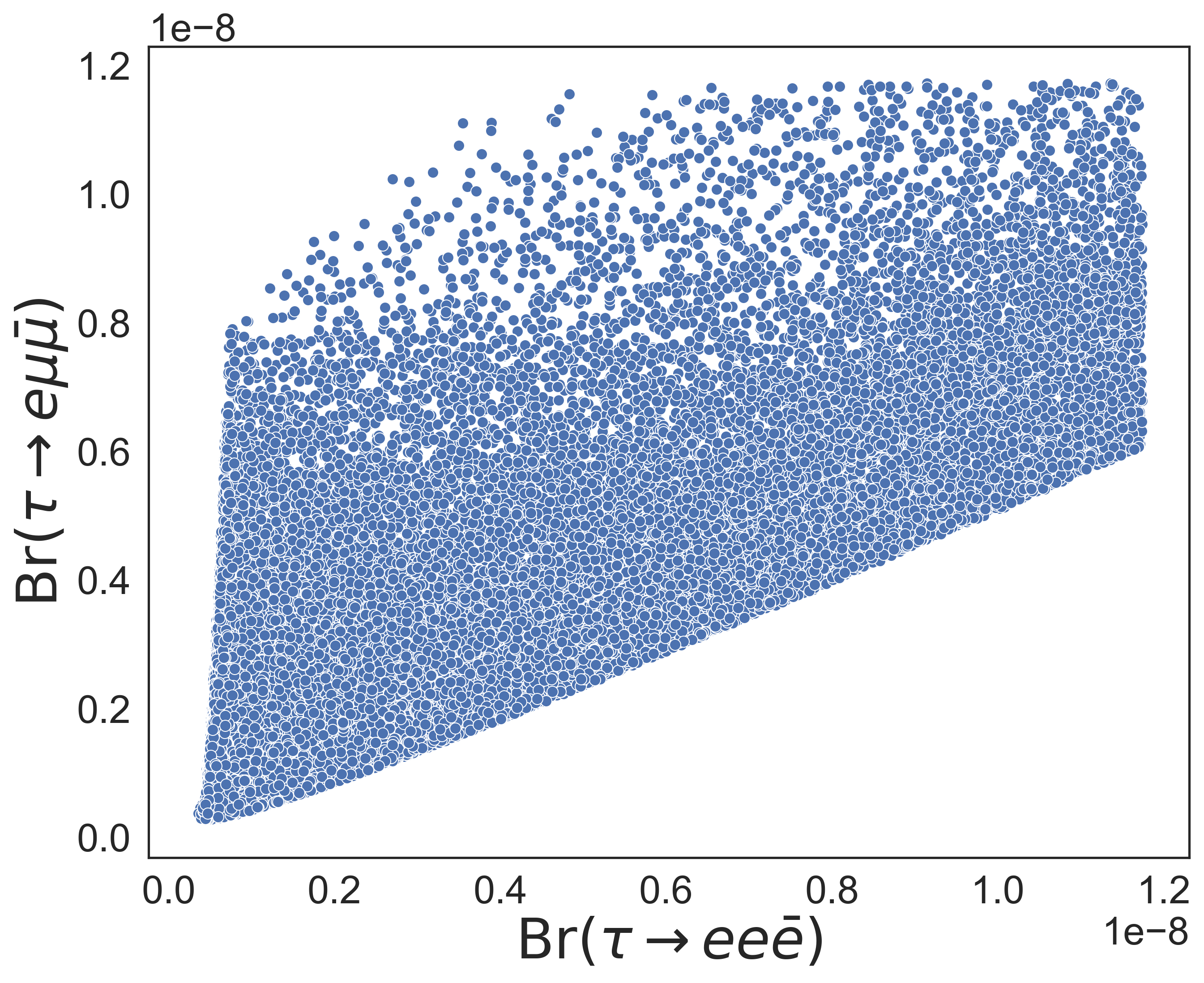}
    \caption{Scatter plot $\mathrm{Br}(Z \rightarrow ee \bar{e})$ vs. $\mathrm{Br}(\tau \rightarrow e \mu \bar{\mu})$.}
    \label{fig:etau_3}
\end{figure}

\begin{figure}[!ht]
    \centering
    \includegraphics[scale = 0.6]{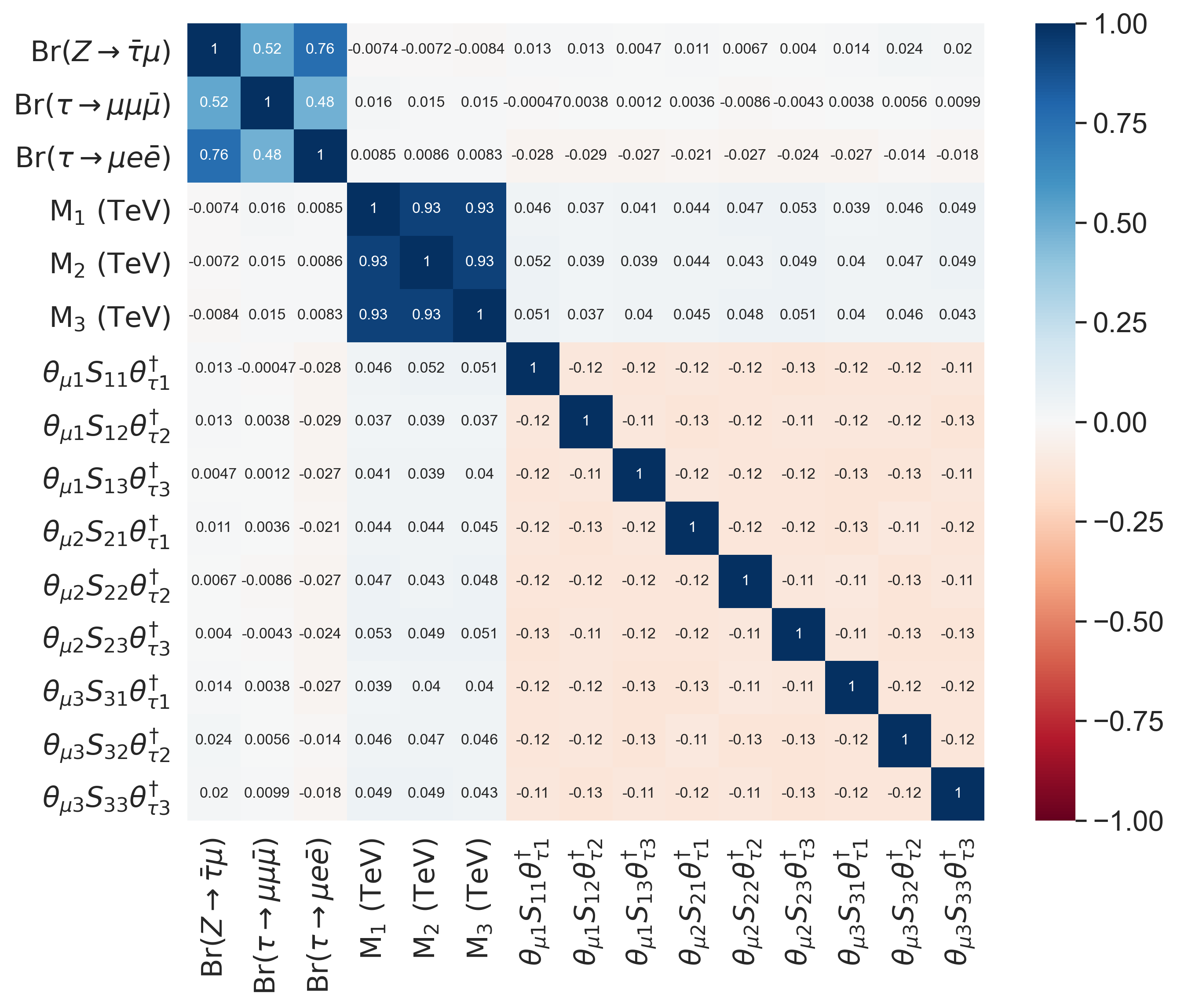}
    \caption{Heat map that stands for the correlation matrix among $(\theta S \theta^{\dagger})_{\mu \tau}-$processes: $Z \rightarrow \bar{\tau}\mu$, $\tau \rightarrow \mu \mu \bar{\mu}$, $\tau \rightarrow \mu e \bar{e}$, and their free parameters.}
    \label{fig:mutau}
\end{figure}

\begin{figure}[!ht]
\centering
    \begin{minipage}{0.49\linewidth}
    \centering
        \includegraphics[width=\linewidth]{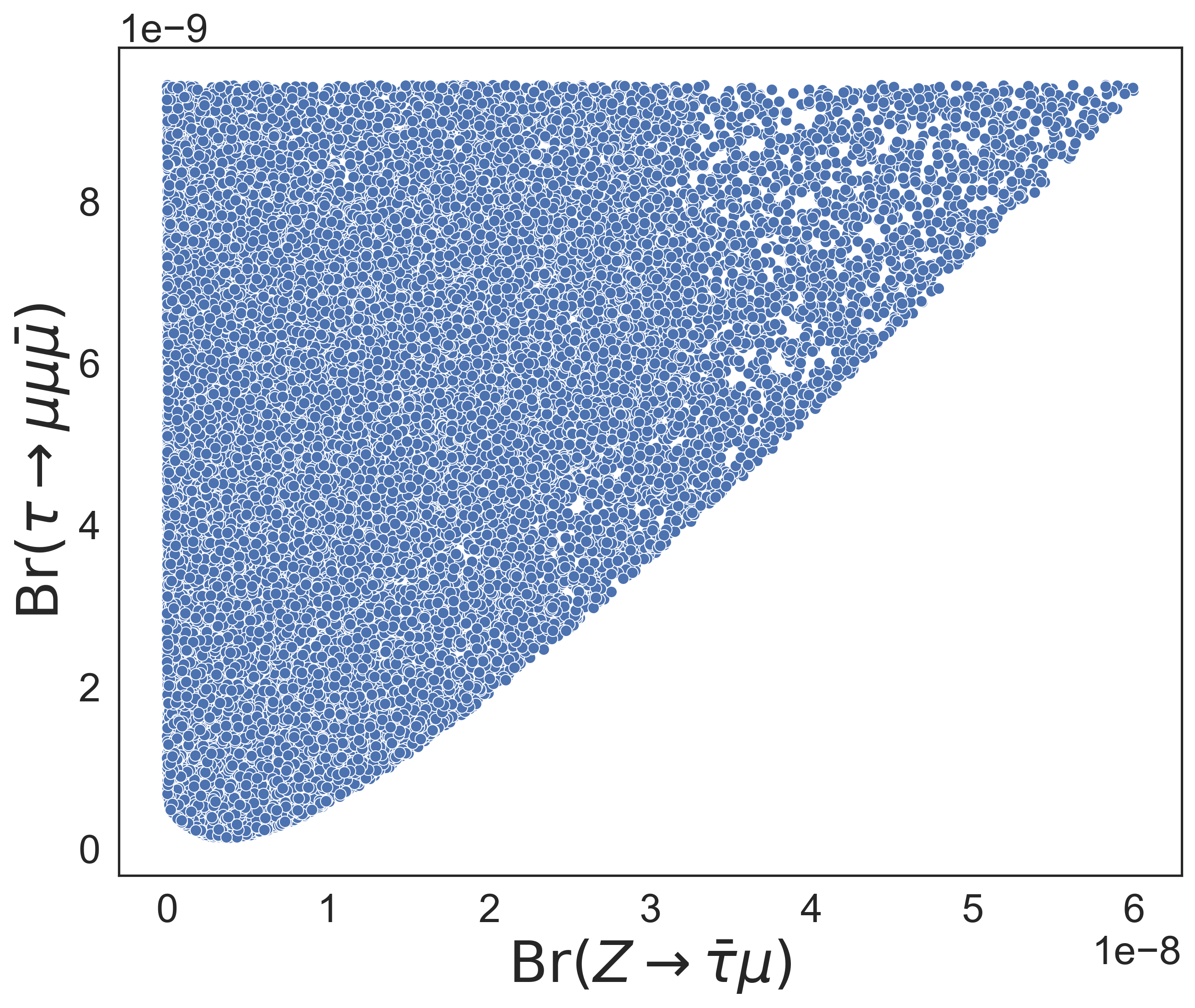}
        \caption{Scatter plot $\mathrm{Br}(Z \rightarrow \bar{\tau}\mu)$ vs. $\mathrm{Br}(\tau \rightarrow \mu \mu \bar{\mu})$.}
        \label{fig:mutau_1}
    \end{minipage}
    \hspace{1cm}
    \begin{minipage}{0.49\linewidth}
        \centering
        \includegraphics[width=\linewidth]{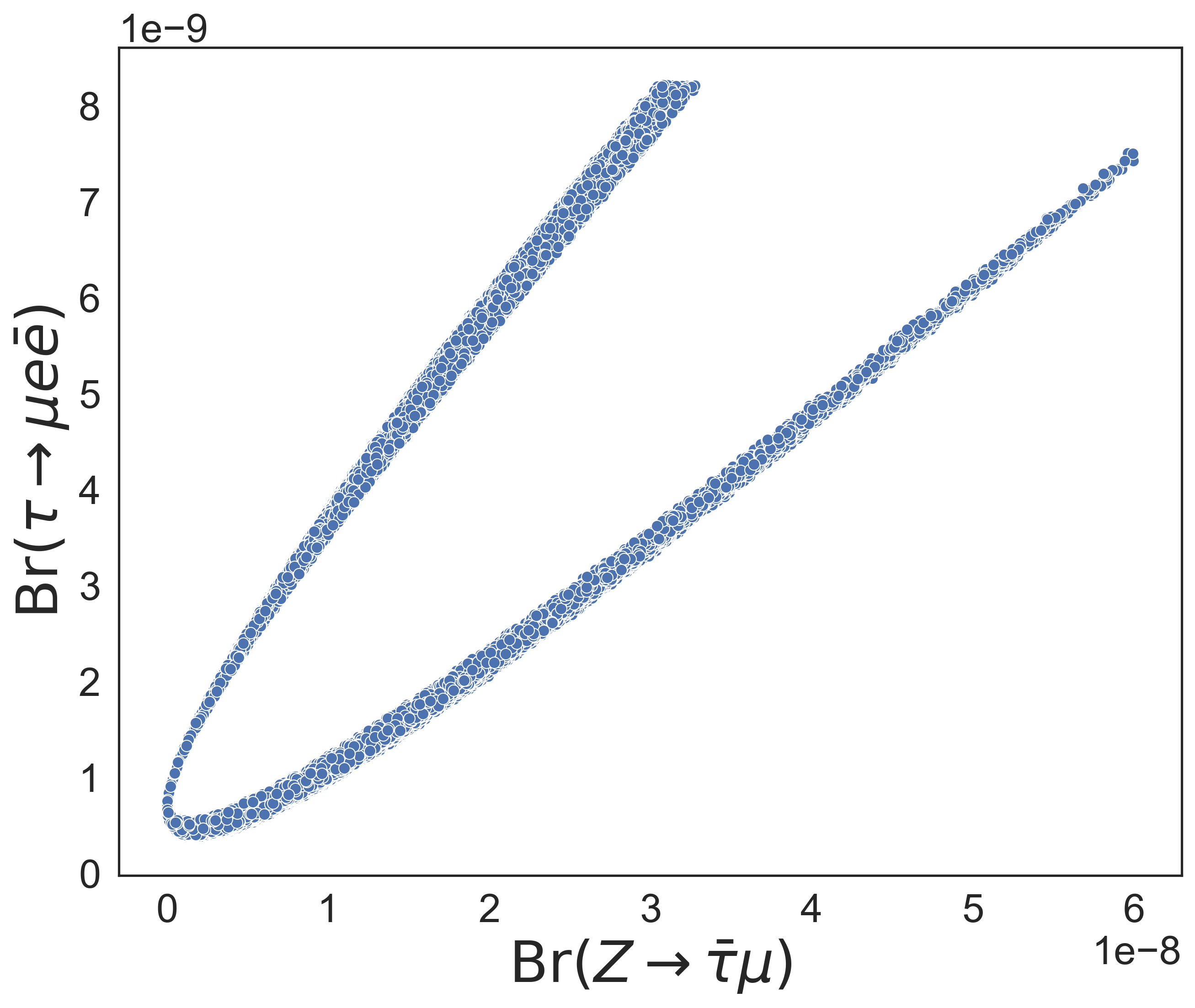}
        \caption{$\mathrm{Br}(Z \rightarrow \bar{\tau}\mu)$ vs. $\mathrm{Br}(\tau \rightarrow \mu e \bar{e})$.}
        \label{fig:mutau_2}
    \end{minipage}
    \hspace{1cm}
    \begin{minipage}{0.49\linewidth}
        \centering
        \includegraphics[width=\linewidth]{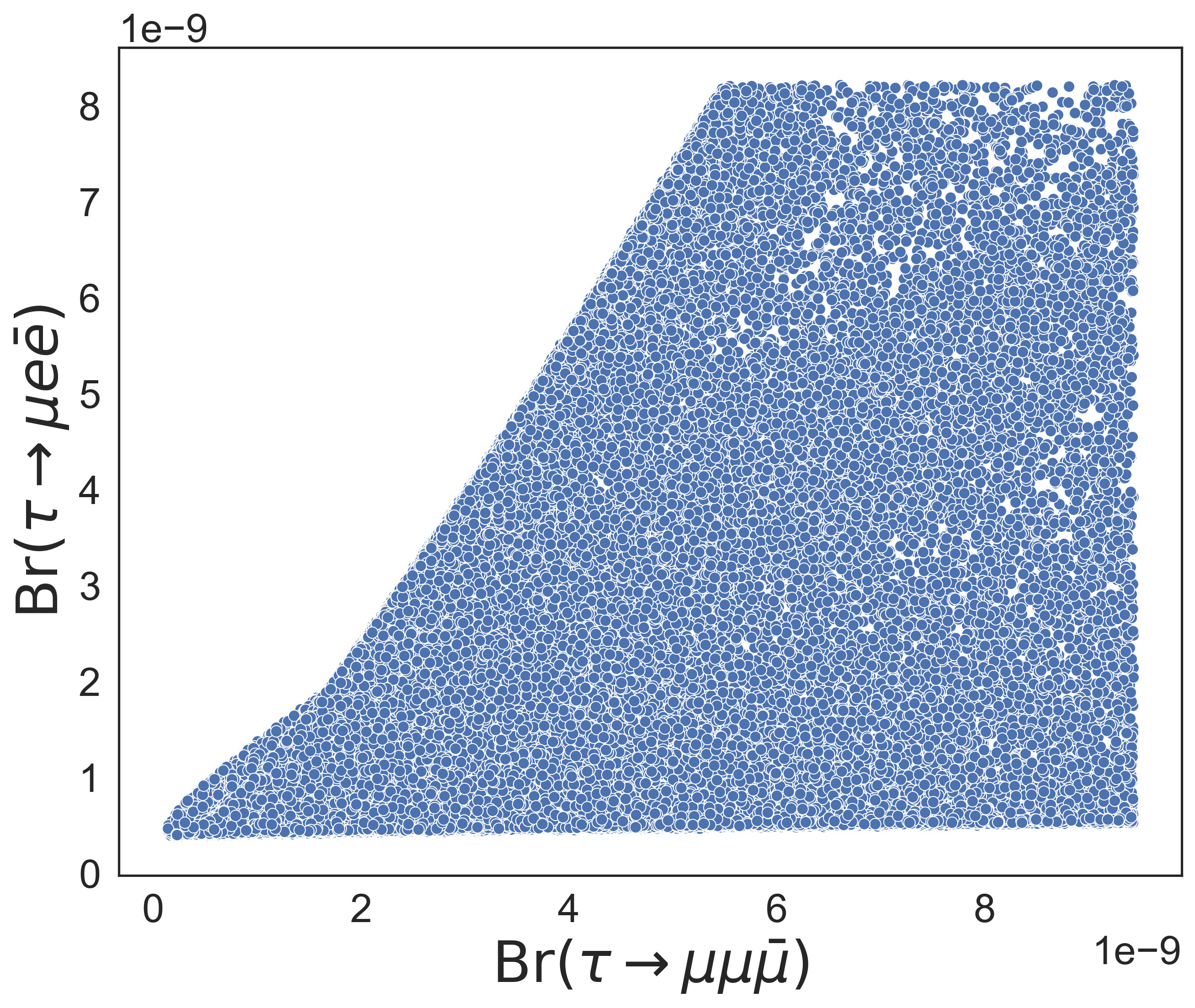}
        \caption{$\mathrm{Br}(\tau \rightarrow \mu \mu \bar{\mu})$ vs. $\mathrm{Br}(\tau \rightarrow \mu e \bar{e})$.}
        \label{fig:mutau_3}
    \end{minipage}
\end{figure}

\begin{figure}[!ht]
    \centering
    \includegraphics[scale =0.5]{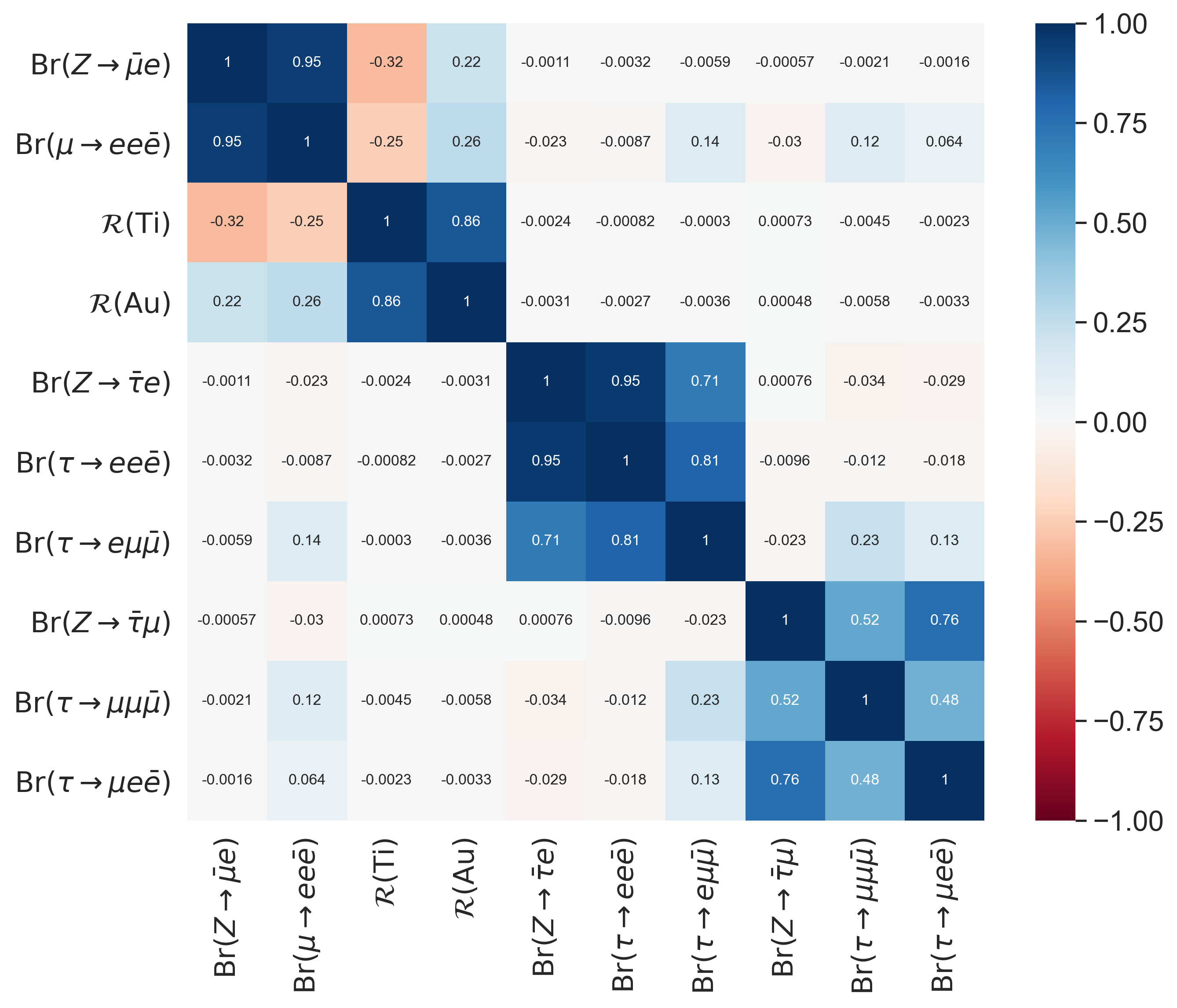}
    \caption{Heat map that stands for the correlation matrix exclusively among the 10 processes analyzed in this section. We observe that this matrix seems a block matrix representation where each block corresponds to each neutral coupling category.}
    \label{fig:global_processes}
\end{figure}

\begin{figure}[!ht]
\centering
    \begin{minipage}{0.45\linewidth}
    \centering
        \includegraphics[width=\linewidth]{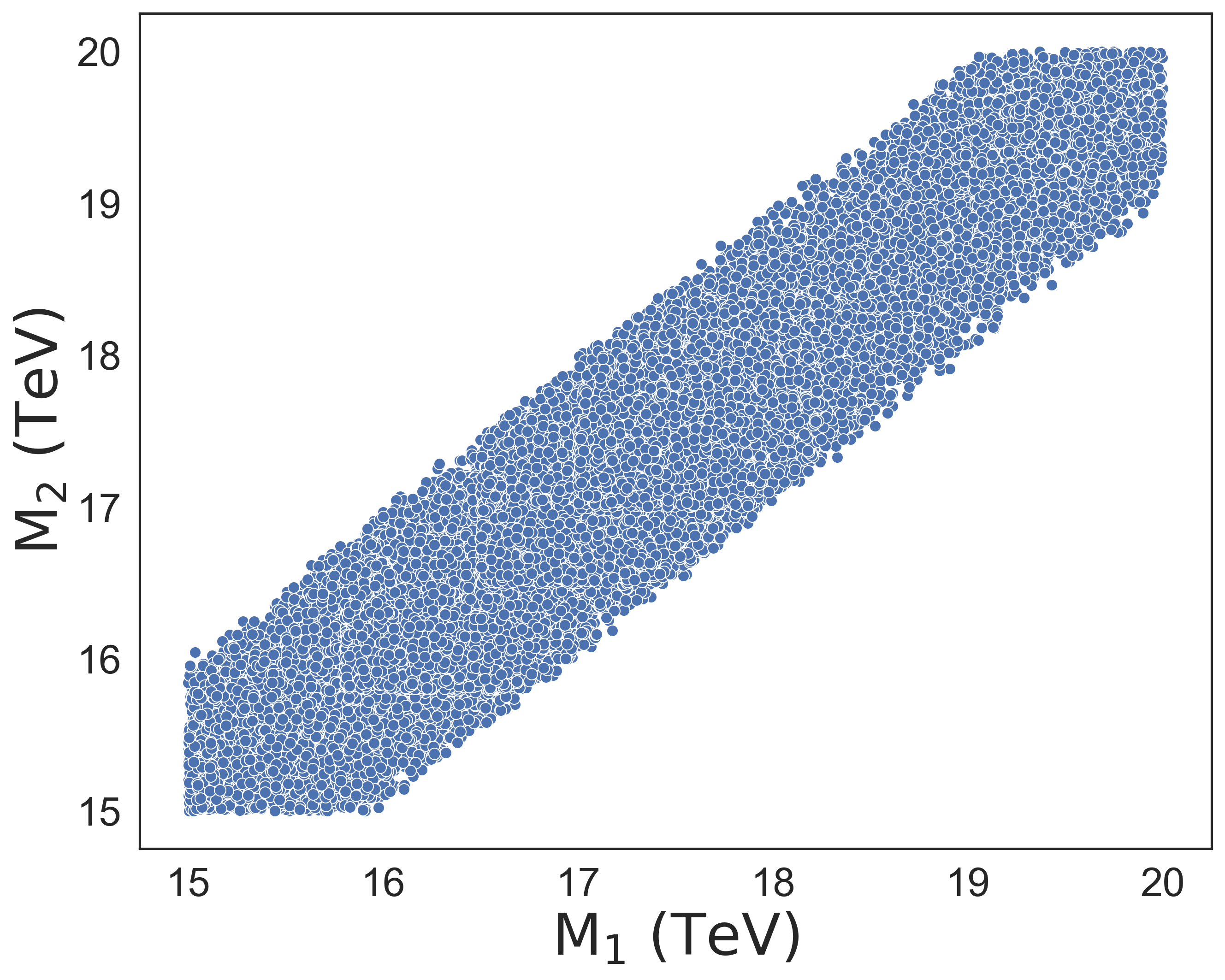}
        \caption{Scatter plot $\mathrm{M_{1}}$ vs. $\mathrm{M_{2}}$.}
        \label{fig:m1vsm2}
    \end{minipage}
    \hspace{1cm}
    \begin{minipage}{0.45\linewidth}
        \centering
        \includegraphics[width=\linewidth]{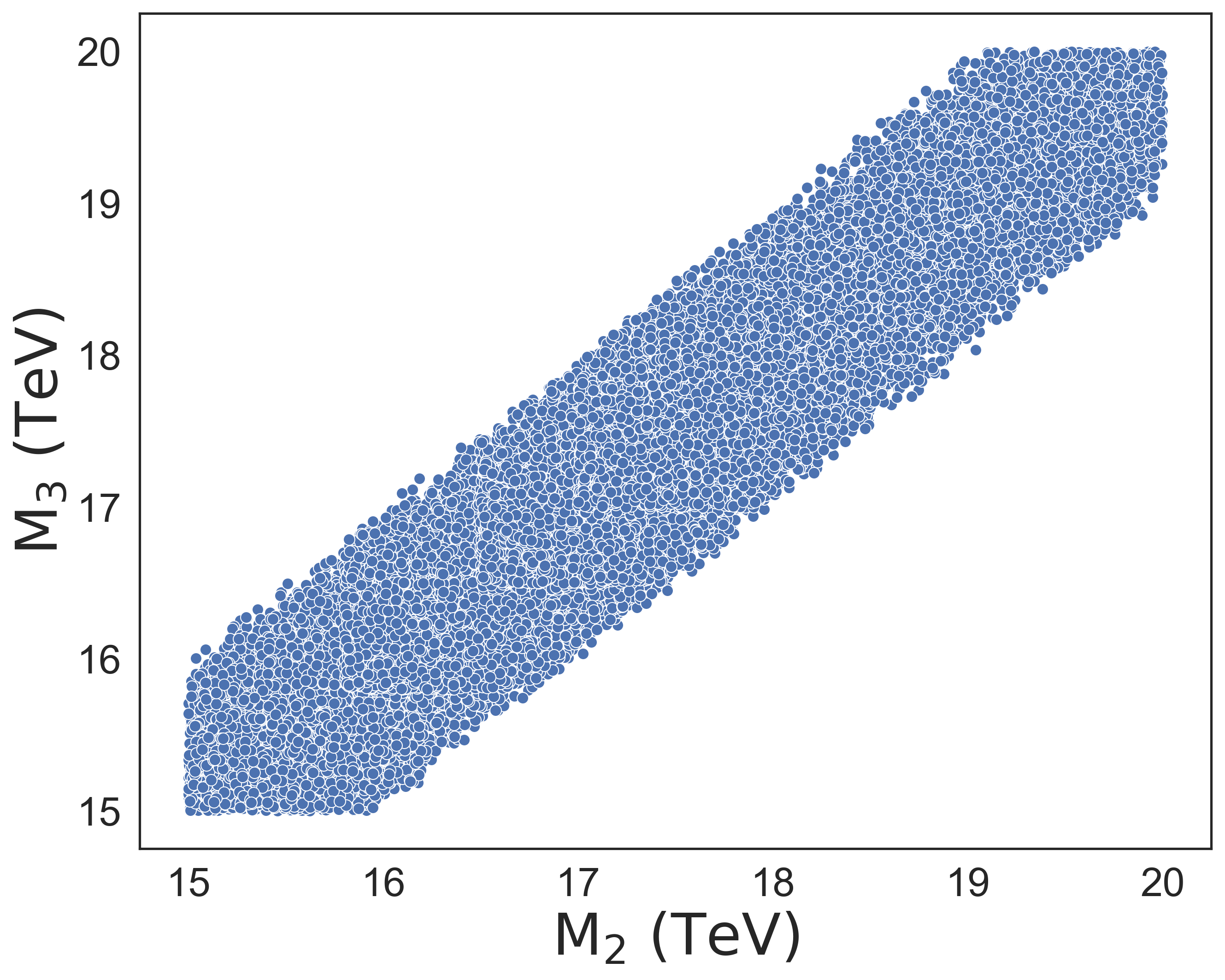}
        \caption{Scatter plot $\mathrm{M_{2}}$ vs. $\mathrm{M_{3}}$.}
        \label{fig:m2vsm3}
    \end{minipage}
\end{figure}

In order to find relations among the above processes we group them into 3 categories based on their neutral couplings: $(\theta S \theta^{\dagger})_{e \mu}$, $(\theta S \theta^{\dagger})_{e \tau}$, and $(\theta S \theta^{\dagger})_{\mu \tau}$.
\begin{itemize}
    \item $(\theta S \theta^{\dagger})_{e \mu}-$processes: $Z \rightarrow \bar{\mu}e$, $\mu \rightarrow ee\bar{e}$, and $\mu - e$ conversion in nuclei $^{48}_{22}$Ti and $^{197}_{79}$Au. 
    \item $(\theta S \theta^{\dagger})_{e \tau}-$processes: $Z \rightarrow \bar{\tau}e$, $\tau \rightarrow ee\bar{e}$, and $\tau \rightarrow e \mu \bar{\mu}$.
    \item$(\theta S \theta^{\dagger})_{\mu \tau}-$processes: $Z \rightarrow \bar{\tau}\mu$, $\tau \rightarrow \mu \mu \bar{\mu}$, and $\tau \rightarrow \mu e \bar{e}$.
\end{itemize}
In Figure \ref{fig:emu} a heat map shows  the correlation matrix among $(\theta S \theta^{\dagger})_{e \mu}-$processes and their free parameters. First of all, we see that there is no  sizeable correlation among any process probability and its free parameters. Second, the small correlations among every pair of $(\theta S \theta^{\dagger})_{e \mu}$ matrix elements is negative%, it indicates that while one of them increases the other decreases
. Furthermore, $Z \rightarrow \bar{\mu}e$ decay is strongly correlated with $\mu \rightarrow ee\bar{e}$. Similarly, the conversion rate in $^{48}_{22}$Ti is highly correlated with the one in $^{197}_{79}$Au. In Figure \ref{fig:emu_1} (\ref{fig:emu_2}) the  correlations between $BR(Z\to\bar{\mu}e)$ and $BR(\mu\to e e \bar{e})$ ($\mathcal{R}(\mathrm{Ti})$ and $\mathcal{R}(\mathrm{Au})$) are shown in scatter plots. Although heavy neutrino masses are quite correlated, the solutions are mostly far from quasidegenerate scenarios, which is a natural solution.\\
In Figure \ref{fig:etau} the correlations among $(\theta S \theta^{\dagger})_{e \tau}-$processes and their free parameters are represented. The interpretation of this plot is very similar to Figure \ref{fig:emu}. The branching ratios of these decays have a sizeable correlation to each other, but the predominant one is between $\mathrm{Br}(Z \rightarrow \bar{\tau}e)$ and $\mathrm{Br}(\tau \rightarrow ee\bar{e})$. We show all those behaviors in Figures \ref{fig:etau_1}, \ref{fig:etau_2} and \ref{fig:etau_3}.\\
For $(\theta S \theta^{\dagger})_{\mu \tau}-$processes their branching ratios are not correlated with any free parameter as we can observe in Figure \ref{fig:mutau}. Nevertheless, we can see sizeable correlations among branching ratios, where the largest one is between $\mathrm{Br}(Z \rightarrow \bar{\tau}\mu)$ and $\mathrm{Br}(\tau \rightarrow \mu e\bar{e})$. The correlations among decays are displayed in Figures \ref{fig:mutau_1}, \ref{fig:mutau_2} and \ref{fig:mutau_3}.\\
In the three heat maps for the processes whose behavior involves neutral couplings given by $(\theta S \theta^{\dagger})_{\ell' \ell}$ matrix, the three heavy masses are strongly correlated to each other. Still, solutions span the range \textcolor{black}{$[15,20]$} TeV and do not favor (quasi)degenerate scenarios.\\ %(recall that the values for heavy neutrinos are the same for all processes)\\
Finally, we display a heat map in Figure \ref{fig:global_processes} where only branching ratios and conversion rates are involved. This heat map, that stands for a correlation matrix, seems a block matrix where each block represents a category of $(\theta S \theta^{\dagger})_{\ell' \ell}-$processes, so that we can conclude that processes with different neutral coupling are mildly correlated, as expected.\\
The scatter plots among two  pairs of heavy neutrino masses in Figures \ref{fig:m1vsm2} and \ref{fig:m2vsm3} show neatly that solutions do not restrict to the nearly degenerate case~\footnote{This is of course independent of the mean values for these three masses being very approximately equal in all our runs.}.
    \clearpage
\subsection{Numerical Analysis for $\ell\to\ell'\ell''\bar{\ell}'''$ of type III: Wrong Sign decays}\label{sec:WrongSign}
In this subsection we  study two tau decays which are known as wrong-sign processes: $\tau \rightarrow ee \bar{\mu}$ and $\tau \rightarrow \mu \mu \bar{e}$. We analyze them assuming that the terms associated with LNV vertices are free parameters, thus we are able to bind these couplings. So the free parameters to each wrong sign processes are going to be
\begin{itemize}
    \item $\tau \rightarrow ee \bar{\mu}$: the masses of heavy neutrinos: $\mathrm{M_{i}}$; LNV couplings: $(\theta_{\mu i}\theta_{\tau i})^{\dagger}$, and $\theta_{e i}\theta_{ei}$ with $i = 1,2,3$. We bind the couplings as follows (see also \cite{Enrique Fernandez})
\begin{eqnarray}
    %\begin{split}
    |\theta_{\mu 1}\theta_{\tau 1}| + |\theta_{\mu 2}\theta_{\tau 2}|+|\theta_{\mu 3}\theta_{\tau 3}| &<& %(1.225) 
    %\textcolor{blue}{(0.45)} \textcolor{cyan}{(0.4)} \textcolor{red}{(0.32)} \times 10^{-3}
    \textcolor{black}{0.32\times 10^{-3}}, \nonumber \\
    |\theta_{e 1}\theta_{e 1}| + |\theta_{e 2}\theta_{e 2}|+|\theta_{e 3}\theta_{e 3}| &<& \textcolor{black}{0.01},
    %\end{split}
    \label{eq294}
\end{eqnarray}
and their product must satisfy (see also \cite{Enrique Fernandez})
\begin{equation}
    |\theta_{\mu i}\theta_{\tau i}||\theta_{e j}\theta_{e j}| < \textcolor{black}{0.32} \times 10^{-5}.
    \label{eq296}
\end{equation}
    \item $\tau \rightarrow \mu \mu \bar{e}$: the masses of heavy neutrinos: $\mathrm{M_{i}}$; LNV couplings: $(\theta_{e i}\theta_{\tau i})^{\dagger}$, and $(\theta_{\mu i} \theta_{\mu  i})$ with $i = 1,2,3$. Bounds on the couplings are (see also \cite{Enrique Fernandez})
\begin{eqnarray}
    %\begin{split}
    |\theta_{e 1}\theta_{\tau 1}| + |\theta_{e 2}\theta_{\tau 2}|+|\theta_{e 3}\theta_{\tau 3}| &<& \textcolor{black}{0.9} \times 10^{-3},\nonumber\\
    |\theta_{\mu 1}\theta_{\mu 1}| + |\theta_{\mu 2}\theta_{\mu 2}|+|\theta_{\mu 3}\theta_{\mu 3}| &<& \textcolor{black}{0.0075},
    %\end{split}
    \label{eq307}
\end{eqnarray}
with their product fulfilling (see also \cite{Enrique Fernandez})
\begin{equation}
    |\theta_{e i}\theta_{\tau i}||\theta_{\mu j}\theta_{\mu j}| < \textcolor{black}{0.68} \times 10^{-5}.
    \label{eq309}
\end{equation}
\end{itemize}
The heavy neutrino masses $\mathrm{M_{i}} \ (i = 1,2,3)$ run from %$3$ (
$\textcolor{black}{15}$ to $\textcolor{black}{20}$ %)
TeV as the analysis above. In Table \ref{tabla: wrong_sign} we show the mean values for branching ratios of wrong-sign processes, heavy neutrino masses and LNV couplings. Remarkably, these wrong-sign decays typically yield branching ratios only one order of magnitude below the current upper limits, as the processes with the brightest prospects presented in the previous section. For completeness, we quote here that neutrinoless double beta decays \cite{ParticleDataGroup:2020ssz} bind $\sum_{i=1}^3 |\theta_{ei}|^2$ to be smaller than or, at most, of $\mathcal{O}(1\times10^{-5})$. As there can be cancellations among contributions in the previous sum, this limit is not in conflict with our results.

\begin{table}[ht]
\begin{center}
\begin{tabular}{|c||c|c|c|c|}
\hline
Branching Ratios & Our mean values%(C.L. = 90\%) 
\\
\hline
$\mathrm{Br}(\tau \rightarrow ee \bar{\mu})$ & $%1.6%204 
%\times 10^{-9} (
\textcolor{black}{1.8\times 10^{-9}}%) %\pm 2.0197 \times 10^{-11}
$ \\
\hline
$\mathrm{Br}(\tau \rightarrow \mu\mu \bar{e})$& $%1.6%5883
%\times 10^{-9} 
\textcolor{black}{1.9\times 10^{-9}}% \pm 2.0181 \times 10^{-11}
$ \\
\hline
 & \\
 \hline
Heavy neutrino masses & %(C.L. = 90\%) 
\\
\hline
$\mathrm{M}_{1}$ (TeV)& $%3.9 
\textcolor{black}{17.170}%307 \pm 0.0061
$ \\
\hline
$\mathrm{M}_{2}$ (TeV)& $%3.9 
\textcolor{black}{17.166}
%334 \pm 0.0061
$ \\
\hline
$\mathrm{M}_{3}$ (TeV)& $%3.9(
\textcolor{black}{17.166}%328 \pm 0.0061
$ \\
\hline
& \\
\hline
LNV couplings & %(C.L. = 90\%) 
\\
\hline
$(\theta_{e 1}\theta_{\tau 1})^{\dagger}$ & %$(7%6.638 \times 10^{-6} 
%\pm 4)%3.881 
%\times 10^{-6}$ 
$\textcolor{black}{(2.2 \pm 9.6) \times 10^{-7}}$ 
\\
\hline
$(\theta_{e 2}\theta_{\tau 2})^{\dagger}$ & %$(3%2.925 \times 10^{-6} 
%\pm 4)%3.899 
%\times 10^{-6}$ 
$\textcolor{black}{(1.5 \pm 1.0) \times 10^{-6}}$
\\
\hline
$(\theta_{e 3}\theta_{\tau 3})^{\dagger}$ & %$-(2%-1.887 \times 10^{-6}
%\pm 4)% 3.903 
%\times 10^{-6}$ 
$\textcolor{black}{(0.2 \pm 1.0)\times 10^{-6}}$ \\
\hline
$|\theta_{e i}\theta_{\tau i}|$ & %$9%8.763 
%\times 10^{-4} 
%\pm 3)% 2.728 
%\times 10^{-6}$ 
$\textcolor{black}{2.76 \times 10^{-4}}$ \\
\hline
& \\
\hline
$(\theta_{\mu 1}\theta_{\mu 1})$ & %$-(0.8%.318 \times 10^{-5} 
%\pm 1.6) %1.586 
%\times 10^{-4}$ 
$\textcolor{black}{(0.8 \pm 3.0)\times 10^{-5}}$ \\
\hline
$(\theta_{\mu 2}\theta_{\mu 2})$ & %$-(0.8%7.542 \times 10^{-5} 
%\pm 1.6)%1.586 
%\times 10^{-4}$ 
$\textcolor{black}{-(8.8 \pm 3.0)\times 10^{-5}} $ \\
\hline
$(\theta_{\mu 3}\theta_{\mu 3})$ & %$-(1.0%17 \times 10^{-4} 
%\pm  1.6)%0 
%\times 10^{-4}$ 
$\textcolor{black}{(1.0 \pm 3.0)\times 10^{-5}} $\\
\hline
$|\theta_{\mu i}\theta_{\mu i}|$ & %$4%3.577 
%\times 10^{-2} 
%\pm 1)%.113 
%\times 10^{-4}$ 
$\textcolor{black}{8.5\times 10^{-3}}$
\\
\hline
& \\
\hline
$(\theta_{\mu 1}\theta_{\tau 1})^{\dagger}$ & %$-(1%9.41 \times 10^{-7} 
%\pm 9)%8.609 
%\times 10^{-6}$ 
$\textcolor{black}{(2.1 \pm 2.3)\times 10^{-6}}$ \\
\hline
$(\theta_{\mu 2}\theta_{\tau 2})^{\dagger}$ & %$(1%8.72 \times 10^{-7} 
%\pm 9)%8.614 
%\times 10^{-6}$ 
$\textcolor{black}{-(0.4 \pm 2.3)\times 10^{-6}}$ \\
\hline
$(\theta_{\mu 3}\theta_{\tau 3})^{\dagger}$ & %$(1%.11 \times 10^{-6} 
%\pm 9)% 8.624 
%\times 10^{-6}$ 
$\textcolor{black}{(0.6 \pm 2.3)\times 10^{-6}}$ \\
\hline
$|\theta_{\mu i}\theta_{\tau i}|$ & %$2%1.937 
%\times 10^{-3} %\pm 5.946 \times 10^{-6}$ 
$\textcolor{black}{6.5\times 10^{-4}}$ \\
\hline
& \\
\hline
$(\theta_{e 1}\theta_{e 1})$ & %$-(5%.427 \times 10^{-5} 
%\pm 7)%6.703 
%\times 10^{-5}$ 
$\textcolor{black}{(3.9 \pm 2.0)\times 10^{-5}}$ \\
\hline
$(\theta_{e 2}\theta_{e 2})$ & %$(1%7.711 \times 10^{-6} 
%\pm 7)% 6.718 
%\times 10^{-5}$ 
$\textcolor{black}{(4.6 \pm 2.0)\times 10^{-5}}$ \\
\hline
$(\theta_{e 3}\theta_{e 3})$ & %$-(1%8.585 \times 10^{-6} 
%\pm 7)% 6.673 
%\times 10^{-5}$ 
$\textcolor{black}{-(5.0 \pm 2.0)\times 10^{-5}}$ \\
\hline
$|\theta_{e i}\theta_{e i}|$ & %$2%1.5066 
%\times 10^{-2}%\pm 4.616 \times 10^{-5}$ 
$\textcolor{black}{5.7\times 10^{-3}}$\\
\hline
\end{tabular}
\caption{Mean values for the free parameters and branching ratios in the wrong sign processes considering Majorana neutrinos in the LHT. Statistical errors which are not shown are smaller than the last significant figure. We recall the $90\%$ C.L. limits \cite{ParticleDataGroup:2020ssz}: $1.5\times10^{-8}$ (on $\mathrm{Br}(\tau \rightarrow ee \bar{\mu})$) and $1.7\times10^{-8}$ (on $\mathrm{Br}(\tau \rightarrow \mu\mu \bar{e})$).}
\label{tabla: wrong_sign}
\end{center}
\end{table}

%Unlike the processes discussed in section \ref{sec:Joint}, neither branching ratio of wrong-sign processes $(\tau \rightarrow ee\bar{\mu}$ and $\tau \rightarrow \mu \mu \bar{e})$, nor heavy neutrino masses are correlated to each other as we see in Figure \ref{fig:wrong_sign}. Actually, no free parameter is correlated to each other for wrong sign processes.\\
\textcolor{black}{The heavy neutrino masses ($\mathrm{M}_{i}$) present a sizeable correlation among them as in the previous analysis. Also, LNV couplings, $|\theta_{ei}\theta_{\tau i}|$ and $|\theta_{\mu i}\theta_{\mu i}|$, are moderately correlated with the heavy neutrino masses, while $|\theta_{\mu i}\theta_{\tau i}|$ and $|\theta_{e i}\theta_{e i}|$ have a minimum correlation with them.}\\
LHT does not predict these ``wrong sign" decays through T-odd leptons \cite{delAguila:2019htj}. However, when we extend the LHT model involving Majorana neutrinos in the ISS, the branching ratios are predicted at similar rates, $\sim 10^{-9}$, than the other LFV three lepton tau decays. Within this setting, we can bind the LNV couplings shown in Table \ref{tabla: wrong_sign}, which were not restricted in ref. \cite{Enrique Fernandez}.\\
The mean values for the heavy neutrino masses from the studies in the previous section differ only slightly from the `Wrong Sign' analysis, %$\sim3\%$ 
\textcolor{black}{$\sim0.12\%$} in all cases.
\begin{figure}[!ht]
     \centering
     \includegraphics[scale=0.5]{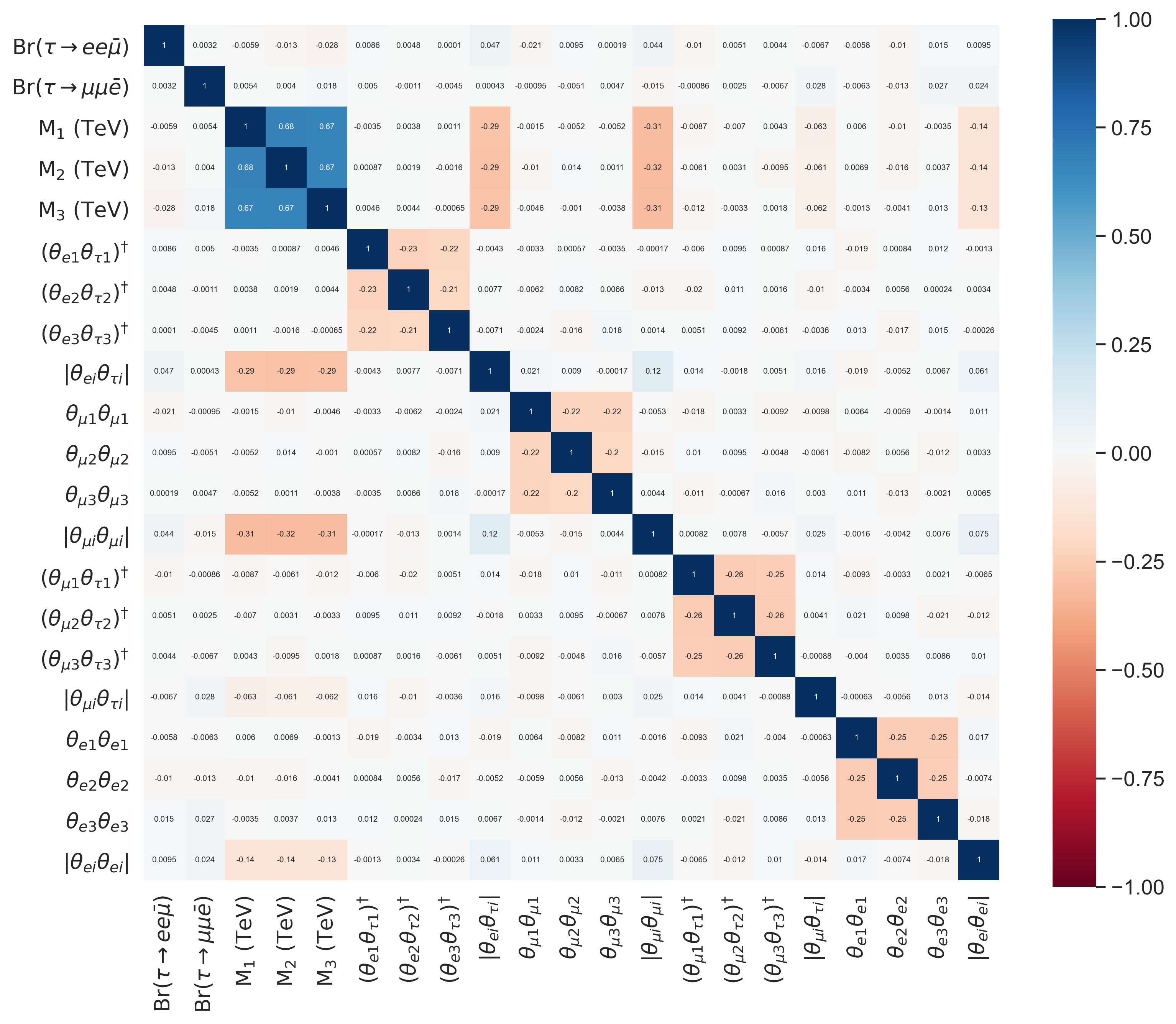}
     \caption{Heat map that stands for a correlation matrix among wrong sign branching ratios and free parameters.}
     \label{fig:wrong_sign}
 \end{figure}

\clearpage
\section{Conclusions}\label{sec:Concl}
The richness of the Littlest Higgs Model with T parity, LHT, allows for understanding light neutrino mass values via an ISS mechanism of Type I. As a consequence, there appears one heavy Majorana neutrino per family, with $\mathcal{O}\textcolor{black}{(10}$ TeV) mass. In this way, this new mass scale is of the order of \textcolor{black}{$4\pi$} times the vacuum expectation value associated to the collective spontaneous symmetry breakdown of the LHT, that produces T odd particles in the TeV region. We have focused in this work in the new contributions given by the heavy Majorana neutrinos to several LFV processes. Our results are encouraging:
\begin{itemize}
\item In all $\tau\to\ell\ell'\bar{\ell}''$ (including wrong-sign) decays and in $\mu\to e$ conversion in Ti, the mean values of our simulated events satisfying all present bounds are only one order of magnitude smaller than current limits. In $\mu\to ee\bar{e}$, $Z\to\bar{\tau}\ell$ and conversion in Au, our mean values are around two orders of magnitude smaller than current limits (only $Z\to\bar{\mu}e$ does not have the potential of probing our results in the near future).
\item The pattern of correlations among processes is completely different to the `traditional' LHT (without heavy Majorana neutrinos), where for instance wrong-sign decays are negligible. It should also be noted that the correlation between $L\to\ell\gamma$ and $L\to\ell\ell'\bar{\ell}''$ decays, which is a celebrated signature distinguishing underlying models producing the LFV, here is broken, as the former decays depend only on the charged current mixings $\theta \theta^\dagger$ and the neutral current ones also on the neutral current admixtures, $\theta S \theta^\dagger$, which reduces sizeably the correlation among both decay modes. Only within the LHT, upon eventual discovery of LFV in charged leptons in several processes, correlations among them would immediately distinguish the usual scenario  \cite{delAguila:2019htj} from the one studied here. If both heavy Majorana neutrinos and T-odd particles (both with $\mathcal{O}$(TeV) masses)  contributed commensurately, a general analysis (that we plan to undertake in future work) would be needed.
\end{itemize}
\section*{Acknowledgements}
We thank J. I. Illana, G. Hernández Tomé and M. A. Arroyo Ureña for useful discussions on this topic. We acknowledge J.~I.~Illana and J.~M.~Pérez Poyatos for their remarks to improve our manuscript \textcolor{black}{and the referee for pointing out  that our initial heavy neutrino masses were too light for the consistency of the model}. P.~R.~ thanks Swagato Banerjee for stressing the interest of wrong-sign LFV tau decays. I. P. acknowledges Conacyt funding  his Ph. D and P. R. the financial support of Cátedras Marcos Moshinsky (Fundación Marcos Moshinsky).

\section*{Appendix: Loop functions}

\subsection*{ Two-point Functions}
\label{app:A}
Considering a diagram with two legs, the general form of the loop integral is \cite{delAguila:2008zu}
\begin{equation}
    \frac{i}{16\pi^{2}}\{B_{0},B^{\mu} \} = \mu^{4-D}\int \frac{d^{D}q}{(2\pi)^{D}} \frac{\{ 1,q^{\mu} \}}{(q^{2}-m_{0}^{2})[(q+p)^{2} - m_{1}^{2}]},
    \label{A01}
\end{equation}
where $m_{0}$ and $m_{1}$ are the internal masses. The corresponding tensor coefficients are functions of the invariant quantities $(args) = (p^{2},m_{0}^{2},m_{1}^{2})$, where $p$ is the momentum of the particle. The functions $B \equiv B(0;M_{1}^{2},M_{2}^{2})$ and $\overline{B} \equiv B(0;M_{2}^{2},M_{1}^{2})$ read
\begin{equation}
    B_{0} = \overline{B}_{0} = \Delta_{\epsilon}+1-\frac{M_{1}^{2}\ln\frac{M_{1}^{2}}{\mu^{2}}-M_{2}^{2}\ln\frac{M_{2}^{2}}{\mu^{2}}}{M_{1}^{2}-M_{2}^{2}},
    \label{A02}
\end{equation}
\begin{eqnarray}
    %\begin{split}
        B_{1} & = &-\frac{\Delta_{\epsilon}}{2} + \frac{4M_{1}^{2}M_{2}^{2}-3M_{1}^{4}-M_{2}^{4}+2M_{1}^{4}\ln\frac{M_{1}^{2}}{\mu^{2}}+2M_{2}^{2}(M_{2}^{2}-2M_{1}^{2})\ln\frac{M_{2}^{2}}{\mu^{2}}}{4(M_{1}^{2}-M_{2}^{2})^{2}} \nonumber \\
        & = & -\overline{B}_{0}-\overline{B}_{1}, \label{B03}
    %\end{split}
\end{eqnarray}
with $\Delta_{\epsilon} \equiv \frac{2}{\epsilon}-\gamma + \ln 4\pi$. These functions are ultraviolet divergent in $D=4$ dimensions. 

\subsection*{Three-point Functions}
\label{app:B}
Appendix C of \cite{delAguila:2008zu} shows the three-point functions that we used. The function's arguments are $(args) = (p_{1}^{2}, Q^{2}, p_{2}^{2}; m_{0}^{2}, m_{1}^{2}, m_{2}^{2})$, with $p_{1}$ and $p_{2}$ the external momenta, $m_{0}$ the mass propagator, $M_{1}$ and $M_{2}$ the masses of particles within the loop and $Q \equiv p_{2}-p_{1}$. Thus,
\begin{equation}
%\begin{split}
    \frac{i}{16\pi^{2}} = \{ C_{0},C^{\mu}, C^{\mu\nu} \}(args) = \\
    \mu^{4-D}\int \frac{d^{D}q}{(2\pi)^{D}}\frac{\{ 1,q^{\mu},q^{\mu}q^{\nu} \}
    }{(q^{2} - m_{0}^{2})[(q+p_{1})^{2} - m_{1}^{2}][(q+p_{2})^{2} - m_{2}^{2}]}.
%\end{split}
    \label{B01}
\end{equation}
The functions $C\equiv C(0,Q^{2},0; M_{1}^{2},M_{2}^{2},M_{2}^{2})$ with $x\equiv M_{2}^{2}/M_{1}^{2}$ are given by
\begin{equation}
    %\begin{split}
        C_{0} = \frac{1}{M_{2}^{2}} \left[ \frac{1-x+\ln x}{(1-x)^{2}} \right. \\
         \left. + \frac{Q^{2}}{M_{1}^{2}}\frac{-2-3x+6x^{2}-x^{3}-6x\ln x}{12x(1-x)^{4}} \right],
    %\end{split}
    \label{B02}
\end{equation}
\begin{equation}
    C_{1} = C_{2} = \frac{1}{M_{1}^{2}}\frac{-3+4x-x^{2}-2\ln x}{4(1-x)^{3}} + \mathcal{O}(Q^{4}),
    \label{B003}
\end{equation}
\begin{equation}
    C_{11} = C_{22} = 2C_{12} = \frac{1}{M_{1}^{2}}\frac{11-18x+9x^{2}-2x^{3}+6\ln x}{18(1-x)^{4}} + \mathcal{O}(Q^{4}),
    \label{B04}
\end{equation}
\begin{equation}
    C_{00} = -\frac{1}{2}B_{1} - \frac{Q^{2}}{M_{1}^{2}}\frac{11-18x+9x^{2}-2x^{3}+6\ln x}{72(1-x)^{4}} + \mathcal{O}(Q^{4}).
    \label{B05}
\end{equation}
Defining $\overline{C} \equiv C(0,Q^{2},0; M_{2}^{2},M_{1}^{2},M_{1}^{2})$,
\begin{equation}
    %\begin{split}
        \overline{C}_{0} = \frac{1}{M_{2}^{2}} \left[ \frac{-1+x-\ln x}{(1-x)^{2}}  + \frac{Q^{2}}{M_{1}^{2}}\frac{-1+x-3x^{2}-2x^{3}+6x^{2}\ln x}{12x(1-x)^{4}} \right] + \mathcal{O}(Q^{4}),
    %\end{split}
    \label{B06}
\end{equation}
\begin{equation}
    \overline{C}_{1} = \overline{C}_{2} = \frac{1}{M_{1}^{2}}\frac{1-4x+3x^{2}-2x^{2}\ln x}{4(1-x)^{3}},
    \label{B07}
\end{equation}
\begin{equation}
    \overline{C}_{11} = \overline{C}_{22} = 2\overline{C}_{12} = \frac{1}{M_{1}^{2}}\frac{-2+9x-18x^{2}+11x^{3}-6x^{3}\ln x}{18(1-x)^{4}},
    \label{B08}
\end{equation}
\begin{equation}
    \overline{C}_{00} = -\frac{1}{2}\overline{B}_{1} - \frac{Q^{2}}{M_{1}^{2}}\frac{-2+9x-18x^{2}+11x^{3}+6x^{3}\ln x}{72(1-x)^{4}} + \mathcal{O}(Q^{4}).
    \label{B09}
\end{equation}
It is important to note that functions $C_{00}$ y $\overline{C}_{00}$ are ultraviolet divergent in $D=4$ dimensions.\\
In the limit $Q^{2}=0$ the following useful relations among two and three point functions hold \begin{equation}
    \overline{B}_{1}+2\overline{C}_{00} = 0,
    \label{B10}
\end{equation}
\begin{equation}
    -\frac{1}{4}+\frac{1}{2}\overline{B}_{1}+C_{00}-\frac{x}{2}M_{1}^{2}C_{0} = 0,
    \label{B11}
\end{equation}
\begin{equation}
    -\frac{1}{2}+\overline{B}_{1}+6\overline{C}_{00}-xM_{1}^{2}\overline{C}_{0} = \Delta_{\epsilon}-\ln\frac{M_{1}^{2}}{\mu^{2}}. 
    \label{B12}
\end{equation}

\subsection*{Four-point Functions}
\label{app:C}
The functions that we used in our development are all ultraviolet finite
\begin{equation}
    \frac{i}{16\pi^{2}}\{D_{0},D^{\mu},D^{\mu\nu} \}(args) = \int \frac{d^{4}q}{(2\pi)^{4}} \frac{\{ 1,q^{\mu},q^{\mu}q^{\nu} \}}{(q^{2}-m_{0}^{2})[(q+k_{1})^{2} - m_{1}^{2}][(q+k_{2})^{2} - m_{2}^{2}][(q+k_{3})^{2} - m_{3}^{2}]},
    \label{C01}
\end{equation}
with $k_{j} = \sum_{i}^{j}p_{i}$ and $(args) = (p_{1}^{2},p_{2}^{2},p_{3}^{2},p_{4}^{2},(p_{1}+p_{2})^{2}, (p_{2}+p_{3})^{2};m_{0}^{2},m_{1}^{2},m_{2}^{2},m_{3}^{2})$. In the limit of zero external momenta, only the following integrals are relevant
\begin{equation}
    \frac{i}{16\pi^{2}}D_{0} = \int \frac{d^{4}q}{(2\pi)^{2}} \frac{1}{(q^{2}-m_{0}^{2})(q^{2}-m_{1}^{2})(q^{2}-m_{2}^{2})(q^{2}-m_{3}^{2})},
    \label{C02}
\end{equation}
\begin{equation}
    \frac{i}{16\pi^{2}}D_{00} = \frac{1}{4}\int \frac{d^{4}q}{(2\pi)^{2}} \frac{q^{2}}{(q^{2}-m_{0}^{2})(q^{2}-m_{1}^{2})(q^{2}-m_{2}^{2})(q^{2}-m_{3}^{2})}.
    \label{C03}
\end{equation}
In terms of the mass ratios $x = m_{1}^{2}/m_{0}^{2}$, $y = m_{2}^{2}/m_{0}^{2}$, $z = m_{3}^{2}/m_{0}^{2}$ the integrals above can be written as \cite{delAguila:2008zu, delAguila:2019htj}
\begin{equation}
    %\begin{split}
        d_{0}(x,y,z) \equiv m_{0}^{4}D_{0} = \left[ \frac{x\ln x}{(1-x)(x-y)(x-z)} - \frac{y\ln y}{(1-y)(x-y)(y-z)}  + \frac{z\ln z}{(1-z)(x-z)(y-z)} \right],
    %\end{split} 
    \label{C04}
\end{equation}
\begin{equation}
    %\begin{split}
        \widetilde{d}_{0}(x,y,z) \equiv 4m_{0}^{2}D_{00} = \left[ \frac{x^{2}\ln x}{(1-x)(x-y)(x-z)} - \frac{y^{2}\ln y}{(1-y)(x-y)(y-z)} + \frac{z^{2}\ln z}{(1-z)(x-z)(y-z)} \right],
    %\end{split} 
    \label{C05}
\end{equation}
\begin{equation}
        \widetilde{d}'_{0}(x,y,z) =  \frac{x^{2}\ln x}{(1-x)(x-y)(z-x)} + \frac{y^{2}\ln y}{(1-y)(x-y)(z-y)} + \frac{z^{2}\ln z}{(1-z)(x-z)(y-z)},
    \label{C06}
\end{equation}
with $\widetilde{d}(x,y) = \widetilde{d}'(x,y,1)$. For two equal masses $(m_{0}=m_{3})$ we get
\begin{equation}
        d_{0}(x,y) = -\left[ \frac{x\ln x}{(1-x)^{2}(x-y)} - \frac{y\ln y}{(1-y)^{2}(x-y)} + \frac{1}{(1-x)(1-y)} \right],
    \label{C07}
\end{equation}
\begin{equation}
        \widetilde{d}_{0}(x,y) =- \left[ \frac{x^{2}\ln x}{(1-x)^{2}(x-y)} - \frac{y^{2}\ln y}{(1-y)^{2}(x-y)} + \frac{1}{(1-x)(1-y)} \right].
    \label{C08}
\end{equation}

\subsection*{Light-Heavy Four-point Functions }
\label{app:D}
The form factors involved in the $\ell \rightarrow \ell' \ell'' \bar{\ell}'''$ decay are given by the eqs. %(\ref{eq214}), 
(\ref{eq215}) and (\ref{eq216}).
%\begin{equation}
%    F_{B}^{\nu_{i}^{l}\nu_{j}^{l}} = \frac{\alpha_{W}}{16 \pi M_{W}^{2} s_{W}^{2}} \sum_{i,j=1}^{3} \{ W_{\ell i}W^{\dagger}_{\ell' i} W_{\ell''' j} W_{\ell''  j}^{\dagger} + (\ell' \leftrightarrow \ell'') \} f_{B}^{l}(y_{i},y_{j}),
 %   \label{eqD1}
%\end{equation}
%\begin{equation}
%    F_{B}^{\nu_{i}^{l}\chi_{j}^{h}} = \frac{\alpha_{W}}{16 \pi M_{W}^{2} s_{W}^{2}} \sum_{i=1}^{3}\sum_{j=7}^{9} \{ W_{\ell i}W^{*}_{\ell' i} \theta^{\dagger}_{\ell''' j} \theta_{\ell''  j} + (\ell' \leftrightarrow \ell'') \} f_{B}^{lh}(y_{i},x_{j}),
 %   \label{eqD2}
%\end{equation}
%\begin{equation}
%    F_{B}^{\chi_{i}^{h}\chi_{j}^{h}} = \frac{\alpha_{W}}{16 \pi M_{W}^{2} s_{W}^{2}} \sum_{i,j=7}^{9} \{ \theta^{\dagger}_{\ell i}\theta_{\ell' i} \theta^{\dagger}_{\ell''' j} \theta_{\ell''  j} + (\ell' \leftrightarrow \ell'') \} f_{B}^{h}(x_{i},x_{j}),
 %   \label{eqD3}
%\end{equation}
%where we have added the superscripts $(l)$, $(lh)$ and $(h)$ to the $f_{B}$ functions to indicate that these functions are composed of light-light, light-heavy and heavy-heavy neutrinos, respectively.\\
Since the masses of light neutrinos satisfy $m_{i} \ll M_{W}$, we find convenient to define the $y_{i}$ variable as $y_{i} = \frac{m_{i}^{2}}{M_{W}^{2}}$, so that $y_{i} \rightarrow 0$. On the other hand, heavy neutrino masses satisfy $M_{W} \ll M_{j}$, hence it is natural to define $x_{j} = \frac{M_{W}^{2}}{M_{j}^{2}} $, for the $x_{j}$ variable to fulfill $x_{j} \rightarrow 0$.\\ \\
The $f_{B}^{l}(y_{i},y_{j})$ function  is formed by the $d_{0}$ and $\bar{d}_{0}$ functions. As just light neutrinos are considered in the $f_{B}$ function, the $d_{0}$ and $\bar{d}_{0}$ ones have $y_{i,j}$ as variables. Therefore, $d_{0}$ and $\bar{d}_{0}$ functions are given from the eqs. (\ref{C07}) and (\ref{C08}). Thus, the $f_{B}^{l}(y_{i},y_{i})$ function can be written 
\begin{equation}
       f_{B}^{l}(y_{i},y_{j}) =   \left( 1 + \frac{1}{4}y_{i}y_{j} \right)\bar{d}_{0}(y_{i},y_{j}) - 2y_{i}y_{j}d_{0}(y_{i},y_{j}),
    \label{eqD4}
\end{equation}
%with $y_{i,j} = \frac{m_{i,j}^{2}}{M_{W}^{2}}$ $(i,j = 1,2,3)$.\\ \\

The $f_{B}^{lh}(y_{i},x_{j})$ function mixes light and heavy neutrinos, then it has $y_{i}$ and $x_{j}$ as variables. The $d_{0}$ and $\bar{d}_{0}$ functions defined in the previous section, have variables which behave as $\frac{m_{i}^{2}}{M_{W}^{2}}$, while for heavy neutrinos variables we have $x_{j} = \frac{M_{W}^{2}}{M_{j}^{2}}$, though. We have to refactor them considering the $y_{i}$ and $x_{j}$ variables for light and heavy neutrinos respectively, 
\begin{equation}
    d_{0}^{lh}(y_{i},x_{j}) = \frac{y_{i}x_{j} \ \mathrm{ln}y_{i}}{(1-y_{i})^{2}(1-y_{i}x_{j})} + \frac{x_{j}^{2} \  \mathrm{ln}x_{j}}{(1-x_{j})^{2}(1-y_{i}x_{j})} + \frac{x_{j}}{(1-y_{i})(1-x_{j})},
    \label{eqD5}
\end{equation}
\begin{equation}
    \bar{d}_{0}^{lh}(y_{i},x_{j}) = \frac{y_{i}^{2}x_{j} \ \mathrm{ln}y_{i}}{(1-y_{i})^{2}(1-y_{i}x_{j})} + \frac{x_{j} \ \mathrm{ln}x_{j}}{(1-x_{j})^{2}(1-y_{i}x_{j})} + \frac{x_{j}}{(1-y_{i})(1-x_{j})},
    \label{eqD6}
\end{equation}
where $y_{i} = m_{i}^{2}/M_{W}^{2}$ $(i = 1,2,3)$ and $x_{j} = M_{W}^{2} / M_{j}^{2}$ $(j = 1,2,3)$. From the equations above the $f_{B}^{lh} (y_{i}, x_{j})$ function reads 
\begin{equation}
        f_{B}^{lh}(y_{i},x_{j}) =   \left( 1 + \frac{1}{4} \frac{y_{i}}{x_{j}} \right)\bar{d}^{lh}_{0}(y_{i},x_{j}) - 2\frac{y_{i}}{x_{j}}d^{lh}_{0}(y_{i},x_{j}).
    \label{eqD7}
\end{equation}
Finally, the $f_{B}^{h}(x_{i}, x_{j})$ function just has heavy neutrino variables $x_{i,j} = M_{W}^{2} / M_{j}^{2}$, hence, we need to refactor the $d_{0}$ and $\bar{d}_{0}$ functions as
\begin{equation}
    d_{0}^{h}(x_{i},x_{j}) = - \left[ \frac{x_{i}^{2}x_{j} \ \mathrm{ln}x_{i}}{(1-x_{i})^{2}(x_{i}-x_{j})} - \frac{x_{i}x_{j}^{2} \ \mathrm{ln}x_{j}}{(1-x_{j})^{2}(x_{i}-x_{j})} + \frac{x_{i}x_{j}}{(1-x_{i})(1-x_{j})} \right],
    \label{eqD8}
\end{equation}
\begin{equation}
    \bar{d}_{0}^{h}(x_{i},x_{j}) = - \left[ \frac{x_{i}x_{j} \ \mathrm{ln}x_{i}}{(1-x_{i})^{2}(x_{i}-x_{j})} - \frac{x_{i}x_{j} \ \mathrm{ln}x_{j}}{(1-x_{j})^{2}(x_{i}-x_{j})} + \frac{x_{i}x_{j}}{(1-x_{i})(1-x_{j})} \right],
    \label{eqD9}
\end{equation}
with $i,j = 1,2,3$. Therefore, the $f_{B}^{h}(x_{i},x_{j})$ function is given by
\begin{equation}
        f_{B}^{h}(x_{i},x_{j}) =   \left( 1 + \frac{1}{4} \frac{1}{x_{i}x_{j}} \right)\bar{d}^{h}_{0}(x_{i},x_{j}) - 2\frac{1}{x_{i}x_{j}}d^{h}_{0}(x_{i},x_{j}).
    \label{eqD10}
\end{equation}
For the functions with a pair of LNV vertices $f_{B}^{(l,lh,h)-LNV}(z_{i},z_{j})$, %(eq. (\ref{eq235}))
we can apply the same arguments as the previous $f_{B}^{l,lh,h}(z_{i},z_{j})$. Therefore
\begin{eqnarray}
     %\begin{split}
         f_{B}^{l-LNV}(y_{i},y_{j}) & = & \sqrt{y_{i}y_{j}} \left( 2 \bar{d}_{0}(y_{i},y_{j}) - (4+y_{i}y_{j})d_{0}(y_{i},y_{j}) \right),\nonumber\\
 f_{B}^{lh-LNV}(y_{i},x_{j}) & = & \sqrt{\frac{y_{i}}{x_{j}}} \left( 2 \bar{d}^{lh}_{0}(y_{i},x_{j}) - (4+\frac{y_{i}}{x_{j}})d^{lh}_{0}(y_{i},x_{j}) \right),\nonumber\\
   f_{B}^{h-LNV}(x_{i},x_{j}) & = & \frac{1}{\sqrt{x_{i}x_{j}}} \left( 2 \bar{d}^{h}_{0}(x_{i},x_{j}) - \left(4+\frac{1}{x_{i}x_{j}} \right)d^{h}_{0}(x_{i},x_{j}) \right).\;\;%,
     %\end{split}
     \label{eqD11}
\end{eqnarray}
%with $y_{i} = m_{i}^{2} / M_{W}^{2}$, $m_{i}$ the mass of light neutrinos; $x_{j} = M_{W}^{2} / m_{j}^{2}$, $m_{j}$ the mass of heavy neutrinos. 


\begin{thebibliography}{}
\bibitem{ATLAS:2012yve}
G.~Aad \textit{et al.} [ATLAS],
%``Observation of a new particle in the search for the Standard Model Higgs boson with the ATLAS detector at the LHC,''
Phys. Lett. B \textbf{716} (2012), 1-29.
%doi:10.1016/j.physletb.2012.08.020
%[arXiv:1207.7214 [hep-ex]].

\bibitem{CMS:2012qbp}
S.~Chatrchyan \textit{et al.} [CMS],
%``Observation of a New Boson at a Mass of 125 GeV with the CMS Experiment at the LHC,''
Phys. Lett. B \textbf{716} (2012), 30-61.
%doi:10.1016/j.physletb.2012.08.021
%[arXiv:1207.7235 [hep-ex]].

\bibitem{ATLAS:2016neq}
G.~Aad \textit{et al.} [ATLAS and CMS],
%``Measurements of the Higgs boson production and decay rates and constraints on its couplings from a combined ATLAS and CMS analysis of the LHC pp collision data at $ \sqrt{s}=7 $ and 8 TeV,''
JHEP \textbf{08} (2016), 045.
%doi:10.1007/JHEP08(2016)045
%[arXiv:1606.02266 [hep-ex]].

\bibitem{CMS:2018uag}
A.~M.~Sirunyan \textit{et al.} [CMS],
%``Combined measurements of Higgs boson couplings in proton\textendash{}proton collisions at $\sqrt{s}=13\,\text {Te}\text {V} $,''
Eur. Phys. J. C \textbf{79} (2019) no.5, 421.
%doi:10.1140/epjc/s10052-019-6909-y
%[arXiv:1809.10733 [hep-ex]].

\bibitem{ATLAS:2019nkf}
G.~Aad \textit{et al.} [ATLAS],
%``Combined measurements of Higgs boson production and decay using up to $80$ fb$^{-1}$ of proton-proton collision data at $\sqrt{s}=$ 13 TeV collected with the ATLAS experiment,''
Phys. Rev. D \textbf{101} (2020) no.1, 012002.
%doi:10.1103/PhysRevD.101.012002
%[arXiv:1909.02845 [hep-ex]].

\bibitem{ParticleDataGroup:2020ssz}
P.~A.~Zyla \textit{et al.} [Particle Data Group],
%``Review of Particle Physics,''
PTEP \textbf{2020} (2020) no.8, 083C01.
%doi:10.1093/ptep/ptaa104

\bibitem{HFLAV:2019otj}
Y.~S.~Amhis \textit{et al.} [HFLAV],
%``Averages of b-hadron, c-hadron, and $\tau $-lepton properties as of 2018,''
Eur. Phys. J. C \textbf{81} (2021) no.3, 226.
%doi:10.1140/epjc/s10052-020-8156-7
%[arXiv:1909.12524 [hep-ex]].

\bibitem{Glashow:1961tr}
S.~L.~Glashow,
%``Partial Symmetries of Weak Interactions,''
Nucl. Phys. \textbf{22} (1961), 579-588.
%doi:10.1016/0029-5582(61)90469-2

\bibitem{Weinberg:1967tq}
S.~Weinberg,
%``A Model of Leptons,''
Phys. Rev. Lett. \textbf{19} (1967), 1264-1266.
%doi:10.1103/PhysRevLett.19.1264

\bibitem{Salam:1968rm}
A.~Salam,
%``Weak and Electromagnetic Interactions,''
Conf. Proc. C \textbf{680519} (1968), 367-377.
%doi:10.1142/9789812795915\_0034

\bibitem{Arkani-Hamed:2002ikv}
N.~Arkani-Hamed, A.~G.~Cohen, E.~Katz and A.~E.~Nelson,
%``The Littlest Higgs,''
JHEP \textbf{07} (2002), 034.
%doi:10.1088/1126-6708/2002/07/034
%[arXiv:hep-ph/0206021 [hep-ph]].

\bibitem{Schmaltz:2005ky}
M.~Schmaltz and D.~Tucker-Smith,
%``Little Higgs review,''
Ann. Rev. Nucl. Part. Sci. \textbf{55} (2005), 229-270.
%doi:10.1146/annurev.nucl.55.090704.151502
%[arXiv:hep-ph/0502182 [hep-ph]].

\bibitem{Perelstein:2005ka}
M.~Perelstein,
%``Little Higgs models and their phenomenology,''
Prog. Part. Nucl. Phys. \textbf{58} (2007), 247-291.
%doi:10.1016/j.ppnp.2006.04.001
%[arXiv:hep-ph/0512128 [hep-ph]].

\bibitem{Panico:2015jxa}
G.~Panico and A.~Wulzer,
%``The Composite Nambu-Goldstone Higgs,''
Lect. Notes Phys. \textbf{913} (2016), pp.1-316.
%doi:10.1007/978-3-319-22617-0
%[arXiv:1506.01961 [hep-ph]].

\bibitem{Arkani-Hamed:2001kyx}
N.~Arkani-Hamed, A.~G.~Cohen and H.~Georgi,
%``(De)constructing dimensions,''
Phys. Rev. Lett. \textbf{86} (2001), 4757-4761.
%doi:10.1103/PhysRevLett.86.4757
%[arXiv:hep-th/0104005 [hep-th]].

\bibitem{Arkani-Hamed:2001nha}
N.~Arkani-Hamed, A.~G.~Cohen and H.~Georgi,
%``Electroweak symmetry breaking from dimensional deconstruction,''
Phys. Lett. B \textbf{513} (2001), 232-240.
%doi:10.1016/S0370-2693(01)00741-9
%[arXiv:hep-ph/0105239 [hep-ph]].

\bibitem{Cheng:2003ju}
H.~C.~Cheng and I.~Low,
%``TeV symmetry and the little hierarchy problem,''
JHEP \textbf{09} (2003), 051.
%doi:10.1088/1126-6708/2003/09/051
%[arXiv:hep-ph/0308199 [hep-ph]].

\bibitem{Cheng:2004yc}
H.~C.~Cheng and I.~Low,
%``Little hierarchy, little Higgses, and a little symmetry,''
JHEP \textbf{08} (2004), 061.
%doi:10.1088/1126-6708/2004/08/061
%[arXiv:hep-ph/0405243 [hep-ph]].

\bibitem{Low:2004xc}
I.~Low,
%``T parity and the littlest Higgs,''
JHEP \textbf{10} (2004), 067.
%doi:10.1088/1126-6708/2004/10/067
%[arXiv:hep-ph/0409025 [hep-ph]].

\bibitem{Cheng:2005as}
H.~C.~Cheng, I.~Low and L.~T.~Wang,
%``Top partners in little Higgs theories with T-parity,''
Phys. Rev. D \textbf{74} (2006), 055001.
%doi:10.1103/PhysRevD.74.055001
%[arXiv:hep-ph/0510225 [hep-ph]].

\bibitem{Hubisz:2004ft}
J.~Hubisz and P.~Meade,
%``Phenomenology of the littlest Higgs with T-parity,''
Phys. Rev. D \textbf{71} (2005), 035016.
%doi:10.1103/PhysRevD.71.035016
%[arXiv:hep-ph/0411264 [hep-ph]].

\bibitem{Hubisz:2005tx}
J.~Hubisz, P.~Meade, A.~Noble and M.~Perelstein,
%``Electroweak precision constraints on the littlest Higgs model with T parity,''
JHEP \textbf{01} (2006), 135.
%doi:10.1088/1126-6708/2006/01/135
%[arXiv:hep-ph/0506042 [hep-ph]].

\bibitem{Hubisz:2005bd}
J.~Hubisz, S.~J.~Lee and G.~Paz,
%``The Flavor of a little Higgs with T-parity,''
JHEP \textbf{06} (2006), 041.
%doi:10.1088/1126-6708/2006/06/041
%[arXiv:hep-ph/0512169 [hep-ph]].

\bibitem{Chen:2006cs}
C.~R.~Chen, K.~Tobe and C.~P.~Yuan,
%``Higgs boson production and decay in little Higgs models with T-parity,''
Phys. Lett. B \textbf{640} (2006), 263-271.
%doi:10.1016/j.physletb.2006.07.053
%[arXiv:hep-ph/0602211 [hep-ph]].

\bibitem{Blanke:2006sb}
M.~Blanke, A.~J.~Buras, A.~Poschenrieder, C.~Tarantino, S.~Uhlig and A.~Weiler,
%``Particle-Antiparticle Mixing, epsilon(K), Delta Gamma(q), A**q(SL), A(CP) (B(d) ---\ensuremath{>} psi K(S)), A(CP) (B(s) ---\ensuremath{>} psi phi) and B ---\ensuremath{>} X(s,d gamma) in the Littlest Higgs Model with T-Parity,''
JHEP \textbf{12} (2006), 003.
%doi:10.1088/1126-6708/2006/12/003
%[arXiv:hep-ph/0605214 [hep-ph]].

\bibitem{Buras:2006wk}
A.~J.~Buras, A.~Poschenrieder, S.~Uhlig and W.~A.~Bardeen,
%``Rare $K$ and $B$ Decays in the Littlest Higgs Model without $T^-$ Parity,''
JHEP \textbf{11} (2006), 062.
%doi:10.1088/1126-6708/2006/11/062
%[arXiv:hep-ph/0607189 [hep-ph]].

\bibitem{Belyaev:2006jh}
A.~Belyaev, C.~R.~Chen, K.~Tobe and C.~P.~Yuan,
%``Phenomenology of littlest Higgs model with $T^-$ parity: including effects of $T^-$ odd fermions,''
Phys. Rev. D \textbf{74} (2006), 115020.
%doi:10.1103/PhysRevD.74.115020
%[arXiv:hep-ph/0609179 [hep-ph]].

\bibitem{Blanke:2006eb}
M.~Blanke, A.~J.~Buras, A.~Poschenrieder, S.~Recksiegel, C.~Tarantino, S.~Uhlig and A.~Weiler,
%``Rare and CP-Violating $K$ and $B$ Decays in the Littlest Higgs Model with $T^-$ Parity,''
JHEP \textbf{01} (2007), 066
%doi:10.1088/1126-6708/2007/01/066
%[arXiv:hep-ph/0610298 [hep-ph]].

\bibitem{Blanke:2007db}
M.~Blanke, A.~J.~Buras, B.~Duling, A.~Poschenrieder and C.~Tarantino,
%``Charged Lepton Flavour Violation and (g-2)(mu) in the Littlest Higgs Model with T-Parity: A Clear Distinction from Supersymmetry,''
JHEP \textbf{05} (2007), 013.
%doi:10.1088/1126-6708/2007/05/013
%[arXiv:hep-ph/0702136 [hep-ph]].

\bibitem{Hill:2007zv}
C.~T.~Hill and R.~J.~Hill,
%``$T^-$ Parity Violation by Anomalies,''
Phys. Rev. D \textbf{76} (2007), 115014.
%doi:10.1103/PhysRevD.76.115014
%[arXiv:0705.0697 [hep-ph]].

\bibitem{Goto:2008fj}
T.~Goto, Y.~Okada and Y.~Yamamoto,
%``Ultraviolet divergences of flavor changing amplitudes in the littlest Higgs model with T-parity,''
Phys. Lett. B \textbf{670} (2009), 378-382
%doi:10.1016/j.physletb.2008.11.022
%[arXiv:0809.4753 [hep-ph]].

\bibitem{delAguila:2008zu}
F.~del Aguila, J.~I.~Illana and M.~D.~Jenkins,
%``Precise limits from lepton flavour violating processes on the Littlest Higgs model with T-parity,''
JHEP \textbf{01} (2009), 080.
%doi:10.1088/1126-6708/2009/01/080
%[arXiv:0811.2891 [hep-ph]].

\bibitem{Blanke:2009am}
M.~Blanke, A.~J.~Buras, B.~Duling, S.~Recksiegel and C.~Tarantino,
%``FCNC Processes in the Littlest Higgs Model with T-Parity: a 2009 Look,''
Acta Phys. Polon. B \textbf{41} (2010), 657-683.
%[arXiv:0906.5454 [hep-ph]].

\bibitem{delAguila:2010nv}
F.~del Aguila, J.~I.~Illana and M.~D.~Jenkins,
%``Muon to electron conversion in the Littlest Higgs model with T-parity,''
JHEP \textbf{09} (2010), 040.
%doi:10.1007/JHEP09(2010)040
%[arXiv:1006.5914 [hep-ph]].

\bibitem{Goto:2010sn}
T.~Goto, Y.~Okada and Y.~Yamamoto,
%``Tau and muon lepton flavor violations in the littlest Higgs model with T-parity,''
Phys. Rev. D \textbf{83} (2011), 053011.
%doi:10.1103/PhysRevD.83.053011
%[arXiv:1012.4385 [hep-ph]].

\bibitem{Han:2013ic}
X.~F.~Han, L.~Wang, J.~M.~Yang and J.~Zhu,
%``Little Higgs theory confronted with the LHC Higgs data,''
Phys. Rev. D \textbf{87} (2013) no.5, 055004.
%doi:10.1103/PhysRevD.87.055004.
%[arXiv:1301.0090 [hep-ph]].

\bibitem{Yang:2013lpa}
B.~Yang, N.~Liu and J.~Han,
%``Top quark flavor-changing neutral-current decay to a 125 GeV Higgs boson in the littlest Higgs model with $T$ parity,''
Phys. Rev. D \textbf{89} (2014) no.3, 034020.
%doi:10.1103/PhysRevD.89.034020
%[arXiv:1308.4852 [hep-ph]].

\bibitem{Yang:2014mba}
B.~Yang, G.~Mi and N.~Liu,
%``Higgs couplings and Naturalness in the littlest Higgs model with T-parity at the LHC and TLEP,''
JHEP \textbf{10} (2014), 047.
%doi:10.1007/JHEP10(2014)047.
%[arXiv:1407.6123 [hep-ph]].

\bibitem{Blanke:2015wba}
M.~Blanke, A.~J.~Buras and S.~Recksiegel,
%``Quark flavour observables in the Littlest Higgs model with T-parity after LHC Run 1,''
Eur. Phys. J. C \textbf{76} (2016) no.4, 182.
%doi:10.1140/epjc/s10052-016-4019-7
%[arXiv:1507.06316 [hep-ph]].

\bibitem{Yang:2016hrh}
B.~Yang, J.~Han and N.~Liu,
%``Lepton flavor violating Higgs boson decay $h\rightarrow \mu\tau$ in the littlest Higgs model with $T$ parity,''
Phys. Rev. D \textbf{95} (2017) no.3, 035010.
%doi:10.1103/PhysRevD.95.035010
%[arXiv:1605.09248 [hep-ph]].

\bibitem{delAguila:2017ugt}
F.~del Aguila, L.~Ametller, J.~I.~Illana, J.~Santiago, P.~Talavera and R.~Vega-Morales,
%``Lepton Flavor Changing Higgs decays in the Littlest Higgs Model with T-parity,''
JHEP \textbf{08} (2017), 028
[erratum: JHEP \textbf{02} (2019), 047].
%doi:10.1007/JHEP08(2017)028
%[arXiv:1705.08827 [hep-ph]].

\bibitem{Husek:2020fru}
T.~Husek, K.~Mons\'alvez-Pozo and J.~Portol\'es,
%``Lepton-flavour violation in hadronic tau decays and $\mu-\tau$ conversion in nuclei,''
JHEP \textbf{01} (2021), 059.
%doi:10.1007/JHEP01(2021)059.
%[arXiv:2009.10428 [hep-ph]].

\bibitem{Cirigliano:2021img}
V.~Cirigliano, K.~Fuyuto, C.~Lee, E.~Mereghetti and B.~Yan,
%``Charged Lepton Flavor Violation at the EIC,''
JHEP \textbf{03} (2021), 256.
%doi:10.1007/JHEP03(2021)256
%[arXiv:2102.06176 [hep-ph]].

\bibitem{Dercks:2018hgz}
D.~Dercks, G.~Moortgat-Pick, J.~Reuter and S.~Y.~Shim,
%``The fate of the Littlest Higgs Model with T-parity under 13 TeV LHC Data,''
JHEP \textbf{05} (2018), 049.
%doi:10.1007/JHEP05(2018)049
%[arXiv:1801.06499 [hep-ph]].

\bibitem{delAguila:2019htj}
F.~del Aguila, L.~Ametller, J.~I.~Illana, J.~Santiago, P.~Talavera and R.~Vega-Morales,
%``The full lepton  avor of the littlest Higgs model with T-parity,''
JHEP \textbf{07} (2019), 154.
%doi:10.1007/JHEP07(2019)154
%[arXiv:1901.07058 [hep-ph]].

\bibitem{DelAguila:2019xec}
F.~Del Aguila, J.~I.~Illana, J.~M.~P\'erez-Poyatos and J.~Santiago,
%``Inverse see-saw neutrino masses in the Littlest Higgs model with T-parity,''
JHEP \textbf{12} (2019), 154.
%doi:10.1007/JHEP12(2019)154
%[arXiv:1910.09569 [hep-ph]].

\bibitem{Illana:2021uwu}
J.~I.~Illana and J.~M.~P\'erez-Poyatos,
%``A new and gauge-invariant Littlest Higgs model with T-parity,''
[arXiv:2103.17078 [hep-ph]].

\bibitem{Mohapatra:1986aw}
R.~N.~Mohapatra,
%``Mechanism for Understanding Small Neutrino Mass in Superstring Theories,''
Phys. Rev. Lett. \textbf{56} (1986), 561-563.
%doi:10.1103/PhysRevLett.56.561

\bibitem{Mohapatra:1986bd}
R.~N.~Mohapatra and J.~W.~F.~Valle,
%``Neutrino Mass and Baryon Number Nonconservation in Superstring Models,''
Phys. Rev. D \textbf{34} (1986), 1642.
%doi:10.1103/PhysRevD.34.1642

\bibitem{Bernabeu:1987gr}
J.~Bernabéu, A.~Santamaría, J.~Vidal, A.~Méndez and J.~W.~F.~Valle,
%``Lepton Flavor Nonconservation at High-Energies in a Superstring Inspired Standard Model,''
Phys. Lett. B \textbf{187} (1987), 303-308.
%doi:10.1016/0370-2693(87)91100-2

\bibitem{Ilakovac:1994kj}
A.~Ilakovac and A.~Pilaftsis,
%``Flavor violating charged lepton decays in seesaw-type models,''
Nucl. Phys. B \textbf{437} (1995), 491.
%doi:10.1016/0550-3213(94)00567-X
%[arXiv:hep-ph/9403398 [hep-ph]].

\bibitem{Illana:2000ic}
J.~I.~Illana and T.~Riemann,
%``Charged lepton flavor violation from massive neutrinos in Z decays,''
Phys. Rev. D \textbf{63} (2001), 053004.
%doi:10.1103/PhysRevD.63.053004
%[arXiv:hep-ph/0010193 [hep-ph]].

\bibitem{Hernandez-Tome:2019lkb}
G.~Hern\'andez-Tom\'e, J.~I.~Illana, M.~Masip, G.~L\'opez Castro and P.~Roig,
%``Effects of heavy Majorana neutrinos on lepton flavor violating processes,''
Phys. Rev. D \textbf{101} (2020) no.7, 075020.
%doi:10.1103/PhysRevD.101.075020
%[arXiv:1912.13327 [hep-ph]].

\bibitem{Petcov:1976ff}
S.~T.~Petcov,
%``The Processes $\mu \rightarrow e + \gamma, \mu \rightarrow e + \overline{e}, \nu' \rightarrow \nu + \gamma$ in the Weinberg-Salam Model with Neutrino Mixing,''
Sov. J. Nucl. Phys. \textbf{25} (1977), 340
[erratum: Sov. J. Nucl. Phys. \textbf{25} (1977), 698; erratum: Yad. Fiz. \textbf{25} (1977), 1336]
JINR-E2-10176.

\bibitem{Hernandez-Tome:2018fbq}
G.~Hern\'andez-Tom\'e, G.~L\'opez Castro and P.~Roig,
%``Flavor violating leptonic decays of $\tau$ and $\mu$ leptons in the Standard Model with massive neutrinos,''
Eur. Phys. J. C \textbf{79} (2019) no.1, 84
[erratum: Eur. Phys. J. C \textbf{80} (2020) no.5, 438].
%doi:10.1140/epjc/s10052-019-6563-4
%[arXiv:1807.06050 [hep-ph]].

\bibitem{Blackstone:2019njl}
P.~Blackstone, M.~Fael and E.~Passemar,
%``$\tau \rightarrow \mu \mu \mu $ at a rate of one out of $10^{14}$ tau decays?,''
Eur. Phys. J. C \textbf{80} (2020) no.6, 506.
%doi:10.1140/epjc/s10052-020-8059-7
%[arXiv:1912.09862 [hep-ph]].

\bibitem{Hernandez-Tome:2020lmh}
G.~Hern\'andez-Tom\'e, J.~I.~Illana and M.~Masip,
%``The $\rho$ parameter and $H^0\to \ell_i \ell_j$ in models with TeV sterile neutrinos,''
Phys. Rev. D \textbf{102} (2020) no.11, 113006.
%doi:10.1103/PhysRevD.102.113006
%[arXiv:2005.11234 [hep-ph]].


\bibitem{Lami:2016mjf}
A.~Lami and P.~Roig,
%``$H\to \ell\ell'$ in the simplest little Higgs model,''
Phys. Rev. D \textbf{94} (2016) no.5, 056001.
%doi:10.1103/PhysRevD.94.056001
%[arXiv:1603.09663 [hep-ph]].

\bibitem{Grimus:2000vj}
W.~Grimus and L.~Lavoura,
%``The Seesaw mechanism at arbitrary order: Disentangling the small scale from the large scale,''
JHEP \textbf{11} (2000), 042.
%doi:10.1088/1126-6708/2000/11/042
%[arXiv:hep-ph/0008179 [hep-ph]].

\bibitem{Hettmansperger:2011bt}
H.~Hettmansperger, M.~Lindner and W.~Rodejohann,
%``Phenomenological Consequences of sub-leading Terms in See-Saw Formulas,''
JHEP \textbf{04} (2011), 123.
%doi:10.1007/JHEP04(2011)123
%[arXiv:1102.3432 [hep-ph]].

\bibitem{Nomura:2018ktz}
T.~Nomura and H.~Okada,
%``Inverse seesaw model with large $SU(2)_L$ multiplets and natural mass hierarchy,''
Phys. Lett. B \textbf{792} (2019), 424-429.
%doi:10.1016/j.physletb.2019.04.006
%[arXiv:1809.06039 [hep-ph]].

\bibitem{Pontecorvo:1957qd}
B.~Pontecorvo,
%``Inverse beta processes and nonconservation of lepton charge,''
Zh. Eksp. Teor. Fiz. \textbf{34} (1957), 247.

\bibitem{Maki:1962mu}
Z.~Maki, M.~Nakagawa and S.~Sakata,
%``Remarks on the unified model of elementary particles,''
Prog. Theor. Phys. \textbf{28} (1962), 870-880.
%doi:10.1143/PTP.28.870

\bibitem{Hollik:1998vz}
W.~Hollik, J.~I.~Illana, S.~Rigolin, C.~Schappacher and D.~Stockinger,
%``Top dipole form-factors and loop induced CP violation in supersymmetry,''
Nucl. Phys. B \textbf{551} (1999), 3-40
[erratum: Nucl. Phys. B \textbf{557} (1999), 407-409].
%doi:10.1016/S0550-3213(99)00396-X
%[arXiv:hep-ph/9812298 [hep-ph]].

\bibitem{Bilenky:1977du}
S.~M.~Bilenky, S.~T.~Petcov and B.~Pontecorvo,
%``Lepton Mixing, mu --\ensuremath{>} e + gamma Decay and Neutrino Oscillations,''
Phys. Lett. B \textbf{67} (1977), 309.
%doi:10.1016/0370-2693(77)90379-3

\bibitem{Cheng:1977nv}
T.~P.~Cheng and L.~F.~Li,
%``Muon Number Nonconservation Effects in a Gauge Theory with V A Currents and Heavy Neutral Leptons,''
Phys. Rev. D \textbf{16} (1977), 1425.
%doi:10.1103/PhysRevD.16.1425

\bibitem{Arganda:2008jj}
E.~Arganda, M.~J.~Herrero and J.~Portol\'es,
%``Lepton flavour violating semileptonic tau decays in constrained MSSM-seesaw scenarios,''
JHEP \textbf{06} (2008), 079.
%doi:10.1088/1126-6708/2008/06/079
%[arXiv:0803.2039 [hep-ph]].

\bibitem{Celis:2013xja}
A.~Celis, V.~Cirigliano and E.~Passemar,
%``Lepton flavor violation in the Higgs sector and the role of hadronic $\tau$-lepton decays,''
Phys. Rev. D \textbf{89} (2014), 013008.
%doi:10.1103/PhysRevD.89.013008
%[arXiv:1309.3564 [hep-ph]].


\bibitem{Lami:2016vrs}
A.~Lami, J.~Portol\'es and P.~Roig,
%``Lepton flavor violation in hadronic decays of the tau lepton in the simplest little Higgs model,''
Phys. Rev. D \textbf{93} (2016) no.7, 076008.
%doi:10.1103/PhysRevD.93.076008
%[arXiv:1601.07391 [hep-ph]].

\bibitem{Belle:2021ysv}
A.~Abdesselam \textit{et al.} [Belle],
%``Search for lepton-flavor-violating tau decays to $\ell\gamma$ modes at Belle,''
[arXiv:2103.12994 [hep-ex]].

\bibitem{deBlas:2013gla}
J.~de Blas,
%``Electroweak limits on physics beyond the Standard Model,''
EPJ Web Conf. \textbf{60} (2013), 19008.
%doi:10.1051/epjconf/20136019008.
%[arXiv:1307.6173 [hep-ph]].

\bibitem{DeRomeri:2016gum}
V.~De Romeri, M.~J.~Herrero, X.~Marcano and F.~Scarcella,
%``Lepton flavor violating Z decays: A promising window to low scale seesaw neutrinos,''
Phys. Rev. D \textbf{95} (2017) no.7, 075028.
%doi:10.1103/PhysRevD.95.075028
%[arXiv:1607.05257 [hep-ph]].

\bibitem{R Kitano}
R.~Kitano, M.~Koike and Y.~Okada,
Phys. Rev. D \textbf{66} (2002), 096002.
Erratum: [Phys. Rev. D \textbf{76} (2007), 059902]

\bibitem{T. Suzuki}
T.~Suzuki, D.~F.~Measday and J.~F.~Roalsvig,
Phys. Rev.
C \textbf{35} (1987), 2212.

\bibitem{Calibbi:2021pyh}
L.~Calibbi, X.~Marcano and J.~Roy,
%``Z lepton flavour violation as a probe for new physics at future $e^+e^-$ colliders,''
[arXiv:2107.10273 [hep-ph]].

\bibitem{Enrique Fernandez}
Enrique~ Fern\'andez-Mart\'inez, Josu~Hern\'andez-Garc\'ia and Jacobo~L\'opez-Pav\'on, %Global constraints on heavy neutrino mixing
JHEP 1608 \textbf{033} (2016), 2212.

\end{thebibliography}
\end{document}